\documentclass{emulateapj}
\usepackage{lscape}

\newcommand\kms{\ifmmode{\rm km\thinspace s^{-1}}\else km\thinspace s$^{-1}$\fi}
\newcommand\hip{{\it Hipparcos\/}}

\shortauthors{Torres et al.}
\shorttitle{Capella}

\begin{document}

\journalinfo{Accepted for publication in The Astrophysical Journal}

\title{Binary orbit, physical properties, and evolutionary state of
Capella ($\alpha$ Aurigae)}

\author{
Guillermo Torres\altaffilmark{1},
Antonio Claret\altaffilmark{2},
Patrick A.\ Young\altaffilmark{3}
}

\altaffiltext{1}{Harvard-Smithsonian Center for Astrophysics, 60
Garden St., Cambridge, MA 02138, USA; e-mail: gtorres@cfa.harvard.edu}

\altaffiltext{2}{Instituto de Astrof\'\i sica de Andaluc\'\i a, CSIC,
Apartado 3004, 18080 Granada, Spain; e-mail: claret@iaa.es}

\altaffiltext{3}{School of Earth and Space Exploration, Arizona State
University, P.O.\ Box 871404, Tempe, AZ 85287; e-mail:
patrick.young.1@asu.edu}

\begin{abstract} 

We report extensive radial-velocity measurements of the two giant
components of the detached, 104-day period binary system of Capella.
Our highly accurate three-dimensional orbital solution based on all
existing spectroscopic and astrometric observations including our own
yields much improved masses for the primary and secondary of $2.466
\pm 0.018~M_{\sun}$ and $2.443 \pm 0.013~M_{\sun}$, with relative
errors of only 0.7\% and 0.5\%, respectively. The mass ratio is
considerably closer to unity than previously believed, which has an
impact on assessing the evolutionary state of the system.  Improved
values are presented also for the radii ($11.87 \pm 0.56~R_{\sun}$ and
$8.75 \pm 0.32~R_{\sun}$), effective temperatures ($4920 \pm 70$~K and
$5680 \pm 70$~K), and luminosities ($79.5 \pm 4.8~L_{\sun}$ and $72.1
\pm 3.6~L_{\sun}$). The distance is determined to be $13.042 \pm
0.028$~pc, based on the accurate orbital parallax. The projected
rotational velocities and individual rotation periods are also
known. Capella is unique among evolved stars in that, in addition to
all of the above, the chemical composition is known as well.  This
includes the overall metallicity [m/H], the carbon isotope ratio
$^{12}$C/$^{13}$C for the primary, and the lithium abundance and C/N
ratios for both components. We present new or revised values for some
of these.  The latter three quantities are sensitive diagnostics of
evolution, and change drastically for giants as a result of the
deepening of the convective envelope during the first dredge-up. The
secondary is crossing the Hertzprung gap, while the primary is
believed to be in the longer-lived core-helium burning phase.
Previous studies using only the masses, temperatures, and luminosities
have found good agreement with stellar evolution models placing the
primary in the clump. Here we compare all of the constraints
simultaneously against three sets of current models.  We find that
they are unable to match all of the observations for both components
at the same time, and at a single age, for any evolutionary state of
the primary. This shows the great importance of chemical information
for assessing the evolutionary state of giant stars. A comparison with
models of tidal evolution yields similarly disappointing results, when
tested against the fact that the orbit is circular, the primary is
rotating synchronously, the secondary $\sim$12 times faster than
synchronous, and the spin axes are apparently aligned with the axis of
the orbit. When confronted in detail, our understanding of the
advanced stages of stellar evolution is thus still very incomplete.

\end{abstract}


\section{Introduction}
\label{sec:introduction}

Our knowledge of stellar evolution relies heavily on constraints
provided by measurements of the physical properties of stars,
particularly the mass. Eclipsing binaries have been our main source of
accurate information for stars on the main sequence (masses, radii,
effective temperatures, luminosities).  The more advanced stages of
evolution, after the central hydrogen is exhausted, are much less well
understood than the main-sequence phase primarily because of poorer or
less stringent constraints. Accurate masses are rarely available for
normal giant stars, for practical reasons that are well known: it
requires them to be in binaries, and it requires the orbits to be wide
enough so that the components are detached (i.e., that they do not
interact with each other in the sense of mass transfer). This in turn
calls for sustained observations over long periods of time, often
several years.  The radial-velocity amplitudes are typically low,
making the masses more difficult to measure accurately.  Eclipses are
also much less likely, robbing us in most cases of the possibility of
determining the absolute radii directly.

While precious few normal giants or subgiants are known to be in
eclipsing binaries \citep[examples include AI~Phe, TZ~For, 
and the system OGLE 051019.64$-$685812.3 in the LMC;][]{Andersen:88,
Andersen.etal:91, Pietrzynski:09}, non-eclipsing systems
amenable to both spectroscopic and astrometric studies enabling
accurate dynamical mass determinations are somewhat more common.  A
very prominent example of this class of objects is Capella
($\alpha$~Aurigae, HD~34029, HR~1708, HIP~24608,
\ion{G8}{3}+\ion{G0}{3}, $P = 104$ days), the 6th brightest star in
the sky ($V = 0.07$), and the first binary for which an astrometric
orbit was determined interferometrically \citep{Anderson:20,
Merrill:22}. Although other astrometric measurements have been
gathered for Capella over the decades using a variety of techniques,
those pioneering observations with the original Michelson
interferometer on Mount Wilson have remained a critical component of
the astrometric solution of the orbit, which until about 15 years ago
contributed the dominant share of the uncertainty in the resulting
masses. At that time \cite{Hummel:94} used the Mark~III
interferometer, also on Mount Wilson, to obtain measurements that are
an order of magnitude more precise. As a result, our knowledge of the
masses is now limited by the best-available spectroscopy
\citep{Barlow:93} rather than the astrometry, a somewhat embarrassing
situation given the brightness of the object and its 100-year
observational history.  The masses are currently estimated to be $2.69
\pm 0.06~M_{\sun}$ and $2.56 \pm 0.04~M_{\sun}$ for the cooler primary
and the secondary, respectively. Although these are seemingly quite
precise ($\sim$2\% relative errors), the \emph{accuracy} is just as
important, especially for giants. The predicted properties of stars
from stellar evolution models in these rapid evolutionary stages are
extremely sensitive to mass, and a 2\% error makes a much larger
difference than on the main sequence.

Systematic errors in the radial velocity measurements for the
rapidly-rotating secondary of Capella are not easy to avoid, and as a
result there have been persistent concerns in the literature about the
accuracy of the mass ratio, which is very close to unity.  This
quantity has a significant impact on the precise evolutionary state
inferred for the components.  One of our motivations here is thus to
provide a new, high-quality set of spectroscopic observations that
more nearly matches the precision of the astrometric data of
\cite{Hummel:94}. We focus especially on the accuracy of the mass
determinations, for the purpose of comparing with state-of-the-art
stellar evolution models. Considerable effort is therefore invested in
the inter-comparison of all available astrometric and spectroscopic
observations, in order to properly understand the systematics.  An
additional goal of this work is to carry out a comprehensive analysis
of all the data relevant for the determination of the evolutionary
state of Capella, which has been a subject of debate for decades. The
secondary is clearly crossing the Hertzprung gap, but the primary has
been suggested to be either in the helium-burning clump or on the
first ascent of the giant branch, usually based on partial
information.  This system is perhaps unique in that, in addition to
the masses, effective temperatures, and luminosities that have been
used previously for that purpose, a wealth of other information is
available. This includes direct measurements of the angular diameters,
various activity indicators in the optical, ultraviolet, and X rays,
the projected rotational velocities as well as the rotation periods,
the overall metallicity, and particularly the surface lithium
abundance of both stars, the $^{12}$C/$^{13}$C isotope ratio for the
primary star, and the C/N abundance ratios.  Chemical indicators such
as these are crucial diagnostics of evolution because they change
significantly in the giant phase, mainly during the first dredge-up.
We bring all of these to bear here, for the first time. We wish to
examine the effects of convection prescriptions in the models (mixing
length, overshooting) as well as rotation, which has not previously
been considered for this system.  Furthermore, Capella is an important
point of comparison with tidal evolution theory for evolved stars.
This is because the primary has its rotation synchronized with the
orbital motion while the secondary rotates 12 times faster than
synchronous, despite the nearly identical masses, which are within 1\%
of each other.  Thus another goal of this work is to use our newly
derived accurate dimensions for the components to carry out a detailed
check of current tidal theories and gauge our understanding of these
processes.

Our new spectroscopic observations are presented in
\S\,\ref{sec:spectroscopy}, along with all historical radial-velocity
measurements. The astrometric observations are discussed in
\S\,\ref{sec:astrometry}, and include long-baseline interferometry,
speckle interferometry, and {\it Hipparcos\/} measurements. Our
simultaneous orbital solution for Capella based on these two kinds of
data is documented in \S\,\ref{sec:orbit}, with particular care given
to possible systematic errors that might bias the masses. In
\S\,\ref{sec:lightratio} and \S\,\ref{sec:angdiam} we collect all the
information available on the relative brightness of the components and
the angular diameters, which we use later to estimate effective
temperatures and absolute radii. The chemical composition of Capella
is discussed in \S\,\ref{sec:abundances}, and proves to be of critical
importance for the analysis. The physical properties of the stars are
then described in \S\,\ref{sec:dimensions}. We present a detailed
comparison of the absolute dimensions of Capella with stellar
evolution theory in \S\,\ref{sec:discussion}, focusing on the
determination of the evolutionary state of the stars. In the same
section we carry out tests of current tidal theories.  Finally,
\S\,\ref{sec:finalremarks} summarizes our main conclusions. Two
appendices collect notes of interest on the astrometric observations
for the benefit of future users, and a third contains a discussion of
the coronal abundances of Capella that support the photospheric
determinations.

\section{Spectroscopic observations}
\label{sec:spectroscopy}

The rich history of the radial-velocity measurements of Capella began
more than a century ago with the pioneering efforts by the Greenwich
and Potsdam astronomers \citep{Gill:91, Vogel:91}, and has been
recounted previously by other authors \citep[see, e.g.,][]{Barlow:93}.
More than a dozen semi-independent spectroscopic orbital solutions
have been reported over the decades, based on data sets of greatly
varying quality.  Despite the brightness of the object, the radial
velocity measurements of the rotationally broadened secondary
component are not particularly easy, and have not always been possible
in the past. As a result, there has been considerable debate about the
mass ratio \citep[see, e.g.,][]{Batten:91}, which our analysis
conclusively shows is near unity.  Nevertheless, some of these
historical data are still of value to improve the orbital period, so
we describe them in some detail below and make use of them later. Our
main observational contribution here is a large new set of
high-quality velocity measurements for both components that provides
more than a two-fold improvement in the precision of the velocity
semi-amplitude of the primary, and a four-fold improvement for the
secondary compared to the best existing determinations, by
\cite{Barlow:93}. These new data are described first.

\subsection{New radial-velocity measurements}
\label{sec:specnew}

Spectroscopic observations of Capella were conducted at the
Harvard-Smithsonian Center for Astrophysics (CfA) using the 1.5m Wyeth
reflector at the Oak Ridge Observatory (Harvard, Massachusetts),
beginning in 1996 October and continuing through 1999 November. An
echelle spectrograph with a photon-counting intensified Reticon
detector \citep[Digital Speedometer;][]{Latham:85, Latham:92} was used
to record a single 45\,\AA\ echelle order centered at a wavelength of
5188.5\,\AA, featuring the gravity-sensitive lines of the
\ion{Mg}{1}~b triplet. The resolving power provided by this setup is
$\lambda/\Delta\lambda\approx 35,\!000$. One additional observation
was gathered with a nearly identical system on the 1.5m Tillinghast
reflector at the F.\ L.\ Whipple Observatory (Mount Hopkins, Arizona).
Nominal signal-to-noise ratios per resolution element of 8.5~\kms\
range from about 20 to 90, although for values much higher than 50 the
limit is set by systematics from flat-fielding and not photon
noise. With the inclusion of two archival observations made in 1986
February and March with the instrument at Oak Ridge, the total number
of usable spectra is 162, collected over an interval of 13.8 years.

Radial velocities for both stars were derived using TODCOR, a
two-dimensional cross-correlation technique introduced by
\cite{Zucker:94}. This method uses two templates, one for each
component of the binary, which we selected from a large library of
synthetic spectra based on model atmospheres by R.\ L.\ Kurucz
\citep[see][]{Latham:02}. These templates have been calculated for a
wide range of effective temperatures ($T_{\rm eff}$), surface
gravities ($\log g$), rotational velocities ($v \sin i$ when seen in
projection), and metallicities ([m/H]). Following \cite{Torres:02} the
optimum templates for Capella were determined by means of extensive
grids of cross-correlations with TODCOR, seeking to maximize the
average correlation weighted by the strength of each exposure. The
surface gravities were held fixed at preliminary values of $\log g =
2.5$ and 3.0 for the primary and secondary, and the metallicity was
initially assumed to be solar. As a result of this optimization we
obtained by interpolation effective temperatures of $T_{\rm eff}^{\rm
A} = 4900 \pm 100$~K and $T_{\rm eff}^{\rm B} = 5710 \pm 100$~K for
the primary and secondary, respectively, along with projected
rotational velocities of $v_{\rm A} \sin i = 6.5 \pm 1.0$ \kms\ and
$v_{\rm B} \sin i = 36.0 \pm 1.5$ \kms. Strictly speaking, the latter
values are a measure of the total broadening of the spectral lines. We
discuss these measures and compare them with others in
\S\,\ref{sec:dimensions}.  We repeated these determinations assuming
metallicities of [m/H] $= -1.0$, $-$0.5, and +0.5, but we found the
average correlation values to be slightly lower than with solar
metallicity, indicating a poorer match to the observed spectra. The
templates adopted here are the ones in our library with parameters
nearest to the values above: $T_{\rm eff} = 5000$~K and 5750~K for the
primary and secondary, and rotational velocities of 6~\kms\ and
35~\kms, respectively. The radial velocities derived in this way have
internal errors averaging 0.5 \kms\ and 1.0 \kms\ for the primary and
secondary, but vary individually depending on the signal-to-noise
ratio.  In addition to the radial velocities, our spectra yield the
light ratio at the mean wavelength of our spectra, $\ell_{\rm
B}/\ell_{\rm A} = 1.48 \pm 0.05$. The difference in line blocking
between the components has been accounted for, so that this represents
a true flux ratio rather than a ratio between the continuum
levels. The hotter star is thus brighter at $\sim$5200\,\AA.

The stability of the zero-point of the CfA velocity system was
monitored by means of exposures of the dusk and dawn sky, and small
systematic run-to-run corrections were applied in the manner described
by \cite{Latham:92}. The zero point of the native CfA velocity system
based on synthetic templates is very close to the absolute frame as
defined by extensive observations of the minor planets in the solar
system. The correction required to place our radial velocities on this
absolute frame is $+0.139~\kms$ \citep{Stefanik:99, Latham:02}, and
has not been applied to the measurements listed below.

\begin{figure} 
\epsscale{1.4} 
{\hskip -0.3in \plotone{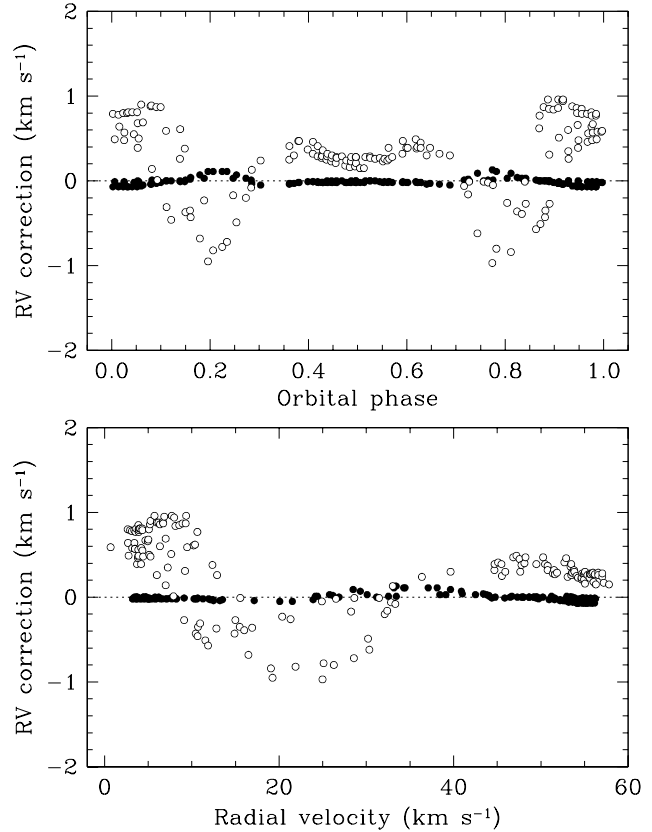}}
\vskip 0.5in 
\figcaption[]{Systematic errors in the raw TODCOR velocities for
Capella as a function of orbital phase and velocity. Filled circles
correspond to the primary, open circles to the secondary. These
differences have been applied to the raw velocities as corrections
(see text).\label{fig:corrections}}
\vskip 0.05in
\end{figure}

One of the main advantages of TODCOR compared to conventional
one-dimensional cross-correlation techniques is that it greatly
reduces the systematic errors in the radial velocities caused by line
blending that have affected many of the previous studies of Capella
(see \S\,\ref{sec:spechist}). Nevertheless, because we are concerned
in this work with the accuracy of the velocities as much as their
precision, we investigated possible systematic effects that may result
from residual blending in our narrow spectral window, and from shifts
of the spectral lines in and out of this window as a function of
orbital phase. Previous experience with similar material has shown
that these effects are sometimes significant, and must be examined on
a case-by-case basis and corrected if necessary \citep[see,
e.g.,][]{Torres:97, Torres:00}.  We performed numerical simulations as
described by \cite{Latham:96} to evaluate these effects. Briefly, we
generated synthetic composite spectra matching our observations by
combining the primary and secondary templates used above (including
rotational broadening), shifted to the appropriate velocities for each
of the exposures as predicted by a preliminary orbital solution, and
scaled to the observed light ratio. These synthetic observations were
then processed with TODCOR in exactly the same way as the real
spectra, and the resulting velocities were compared with the input
shifts. The differences are shown graphically in
Figure~\ref{fig:corrections}, as a function of both velocity and
orbital phase.  The systematic pattern is obvious, and the individual
differences can reach $\pm$1 \kms, which is relatively small in
absolute terms but significant compared to the internal errors. We
therefore applied these differences as corrections to the raw
velocities. The effect on the primary semi-amplitude $K_{\rm A}$ is
negligible, but the change in $K_{\rm B}$ is $\sim$0.5\%, which
translates into a non-negligible change in the derived masses of about
1\% for the primary and 0.6\% for the secondary. The final velocities
in the heliocentric frame are given in Table~\ref{tab:cfarvs}, and
include these corrections.  Similar adjustments based on the same
simulations were applied to the light ratio, and are already included
in the value reported above.

Preliminary single-lined orbital solutions performed separately on the
primary and secondary velocities indicated a slight difference in the
center-of-mass velocities, $\gamma$, of about $0.27 \pm 0.08$~\kms,
with the secondary value being lower. Primary/secondary velocity
differences considerably larger than this are not uncommon in studies
of double-lined eclipsing binaries. This difference $\Delta_{\rm AB}$
persisted in our global solution described later.  Although it is very
small in absolute terms (only about half of the typical uncertainty in
our primary velocities), it is statistically significant due to the
large number of observations in the fit.  Because it may affect the
absolute masses of Capella at some level, we have explored possible
reasons for this shift.  One is the differential gravitational
redshift between the stars, given that our synthetic templates do not
account for this. Estimates based on preliminary values of the masses
and radii of the components indicate that this effect is 0.046 \kms,
but it goes in the wrong direction to explain $\Delta_{\rm AB}$, i.e.,
the secondary redshift is \emph{larger}. It is also possible there are
shifts due to large-scale convective motions \citep{Schwarzschild:75,
Porter:00} that could be different in the two stars, but these are not
well characterized for giants. Given that the stars are observed to be
active, another possibility is the presence of spots on one or both
components (particularly on the rapidly rotating secondary), which can
affect the velocities. A perturbation of this nature was in fact
pointed out by \cite{Hummel:94} for their interferometric visibilities
of Capella (see \S\,\ref{sec:orbit}).  Unfortunately our time sampling
is inadequate to study this in more detail, but unless the spots are
very long-lived we would expect the effect to average out to some
extent over the interval of our observations. A fourth possibility
that cannot be completely rule out is template mismatch \citep[see,
e.g.,][]{Griffin:00}. We have made every effort here to use templates
that maximize the average correlation for all our spectra, and small
differences with the true values of $T_{\rm eff}$, $v \sin i$, $\log
g$ (which we estimate below to be $\log g = 2.68$ for the primary and
2.94 for the secondary) or metallicity compared to what we have
assumed should not have a significant effect on the velocities, in our
experience. However, line broadening from micro- or macro-turbulence
in Capella could be somewhat different from what is assumed in our
library of synthetic spectra (microturbulence $\xi_{\rm t} = 2$~\kms\
and macroturbulence $\zeta_{\rm RT} = 1.5$~\kms), although this is
unlikely to affect the secondary much due to the overwhelming effect
of rotational broadening in that star (36 \kms). We discuss this
effect further in \S\,\ref{sec:tidal} in connection with the accuracy
of the $v \sin i$ measurements. In the absence of a definitive
explanation, we have chosen below to correct for the primary/secondary
shift by solving for the offset and applying it to the secondary
velocities. Not correcting for the shift would affect the
semi-amplitudes at the level of 0.03\% for the primary and 0.14\% for
the secondary, and the final masses at the level of 0.31\% and 0.22\%,
which correspond to less than half of the formal uncertainties in our
final determination of those quantities (see
\S\,\ref{sec:dimensions}).

\begin{figure} 
\epsscale{1.4} 
{\hskip -0.35in \plotone{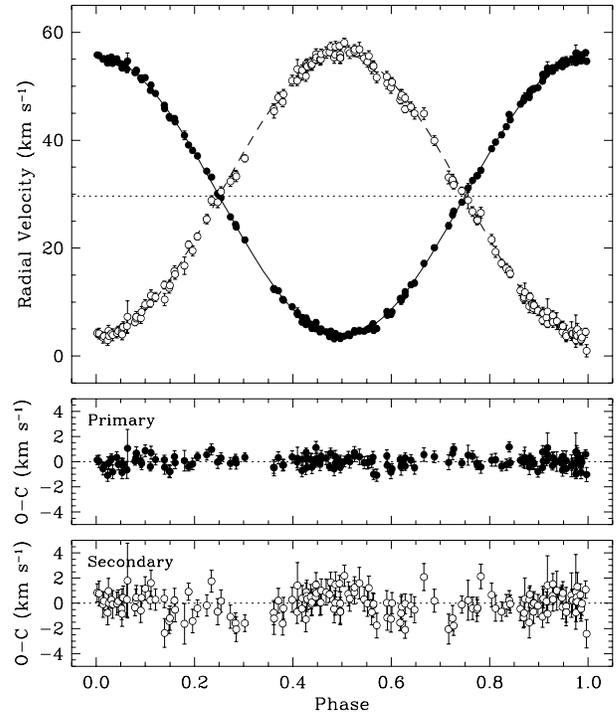}}
\vskip -0.1in 
\figcaption[]{CfA radial-velocity measurements for Capella (filled
circles for the primary, open for the secondary) along with the curves
computed from our combined orbital solution described in
\S\,\ref{sec:orbit}. The dotted line represents the velocity of the
center of mass. $O\!-\!C$ residuals for the primary and secondary are
shown in the bottom panels.\label{fig:cfaorb}}
\vskip 0.1in
\end{figure}

Figure~\ref{fig:cfaorb} displays the CfA observations along with an
orbital solution described later, as well as the $O\!-\!C$
residuals. Those of the secondary show a small residual pattern which
we believe to be of a similar nature as the $\Delta_{\rm AB}$ shift
discussed above. We return to this later in connection with the
orbital solution.

\subsection{Historical radial-velocity measurements}
\label{sec:spechist}

The discovery of the binary character of Capella was announced
independently by \cite{Campbell:99} and \cite{Newall:99} from
photographic spectra taken at Lick Observatory and Cambridge
Observatory (England), respectively. Both investigators noted the
composite nature of the spectrum, but published velocities only for
the component Newall referred to as being of ``solar type''. In our
nomenclature this is the cooler, slightly more massive star we refer
to as the ``primary'' (star A), which has relatively sharp lines.  The
other star (of ``Procyon type'', ``secondary'', or star B) has much
more diffuse lines.  \cite{Campbell:01} reported only that the
velocity of the secondary varies between $-3$ \kms\ and +63 \kms.  His
31 measurements for the primary are of excellent quality ($\sigma_{\rm
RV} \sim 0.8$ \kms), and were used by \cite{Reese:00} to establish the
first reliable spectroscopic orbit. The measurements by
\cite{Newall:00} are somewhat poorer ($\sigma_{\rm RV} \sim 2.1$
\kms), but still potentially useful.

The first published measurements of the secondary velocity appear to
be those by \cite{Goos:08}, who succeeded in detecting it in 19 of his
35 photographic plates taken with a 0.3m refractor at Bonn. All plates
yielded good measurements for the primary. Further velocities for both
stars were reported by \cite{Sanford:22} from Mount Wilson, and
\cite{Struve:39} from Babelsberg. \cite{Struve:53} published a further
series of velocities from Mount Wilson and Lick, though only for the
primary star. Measurements of the velocity \emph{difference} between
the components made at the Dominion Astrophysical Observatory (DAO)
were reported by \cite{Wright:54}, along with a careful study of the
secondary spectrum and the relative brightness of that star compared
to the primary. The velocity measurements are averages from plates
taken at similar orbital phases, so unique dates cannot be assigned to
them and for this reason we do not use these data here.  The
brightness ratio as well as the mass ratio estimate by
\cite{Wright:54}, derived by adopting the primary orbit from
\cite{Struve:53}, were quite influential over the following decades,
although the light ratio is now known to be incorrect (or at least
misleading; see \S\,\ref{sec:lightratio} and
\S\,\ref{sec:dimensions}). More recently \cite{Batten:75} published 18
velocities for the primary from plates obtained at DAO, most of which
were later re-measured by \cite{Batten:91}, superseding the original
determinations. Further measurements for both components obtained at
the Fick Observatory were reported by \cite{Shen:85}. These data were
in turn superseded and significantly expanded by \cite{Beavers:86},
who published the largest set of velocities for Capella aside from our
own. Additional measurements of the primary only were obtained by
\cite{Shcherbakov:90}, including a set based on photospheric lines and
another set from the chromospheric \ion{He}{1}~$\lambda$10830 line. We
do not use the latter because they may not correspond to the true
center of mass of the star, nor do we consider a similar list of
velocities for both components by \cite{Katsova:98}, also from the
\ion{He}{1}~$\lambda$10830 line. Finally, high-quality measurements
for both stars from the McDonald and Kitt Peak Observatories were
published by \cite{Barlow:93}.

The sources above represent the most important velocity data sets
published for Capella in the century since its discovery as a
binary. Even though some of them may have considerably larger
uncertainties (less weight) than the CfA velocities, in principle
there is no reason why they cannot be properly combined with ours to
strengthen the solution, which is our goal in \S\,\ref{sec:orbit}. A
number of smaller lists of less than half a dozen measurements each
have also appeared over the decades, but are ignored here for being
much less significant and more difficult to use because of the poorly
determined zero-point offsets. The richer sources are summarized in
Table~\ref{tab:historicalorbits}, where the last entry corresponds to
our own contribution.

\begin{figure} 
\epsscale{1.35} 
\vskip -0.7in
{\hskip -0.35in\plotone{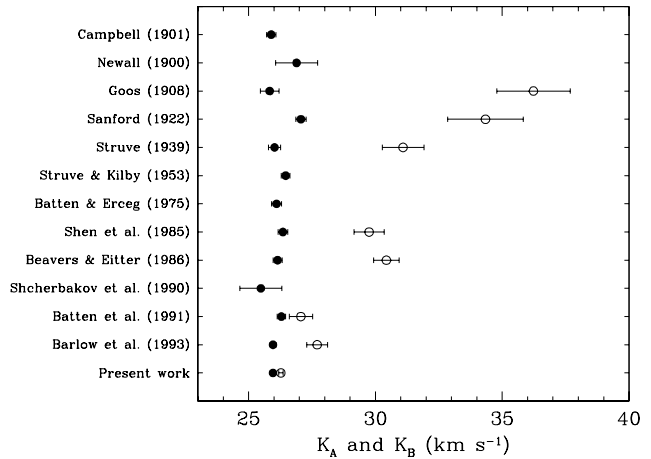}}
\vskip -0.65in 
\figcaption[]{Radial-velocity semi-amplitudes for the primary and
secondary of Capella (filled and open circles, respectively)
throughout the last century, based on our solutions in
Table~\ref{tab:historicalorbits}.\label{fig:k}}
\end{figure}

The potential usefulness of these historical data sets depends on
whether they can be shown to be sufficiently free from systematic
errors. To this end, we have examined each of the sources by computing
separate orbital solutions from the original velocities with the same
fitting code, and comparing them to one based on the CfA data. These
solutions can be found in Table~\ref{tab:historicalorbits}. We list
also the number of observations, their time span, and the
root-mean-square (RMS) residual from the fit in each case, which is
representative of the typical error of the velocities. Because of the
limited duration of some of these studies, the period has been held
fixed at the value $P = 104.022$ days determined from a preliminary
fit to our own observations, and the orbit has been assumed to be
circular.  The center-of-mass velocities $\gamma$ in the second column
show that there are occasional differences in the instrumental zero
points, although these can easily be corrected in a combined solution
by solving for the offsets simultaneously with the other adjustable
parameters. The same holds for the primary/secondary offsets listed in
the third column (see also \S\,\ref{sec:specnew}), which have been set
to zero when a preliminary fit indicated the shift was not
significant. We are more concerned here with the velocity
semi-amplitudes $K_{\rm A}$ and $K_{\rm B}$, which determine the
masses of the components. Excluding the two data sets that rely on
chromospheric lines \citep{Shcherbakov:90, Katsova:98}, the primary
semi-amplitudes from all the others agree reasonably well with
ours. The only exception is the data set by \cite{Sanford:22}, which
is also the smallest. The secondary semi-amplitudes, however, show
significant systematic differences with the CfA value of $K_{\rm B}$,
and if we restrict ourselves to velocities based on photospheric
lines, there appears to be a trend of decreasing amplitudes as a
function of time over the last century, leading up to our own
determination (see Figure~\ref{fig:k}).  We suspect these differences
have to do with systematic effects associated with line blending and
the difficulty of measuring the broad spectral features of the
secondary, particularly in the older studies, a problem that has been
pointed out repeatedly over the years. For this reason, we have chosen
not to use any of the historical secondary velocities here, relying
only on our own. The primary velocities, on the other hand, appear
reasonably free from systematics \citep[save those of][which we
exclude]{Sanford:22}, and add up to more than twice the number of our
own observations although the combined weight is actually $\sim$50\%
lower.

\begin{figure} 
\epsscale{1.35} 
\vskip -0.05in
{\hskip -0.25in \plotone{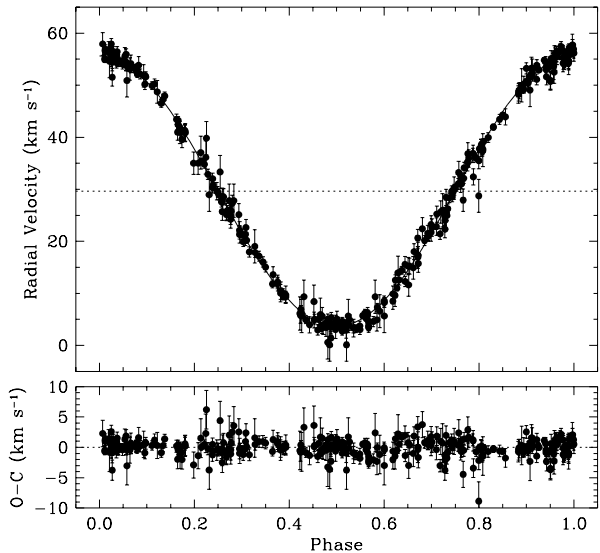}}
\vskip -0.8in 
\figcaption[]{Historical radial-velocity measurements for the primary
component of Capella, along with the curve computed from our combined
orbital solution described in \S\,\ref{sec:orbit}. Individual data
sets have been placed on the same zero-point as the CfA velocities by
applying the offsets described in \S\,\ref{sec:orbit}. The dotted line
represents the velocity of the center of mass. $O\!-\!C$ residuals are
shown in the bottom panel.\label{fig:historicalorb}}
\vskip 0.1in
\end{figure}

All of these velocities are listed in Table~\ref{tab:historicalrvs} on
their original scales (i.e., without the application of any offsets,
to be described below).\footnote{The velocities by \cite{Campbell:01}
used here include small adjustments later determined by
\cite{Campbell:28} to be required in order to place them on the scale
of the homogenized catalog of 1896--1921 Lick velocities they
published.  These adjustments are specific to the person who measured
the plates, and for this case are +0.1 \kms\ (Campbell) and $-0.3$
\kms\ (Wright).}  Individual uncertainties are described in
\S\,\ref{sec:orbit}.  These observations are shown graphically in
Figure~\ref{fig:historicalorb}, along with the same curve for the
primary from Figure~\ref{fig:cfaorb}.
	
\section{Astrometric observations}
\label{sec:astrometry}

Soon after the discovery that Capella is a spectroscopic binary, some
of the most skilled double-star observers of the day attempted to
resolve the pair at the times predicted from the spectroscopic orbit
to be the most favorable, but they were unsuccessful \citep{Hussey:00,
Aitken:00, Hussey:01}. At about the same time, an intriguing series of
visual measurements was made at the Greenwich Observatory that did
appear to barely resolve the object: the observers reported elongated
images with the 28-inch refractor.  Systematic measurements of the
position angle of the binary from these elongated images were carried
out over an interval of about two years, and rough estimates of the
angular separation were also made on a few occasions even though it
was far smaller than the resolving power of the telescope.  However,
these observations were never confirmed and have been called into
question, so we do not use them in our analysis.  Nevertheless, a
number of fascinating aspects of this puzzling data set are worth
noting and are described in more detail in
Appendix~\ref{sec:greenwich}.

It was not until 1919 that Capella was resolved in earnest, with the
6-m baseline Michelson interferometer on the 100-inch telescope on
Mount Wilson \citep{Anderson:20, Merrill:22}. These pioneering
observations are of high quality and internal consistency and have
since been used in nearly all of the astrometric orbital solutions
published for the system. They are valuable because of the extended
time baseline they afford. We incorporate them into our own analysis
as well, although they do contain some systematic errors that we
address later. Except for two more recorded attempts by
\cite{Wilson:39, Wilson:41} to resolve the pair visually, almost 50
years elapsed until the next astrometric observations were made at
Pulkovo Observatory by \cite{Kulagin:70}, with a similar
interferometer also using a 6-m baseline. Additional long-baseline
interferometric observations have been reported by \cite{Blazit:77a}
(baseline 12--20 m), \cite{Koechlin:79} (baseline 13.8 m),
\cite{Koechlin:83} (baseline 5.5--35 m), \cite{Baldwin:96}
(three-element Cambridge Optical Aperture Synthesis Telescope, COAST,
using baselines up to 6.1 m), and more recently by \cite{Kraus:05}
(three-element Infrared Optical Telescope Array, IOTA, using baselines
up to 38 m). By far the most precise interferometric observations of
Capella are those of \cite{Hummel:94} with the Mark~III interferometer
on Mount Wilson, using baselines of 3.0 to 23.6 m. These observations
improved the uncertainties in both the position angle and the angular
separation by about an order of magnitude compared to previous
measures. They are also the only ones, aside from those obtained in
1919--1921, that provide full phase coverage of the orbit.  All
interferometric measurements are listed in
Table~\ref{tab:interferometry_polar} (for those published in polar
coordinates) and Table~\ref{tab:interferometry_xy} (measures published
in Cartesian coordinates).

Because of its brightness and convenient angular separation, Capella
has served for decades as an ideal calibration object for
long-baseline interferometry, and has been referred to as ``an
interferometrist's friend'' \citep{Hartkopf:01}.  Beginning in the
1970s Capella was observed also with the speckle interferometry
technique by a large number of investigators. Though typically less
precise that the long-baseline interferometry results, these measures
are still useful and are folded into our solution below. They are
collected in Table~\ref{tab:speckle}.

Capella was also a target of the \hip\ mission \citep{Perryman:97}.
It was observed under the designation HIP~24608 a total of 43 times
over a 3-yr interval (1990.08--1993.15), corresponding to nearly 11
orbital cycles of the binary.  Each measurement consisted of a
one-dimensional position (`abscissa', $v$) along a great circle
representing the scanning direction of the satellite, tied to an
absolute frame of reference known as the International Celestial
Reference System (ICRS). The typical precision of these measurements
is about 2.3 milli-arc seconds (mas) for Capella. The data were used
by the \hip\ team to solve for the five basic astrometric parameters
of the star, which are the position and proper motion components, and
the parallax.  Although the satellite measurements did not actually
resolve the pair (separation $\sim$56 mas), the motion of the center
of light is large enough that it was clearly detected. Consequently,
extra terms were added during the original reductions by the \hip\
team to model this orbital motion and avoid biases. Several of the
orbital elements were held fixed at the values from the
\cite{Hummel:94} study (period, epoch of nodal passage, inclination
angle, position angle of the node), and the orbit was assumed to be
circular. The semimajor axis of the photocentric motion reported in
the catalog is $a''_{\rm phot} = 2.16 \pm 0.60$ mas. The $O\!-\!C$
residuals from the 5-parameter solution, referred to as `abscissa
residuals' $\Delta v$, are provided with the catalog and together with
the five standard parameters they allow the original measurements to
be reconstructed. In this way, these $\Delta v$ measurements can be
used in principle for further improvements in the overall astrometric
solution if a better visual orbit for Capella were to become
available. In practice they contribute relatively little to the orbit
of Capella, but they do provide a useful check on the secondary
velocity amplitude, to be discussed later. Furthermore, they allow an
independent estimate of the brightness ratio (\S\,\ref{sec:orbit}), so
we have incorporated these measurements into our global solution
described in the next section. They are listed in
Table~\ref{tab:hipparcos}.

Finally, Capella was spatially resolved by direct imaging by
\cite{Young:02}, using the Faint Object Camera aboard HST at
ultraviolet wavelengths (1300--3000~\AA). These measurements are
included in Table~\ref{tab:speckle}.

In many of the interferometric and speckle observations the quadrant
of the position angles has an ambiguity of $\pm180\arcdeg$ due to the
nature of the measurement. Even in cases where the analysis is able to
establish the correct quadrant, that determination is made more
difficult for Capella because the stars are so nearly equal in
brightness, as we discuss in \S\,\ref{sec:lightratio}, and because the
brightness ratio depends on the wavelength of the observation and
reverses around 7000\,\AA. Here we have adjusted the angles where
necessary to be consistent with the usual convention for visual
binaries, in which the position angles are measured from the brighter
star to the fainter one \emph{in the V band}.

Although many of the above astrometric measurements have been used
previously by others to model the orbit of Capella, careful
examination during the present work of the original references and
other bibliographic sources making use of them revealed a number of
inconsistencies, misprints, or mistakes that appear not to have been
noticed before. As a result, the data used here differ slightly from a
listing of the measurements contained in the Washington Double Star
Catalog \citep{Mason:01} provided by the U.S.\ Naval Observatory.  We
document these details in Appendix~\ref{sec:astrometrynotes} for the
benefit of future users.

\section{Orbital solution}
\label{sec:orbit}

The many data sets described above constrain the parameters of
Capella's orbit in different ways. While it is true that in this case
the interferometric observations by \cite{Hummel:94} and our own
radial-velocity measurements carry much more weight than other data
sets, the optimal procedure for obtaining the orbital parameters is
usually to account for the different weights and combine all
observations into a single fit, provided they are sufficiently free
from systematic errors. This is the approach we adopt here. The
observations consist of position angles ($\theta$) and angular
separations ($\rho$), measures of the relative separation in
rectangular coordinates ($\Delta x$ and $\Delta y$), radial velocities
for the primary and secondary, and the \hip\ measurements $\Delta v$.
We solve for the usual orbital elements of a visual-spectroscopic
binary, which are the orbital period ($P$), relative angular
semi-major axis ($a\arcsec$), inclination angle ($i$), eccentricity
($e$), longitude of periastron of the secondary ($\omega$), position
angle of the ascending node for the equinox J2000.0 ($\Omega$), time
of periastron passage ($T$), center-of-mass velocity ($\gamma$), and
the velocity semi-amplitudes for each star ($K_{\rm A}$ and $K_{\rm
B}$).

The use of the \hip\ measurements introduces several additional
parameters that must also be solved for. These are the angular
semimajor axis of the photocenter ($a''_{\rm phot}$), corrections to
the catalog values of the position of the barycenter
($\Delta\alpha^*$, $\Delta\delta$) at the mean catalog reference epoch
of 1991.25, corrections to the proper motion components
($\Delta\mu_{\alpha}^*$, $\Delta\mu_{\delta}$), and a correction to
the \hip\ parallax.\footnote{Following the practice in the \hip\
catalog, we define $\Delta\alpha^* \equiv \Delta\alpha \cos\delta$ and
$\Delta\mu_{\alpha}^* \equiv \Delta\mu_{\alpha} \cos\delta$.} In this
case, however, the fact that the spectroscopic elements $K_{\rm A}$
and $K_{\rm B}$ are obtained in the same solution introduces a
redundancy, and the parallax (referred to here as the ``orbital''
parallax) can be expressed in terms of other elements as
\begin{equation}
\pi_{\rm orb} = 1.0879 \times 10^4 {a'' \sin i \over P (K_{\rm
A}+K_{\rm B}) \sqrt{1 - e^2}}~.
\label{eq:pxorb}
\end{equation}
The numerical constant is such that the result is in the same units as
$a\arcsec$ (typically mas) when the period is given in days and
$K_{\rm A}$ and $K_{\rm B}$ in \kms. We have therefore chosen to
eliminate the parallax correction as an adjustable parameter in the
fit.  The mathematical formalism for modeling the \hip\ abscissa
residuals follows closely that described by \cite{vanLeeuwen:98},
\cite{Pourbaix:00}, and \cite{Jancart:05}, including the correlations
between measurements from the two independent data reduction consortia
that processed the original \hip\ observations
\citep[see][]{Perryman:97}. Full details along with another example of
the application of this technique may be found in \cite{Torres:07}.

As noted earlier (\S\,\ref{sec:spechist}), instrumental effects in
spectroscopy often cause the zero points of the radial velocity
measurements to be different for different observers. These shifts are
accounted for here by solving for an additional offset between each of
the historical RV data sets and our own, which we take as the
reference because it is the largest. We solve for these offsets
$\Delta_i$ ($i = 1,\ldots,9$) in the sense $\langle$other \emph{minus}
CfA$\rangle$ simultaneously with the orbital elements. Additionally,
we solve for a primary/secondary offset $\Delta_{\rm AB}$ for the CfA
velocities themselves, to correct for the small shift described in
\S\,\ref{sec:specnew}. Finally, one more adjustable parameter
$f_{\rho}$ is included as a correction to the scale of the angular
separation measurements of \cite{Merrill:22} and \cite{Kulagin:70}, to
be described below.  Position angles have been precessed from the
original epoch of each observation to the standard epoch
J2000.0. Those of \cite{Hummel:94} have been precessed from their
reference epoch of 1991.9. For consistency we have also applied
precession corrections to the $\Delta x$ and $\Delta y$ measurements,
although they are hardly significant. The \hip\ observations are
already referred to J2000.0.

Altogether there are 26 adjustable parameters, which we determined
simultaneously using standard non-linear least-squares techniques
\citep[see][p.\ 650]{Press:92}.  A total of 1015 individual
observations were used in the fit. A summary of the different data
sets can be found in Table~\ref{tab:datasets}. For approximately half
of the astrometric observations it was necessary to reverse the
quadrants of the position angles or the sign of the $\Delta x$ and
$\Delta y$ measurements for consistency. This is hardly surprising
given the small magnitude difference between the components at optical
wavelengths, and the inherent ambiguities in quadrant determination in
some cases. We discuss this further in \S\,\ref{sec:lightratio}.
Uncertainties for the astrometric observations were adopted from the
original sources, when available, and relative weights within each
series were accounted for, if reported.  For some of the speckle
measurements that have no published errors we adopted typical values
of $\sigma_{\theta} = 2\arcdeg$ and $\sigma_{\rho}$ = 3 mas. With few
exceptions historical radial velocities have no published errors. In
those cases we have assumed them to be equal to the RMS scatter from
preliminary orbital fits. Relative weights for the RVs within a given
series were taken into account in cases where they were given.
Because internal uncertainties are often underestimated, and some of
our guesses are necessarily rough, we have re-scaled them by
iterations in the final solution so as to achieve reduced $\chi^2$
values near unity separately for each source, for all astrometric and
spectroscopic data sets having a sufficient number of observations.

\begin{figure} 
\epsscale{1.3} 
\vskip -0.3in
{\hskip -0.11in \plotone{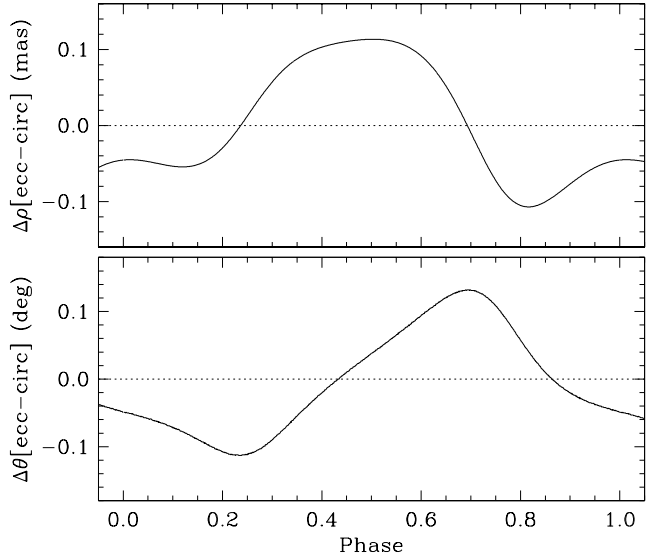}}
\vskip -0.4in 
\figcaption[]{Difference between an eccentric and circular orbital
solution for Capella, using all data sets. The eccentricity is most
likely spurious, and is driven by the measurements of \cite{Hummel:94}
(see text). Phases are counted from the time of nodal passage in the
circular orbit.\label{fig:ecc}}
\end{figure}

All prior studies of Capella based on data sets of sufficient size and
quality have concluded that the eccentricity of the orbit is not
significantly different from zero. We were therefore somewhat
surprised that our initial solutions gave a very small yet
statistically significant value of $e = 0.00087 \pm 0.00021$, with
$\omega = 324\arcdeg\ \pm 14\arcdeg$. Closer examination revealed that
this is driven exclusively by the high-weight \cite{Hummel:94}
observations, which when used alone give $e = 0.00083 \pm 0.00005$ and
$\omega = 334\fdg8 \pm 4\fdg7$. A solution without the Hummel
measurements yields a circular orbit, as does one that uses only the
CfA radial velocities, which carry the largest weight among the
remaining data sets. The CfA primary velocities, when considered
separately, also suggest the orbit is circular, but our secondary
velocities, which have larger uncertainties, prefer $e = 0.018 \pm
0.004$. This result is clearly related to the residual patterns shown
in Figure~\ref{fig:cfaorb}, seen only in the secondary, which we
believe to be most likely of instrumental origin, as discussed in
\S\,\ref{sec:specnew}.  On the basis of this evidence we are inclined
to conclude that the eccentricity we derive from the Hummel et al.\
measurements is spurious. In their own orbital solution those authors
made direct use of the interferometric visibilities ($V^2$) from the
Mark~III instrument, rather than relative positions in polar
coordinates, which are the data finally published. Nightly values for
the latter, condensed from the $V^2$ measures accounting for orbital
motion, were provided by \cite{Hummel:94} for the convenience of the
reader since they are easier to use. Given that Hummel et al.\
reported detecting no significant eccentricity ($e = 0.0000 \pm
0.0002$) in their solutions using the visibilities, we speculate that
our result is due to our use of the published \{$\theta$, $\rho$\}
measurements as opposed to the original $V^2$ values. The translation
from one to the other has apparently introduced very subtle
distortions in the orbit, perhaps related to surface feature
inhomogeneities (spots) or calibration issues, as discussed in some
detail by \cite{Hummel:94}.  In practical terms, the difference
between our eccentric and circular fits using all data sets is very
small, as illustrated in Figure~\ref{fig:ecc}.  The maximum
differences are $\sim$0\fdg1 in position angle and $\sim$0.1 mas in
angular separation. The effect on the absolute masses is considerably
less than their uncertainties ($< 0.5$\%). For the remainder of this
paper we will consider the orbit to be circular. This reduces the
number of adjustable parameters to 24. The epoch $T$ defined above
then refers to the nodal passage (ascending node) rather than
periastron.

Preliminary fits showed a systematic pattern in the residuals of the
interferometric angular separation measurements of \cite{Merrill:22}.
The same pattern is evident in the orbital solutions published by
\cite{McAlister:81} and \cite{Barlow:93}, which show predominantly
negative residuals in $\rho$ from this source.  \cite{Hummel:94} noted
a systematic difference between their semimajor axis for Capella's
orbit and all previous results, beginning with the original study by
\cite{Anderson:20}. They speculated that those early interferometric
measurements have a scale problem, and that the large weight they have
typically received in other studies may have biased previous orbital
solutions. \cite{Hummel:94} also provided a likely explanation for the
scaling problem. It has to do with the adoption by \cite{Merrill:22}
of 5500\,\AA\ as the effective wavelength used for the original Mount
Wilson observations. This adopted wavelength sets the scale of the
angular separations.  They pointed out that while 5500\,\AA\ may be a
suitable value for observations of early G-type stars like the Sun,
the mean temperature of Capella is now known to somewhat cooler than
the Sun's, and therefore a slightly longer effective wavelength would
be more appropriate. In their estimation, the early interferometric
observations should be $\sim$5\% too small. An identical effective
wavelength was adopted in the interferometric observations of
\cite{Kulagin:70}, and in fact those measurements display the same
pattern of negative residuals in the orbital studies of
\cite{McAlister:81} and \cite{Barlow:93}, as well as in our own
preliminary fits. In order to correct for this bias in the angular
separation measurements of \cite{Merrill:22} and \cite{Kulagin:70}, we
have included the scale factor $f_{\rho}$ mentioned earlier as an
additional free parameter in our global solution. Effectively, this
means that those measurements no longer contribute to set the scale of
the orbit, but they still help to constrain the remaining orbital
elements. The result we obtain, $f_{\rho} = 1.0400 \pm 0.0035$,
confirms the significance of the effect, which is nearly of the
magnitude predicted by \cite{Hummel:94}.

In Table~\ref{tab:elements} we present our orbital solution for
Capella. In addition to the adjusted elements, we list a number of
other properties including the absolute masses and the orbital
parallax, inferred from the orbital elements. The uncertainties for
these derived quantities include the contribution from the
off-diagonal terms of the covariance matrix, to account for
correlations among the elements. The determination of the orbital
period has benefited from the century-long baseline afforded by the
observations, and its precision is now 2 parts per million
(corresponding to 19.2 seconds out of 104 days). The orbital parallax we
obtain, $\pi_{\rm orb} = 76.67 \pm 0.17$ mas, is consistent with, but
about 5 times more precise than the value from \hip\ ($\pi_{\rm Hip} =
77.29 \pm 0.89$ mas).\footnote{A recent new reduction of the \hip\
observations by \cite{vanLeeuwen:07} yielded an improved parallax
value for Capella of $\pi_{\rm Hip} = 76.19 \pm 0.47$ mas, which was
subsequently revised in the online version of the catalog to correct
for an error that affected the goodness of fit in some cases. The
updated value, $\pi_{\rm Hip} = 76.20 \pm 0.46$ mas, is still within
1$\sigma$ of our more precise determination.}

\begin{figure}[t!]
\epsscale{1.38} 
\vskip 0.27in
{\hskip -0.35in \plotone{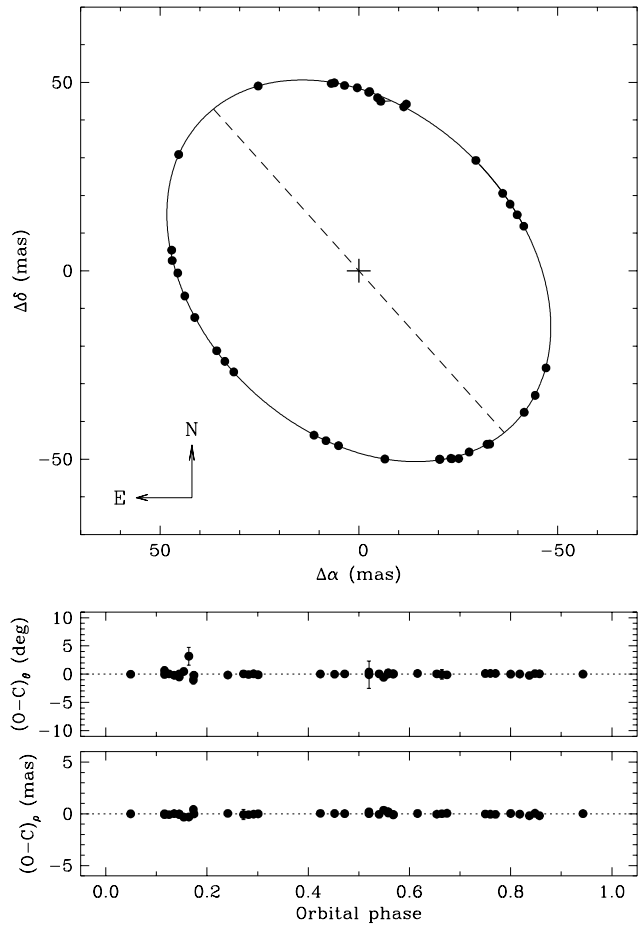}}
\vskip 0.75in 
\figcaption[]{Mark~III interferometric observations of Capella by
\cite{Hummel:94}, together with the orbit computed from our global
solution. The visually brighter component (star B) is at the
center. Solid lines connect the observations with the predicted
position on the orbit. $O\!-\!C$ residuals in position angle and
separation are indicated at the bottom. Error bars are mostly smaller
than the size of the points. The line of nodes is indicated with the
dashed line, and motion on the sky is retrograde (clockwise).
\label{fig:hummel}}
\end{figure}

\begin{figure}[t!]
\epsscale{1.35} 
\vskip 0.23in
{\hskip -0.3in \plotone{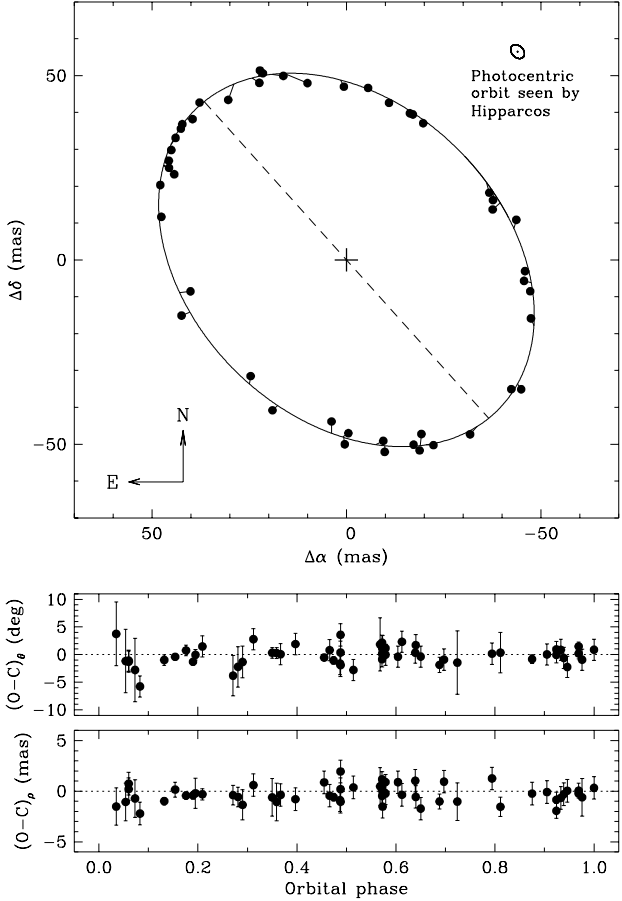}}
\vskip 0.75in 
\figcaption[]{Same as Fig.\,\ref{fig:hummel} for all speckle
observations of Capella. For reference, the orbit of the center of
light of of the binary as seen by \hip\ is shown to scale in the upper
right corner (see also Fig.\,\ref{fig:hipparcos}). $O\!-\!C$ residuals
are indicated at the bottom on the same scale as in
Fig.\,\ref{fig:hummel}, to facilitate the comparison.
\label{fig:speckle}}
\end{figure}

\begin{figure} 
\epsscale{1.9} 
\vskip 0.3in
{\hskip -1.15in\plotone{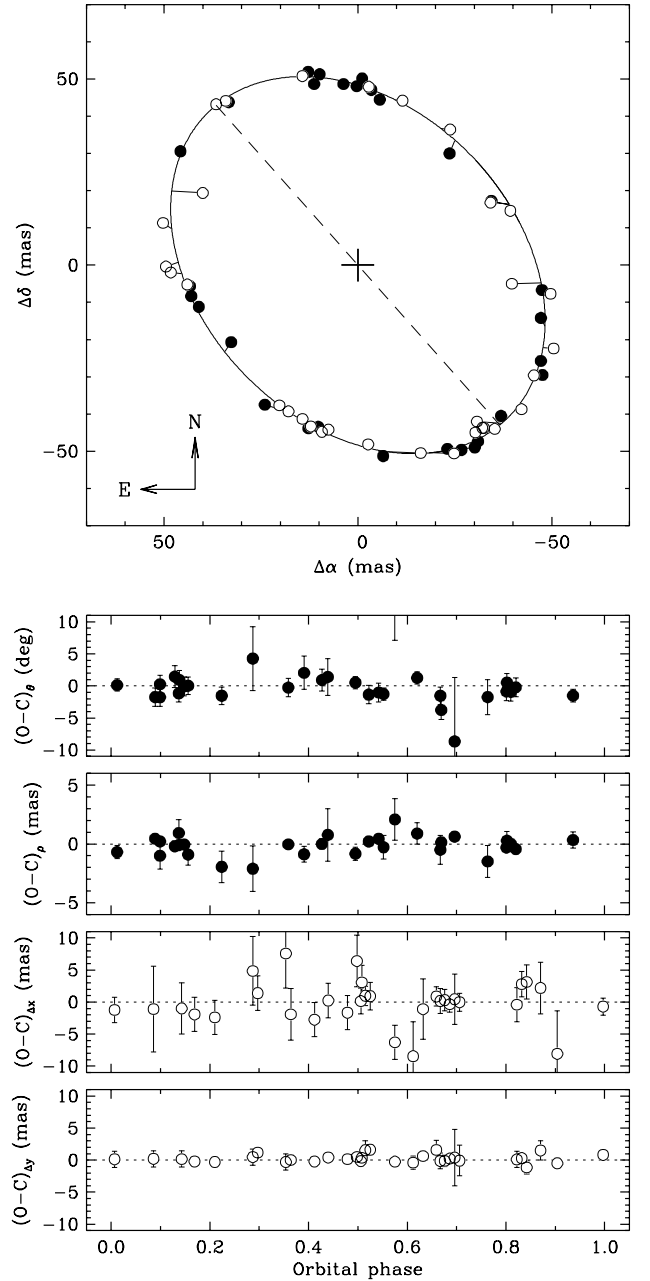}}
\vskip 1.05in 
\figcaption[]{Same as Fig.\,\ref{fig:hummel} and
Fig.\,\ref{fig:speckle} for all long-baseline interferometric
observations of Capella except those of \cite{Hummel:94}. $O\!-\!C$
residuals are shown at the bottom, separately for observations made
originally in polar coordinates ($\theta$, $\rho$ ; filled circles)
and in rectangular coordinates ($\Delta x$, $\Delta y$ ; open
circles). The scale of the position angle and separation residuals is
the same as in Fig.\,\ref{fig:hummel} and Fig.\,\ref{fig:speckle}, to
facilitate the comparison.\label{fig:interf}}
\end{figure}

Residuals from the spectroscopic observations are presented in
Table~\ref{tab:cfarvs} and Table~\ref{tab:historicalrvs}, while those
of the astrometric observations are given in
Table~\ref{tab:interferometry_polar},
Table~\ref{tab:interferometry_xy}, Table~\ref{tab:speckle}, and
Table~\ref{tab:hipparcos}. The typical precision of the measurements
from each source as represented by the RMS residual of unit weight is
given in Table~\ref{tab:datasets}.  As indicated earlier, the Mark~III
observations by \cite{Hummel:94} are by far the most precise of the
astrometric data, and are shown graphically in Figure~\ref{fig:hummel}
separately from the other observations. Residuals in position angle
and separation are also shown, and are typically 0\fdg13 in $\theta$
and 0.11 mas in $\rho$. The speckle observations are displayed in
Figure~\ref{fig:speckle}, with their residuals shown on the same scale
as the previous figure for comparison. All other measurements obtained
by long-baseline interferometry are given in Figure~\ref{fig:interf},
including both those originally made in polar coordinates and those
made in rectangular coordinates. Among the latter, the much larger
residuals in $\Delta x$ (right ascension) than in $\Delta y$
(declination) are due to the north-south orientation of the baseline
of the interferometer used by \cite{Koechlin:79} and
\cite{Koechlin:83}, which is the source of most of those measurements.

Examination of Table~\ref{tab:interferometry_polar} reveals that the
residuals in $\theta$ for the observations by \cite{Merrill:22} show a
tendency toward negative values for the later dates.  The earlier
observations made by \cite{Anderson:20} \citep[and re-reduced
by][]{Merrill:22} show the opposite trend, with the exception of the
very first measurement, which is of much lower quality and has a very
large error. These trends were noticed already by Merrill, who offered
as explanations either an instrumental effect or a real advance of the
node. We find no evidence for a secular change in $\Omega$ in the
other observations, so we tend to agree with \cite{McAlister:81} that
it is most likely an instrumental problem.\footnote{\cite{Merrill:22}
himself pointed out that there was no direct way of checking the
position angle circle of the instrument when attached to the
telescope, so that the actual position angles of the interferometer
slits could have differed by small amounts from the angles as read
from the circle.} As a test, we repeated the orbital solution solving
for two position angle corrections in addition to the other 24
elements. We obtained $-0\fdg5 \pm 0\fdg5$ for the earlier
observations by Anderson, and $+1\fdg4 \pm 0\fdg4$ for the later ones
by Merrill, consistent with expectations. Adjusting the original
values of $\theta$ for these offsets leads to a 1$\sigma$ decrease in
the orbital period of Capella, and a slightly reduced uncertainty in
$P$ of 17 sec. The change in all other elements and derived quantities
is negligible.

\begin{figure}[t!]
\vskip -0.68in
\epsscale{1.42}
{\hskip -0.35in\plotone{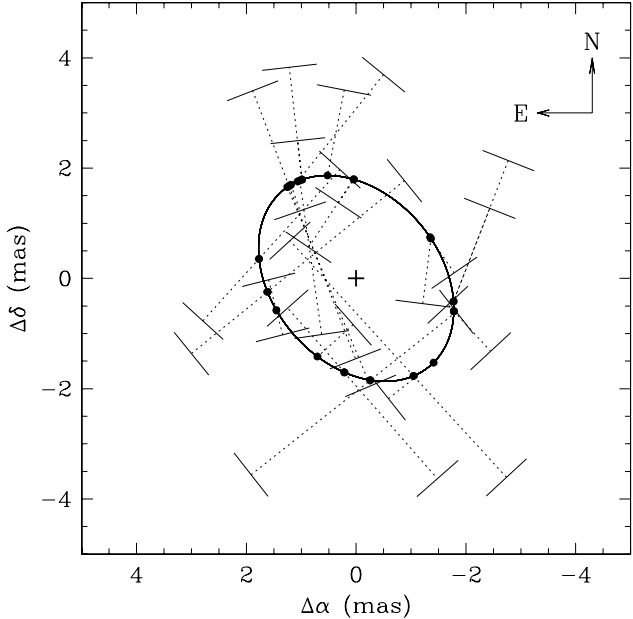}}
\vskip -0.0in

\figcaption[]{Motion of the photocenter of Capella relative to the
center of mass of the binary (indicated by the plus sign) as seen by
\hip.  See Fig.\,\ref{fig:speckle} for a comparison with the size of
the relative orbit.  The solid curve is the computed orbit from our
global solution.  Because these measurements are one-dimensional in
nature, their exact location on the plane of the sky cannot be
displayed graphically.  The abscissa residuals are shown schematically
with a filled circle at the predicted location, dotted lines
representing the scanning direction of the satellite, and short
perpendicular line segments indicating the undetermined location of
the measurement on that line. The length of the dotted lines
represents the magnitude of the $O\!-\!C$ residual from the computed
location.  Measurements with large residuals have been omitted for
clarity. Motion is retrograde (clockwise).\label{fig:hipparcos}}

\end{figure}

Given that the components of Capella are slightly different, the size
of the apparent orbit described by the center of light of the binary
as seen in unresolved observations depends on the wavelength of the
observation. Our inclusion of the \hip\ data in the solution enables
us to derive the brightness ratio $\ell_{\rm B}/\ell_{\rm A}$ between
the stars in the passband of the satellite, denoted $H_p$. For this we
make use of the fact that the semimajor axis of the photocenter and
that of the relative orbit are related by $a''_{\rm phot} = a''
(B-\beta)$, where $B = M_{\rm B}/(M_{\rm A}+M_{\rm B})$ is the
fractional mass and $\beta = \ell_{\rm B}/(\ell_{\rm A}+\ell_{\rm B})$
is the fractional luminosity \citep[see, e.g.,][]{vandeKamp:67}. This
leads to
\begin{equation}
(\ell_{\rm B}/\ell_{\rm A})_{H_p} = \left(\left[{K_{\rm A}\over K_{\rm
A}+K_{\rm B}}-{a''_{\rm phot}\over a''}\right]^{-1} - 1\right)^{-1}~.
\end{equation}
Our resulting light ratio along with other estimates of the relative
brightness are discussed in \S\,\ref{sec:lightratio}.  The projection
of the photocentric orbit of Capella on the plane of the sky along
with a schematic representation the \hip\ measurements is seen in
Figure~\ref{fig:hipparcos}.  The much smaller size of the photocentric
orbit compared to the relative orbit is illustrated in
Figure~\ref{fig:speckle}.

\section{The light ratio}
\label{sec:lightratio}

The near equal brightness of the components of Capella has been a
source of considerable confusion in the past. The quadrant of the
ascending node and the time of nodal passage (or equivalently, the
identity of the brighter star) have been changed more than once since
the publication of the first set of astrometric orbital elements by
\cite{Anderson:20}.\footnote{The choice of quadrant in that work
appears, however, to have been arbitrary \citep[see][]{Finsen:75}.}
The spectrophotometric study by \cite{Wright:54}, in which the author
incorrectly concluded that the cooler star was the brighter one in the
visible by $\sim$0.25 mag, played an important role in our
understanding of the system for several decades, although
unfortunately it also introduced biases in a number of other
investigations that made use of that result. Examples include, among
others, the interferometric study by \cite{Blazit:77a}, who attempted
the first angular diameter measurements of the stars, the Li abundance
determinations by \cite{Wallerstein:66} and \cite{Boesgaard:71}, and
to some extent the $^{12}$C/$^{13}$C ratio estimate by
\cite{Tomkin:76}, all of which adopted Wright's brightness ratio. The
history of this problem has been well summarized by \cite{Griffin:86},
and further discussed by \cite{Barlow:93}. Beginning in the early
1980s a number of authors used long-baseline interferometry and
speckle interferometry techniques to unambiguously identify the hotter
star as the brighter one in the visible (shortward of 7000\,\AA), and
\cite{Griffin:86} provided a reasonable explanation for Wright's
spectroscopic result, which apparently referred to a difference
between continuum heights rather than relative light intensities, and
did not account for the difference in line blocking between the stars.

\begin{figure}[t!]
\epsscale{1.35}
\vskip -0.62in
{\hskip -0.27in\plotone{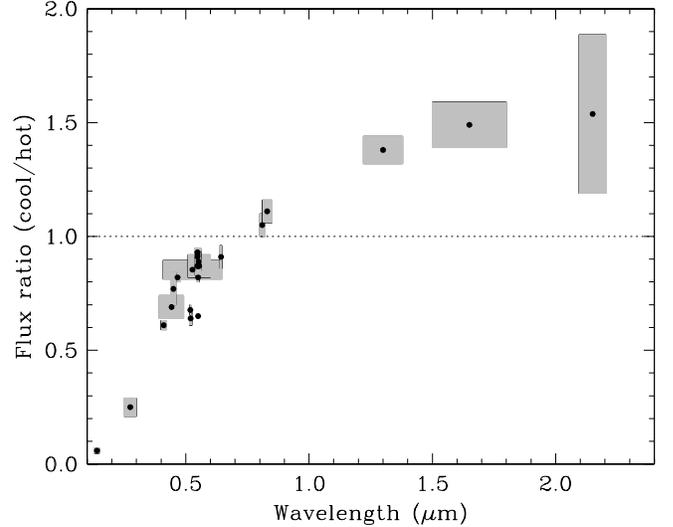}}
\vskip -0.3in

\figcaption[]{Light ratio between the components of Capella as a
function of wavelength. The shaded regions represent the uncertainty
(vertical direction) and the FWHM of the passband (horizontal
direction).\label{fig:lightratio}}

\end{figure}

Nearly two dozen independent measurements of the relative brightness
of the components are now available from the ultraviolet to the
infrared. They are listed in Table~\ref{tab:lightratio}, and are
combined later with absolute photometry to derive effective
temperatures for the components. Included also are our own light-ratio
estimates from the CfA spectroscopy (see \S\,\ref{sec:specnew}) and
from the \hip\ observations (\S\,\ref{sec:orbit}). The spectroscopic
measurement reported by \cite{Strassmeier:90} refers to the ratio of
continuum heights, rather than the intensity ratio. To convert to a
true intensity ratio we have applied a correction for the line
blocking based on appropriate synthetic spectra over the passband of
their observations. We report the corrected value in
Table~\ref{tab:lightratio}.  For uniformity the quantities listed in
the table have all been converted to ratios $\ell_{\rm cool}/\ell_{\rm
hot}$ ($= \ell_{\rm A}/\ell_{\rm B}$) between the cooler, slightly
more massive star and the hotter star. A graphical representation of
these measurements is seen in Figure~\ref{fig:lightratio}. At
wavelengths near the $V$ band the hotter star is slightly brighter,
but around 7000\,\AA\ the ratio reverses, and the cooler, more massive
star becomes dominant. This explains why some of the interferometric
measurements such as those by \cite{Baldwin:96} and \cite{Kraus:05},
which were made at red or near infrared wavelengths, have the
quadrants reversed compared with measurements in the optical.

\section{Angular diameters}
\label{sec:angdiam}

The angular sizes of the components of Capella are large enough that
they have been resolved by long-baseline interferometry on several
occasions. They were first measured by \cite{Blazit:77a}, who obtained
uniform-disk angular diameters of $\Theta_{\rm UD}^{\rm A} = 5.2 \pm
1.0$ mas for the cooler primary star and $\Theta_{\rm UD}^{\rm B} =
4.0 \pm 2.0$ mas for the hotter star. However, these values assumed
that the cooler star is brighter by 0.25 mag, following the results of
\cite{Wright:54}, whereas we now know the cooler star is in fact the
fainter one (see \S\,\ref{sec:lightratio}).  Unfortunately it is not
possible to correct Blazit's original estimates based on the
information reported.  \cite{DiBenedetto:91} obtained a limb-darkened
angular diameter of $\Theta_{\rm LD}^{\rm B} = 6.28 \pm 0.43$ mas for
the secondary, along with a much more uncertain value of $\Theta_{\rm
LD}^{\rm A} = 9.6 \pm 2.3$ mas for the primary, both measured in the
$H$ and $K$ bands. Uniform-disk diameters in the $H$ band were
published by \cite{Kraus:05} as $\Theta_{\rm UD}^{\rm A} = 8.9 \pm
0.6$ mas and $\Theta_{\rm UD}^{\rm B} = 5.8 \pm 0.8$ mas.  The
observations of \cite{Hummel:94} at wavelengths corresponding
approximately to the $B$, $V$, and $I_{\rm C}$ bands gave
limb-darkened angular diameters for Capella of $\Theta_{\rm LD}^{\rm
A} = 8.5 \pm 0.1$ mas and $\Theta_{\rm LD}^{\rm B} = 6.4 \pm 0.3$ mas.

The above measurements are inhomogeneous due to the variety of
limb-darkening corrections used. Those applied by
\cite{DiBenedetto:91} correspond to a scale factor of 1.035 between
$\Theta_{\rm LD}$ and $\Theta_{\rm UD}$. \cite{Hummel:94} used
limb-darkening coefficients from \cite{Manduca:77} and
\cite{Manduca:79}, and \cite{Kraus:05} chose not to apply any
corrections at all.  To place all these measures on the same footing
we have adopted limb-darkening coefficients from the tabulation by
\cite{vanHamme:93}, and computed the $\Theta$ corrections following
\cite{HanburyBrown:74}. The differences in these corrections compared
to the original ones can be as large as 1.7\%.  The homogenized
angular diameters are listed in Table~\ref{tab:angdiam}. The resulting
weighted averages are $\Theta_{\rm LD}^{\rm A} = 8.47 \pm 0.40$ mas
and $\Theta_{\rm LD}^{\rm B} = 6.24 \pm 0.23$ mas. The uncertainties,
which account for the scatter in the individual measurements,
correspond to fractional errors of 4.7\% for the primary and 3.7\% for
the secondary. These angular diameters, combined with the orbital
parallax, yield the absolute radii of the components that are
presented below.

As a check, independent estimates of the angular diameters may be
obtained from the near-infrared surface-brightness relation of
\cite{DiBenedetto:98} for giant stars, which is very tight and has a
scatter of only 1.4\%. The required $V\!-\!K$ indices for the
components of Capella are available from published photometry and are
described in the next section (see also Table~\ref{tab:temps}). After
transformation of the photometry to the standard Johnson system
following \cite{Carpenter:01}, we obtain $\Theta_{\rm LD}^{\rm A} =
8.32 \pm 0.75$ mas and $\Theta_{\rm LD}^{\rm B} = 6.17 \pm 0.65$ mas,
in which the uncertainties include all photometric errors as well as
the scatter of the calibration.  While less precise, these values are
in excellent agreement with the direct measurements from
interferometry.

\section{Chemical composition}
\label{sec:abundances}

Chemical composition plays a very important role in the comparison
with models in the following sections, and provides important clues on
the evolutionary state of the system. In this section we critically
review and discuss all available abundance determinations in some
detail, most of which have never been used before in the analysis of
this binary.

Despite being such a bright star, the determination of the
photospheric chemical composition of Capella has received relatively
little attention by spectroscopists. The only detailed high-resolution
study appears to be that of \cite{McWilliam:90}, in the context of a
survey of 671 G and K giants. The value reported is [Fe/H] $= -0.37
\pm 0.22$ on the scale of \cite{Grevesse:84}, in which the abundance
of iron is $\log N({\rm Fe}) = 7.67$. This result is presumably based
on the sharp lines of the primary.  It does not seem that the study
has accounted for the double-lined nature of the spectrum, which can
influence the metallicity significantly in two ways. On the one hand,
the continuum of the secondary (which has the same brightness as the
other star at the wavelengths of the \cite{McWilliam:90} spectra) will
tend to fill in the lines of the primary at most phases, making them
look weaker. On the other hand, the temperature adopted for Capella in
this analysis (5270~K) was based on the combined-light photometry, and
is too hot if assigned solely to the primary. This will generally
result in abundances that are too high. It may compensate for the
other effect to a certain degree, but the net bias is difficult to
predict. Aside from the particular case of Capella, small systematic
differences in the iron abundances between this work and others have
occasionally been pointed out \citep[e.g.,][]{Luck:95, Zhao:01,
daSilva:06, Liu:07}, and are probably traceable to systematic
differences in the adopted surface gravities or microturbulent
velocities.

Here we place the \cite{McWilliam:90} [Fe/H] determination on the more
recent scale of solar abundances by \cite{Grevesse:98}, used in some
of the models considered later, in which $\log N({\rm Fe}) = 7.50$. We
obtain [Fe/H] $= -0.20 \pm 0.22$, where the error is repeated from
\cite{McWilliam:90} and corresponds to the scatter of the individual
iron line measurements rather than the uncertainty of the mean.  The
abundances of a dozen other elements studied by \cite{McWilliam:90}
were similarly converted to the same scale, and are collected in
Table~\ref{tab:abundances}. For the purpose of comparison with the
models in \S\,\ref{sec:evolution}, which assume solar-scaled
abundances, we follow \cite{Valenti:05} and adopt the average of all
elements as an overall indicator of metallicity: [m/H] $= -0.34 \pm
0.07$.  The uncertainty given here is the error of the mean.  There is
no evidence for enhancement of the $\alpha$ elements. All other
indicators of the photospheric composition of Capella found in the
literature are either circumstantial, contradictory, or
inconclusive.\footnote{\cite{Eggen:60, Eggen:72} regarded Capella as a
member of the Hyades moving group, primarily based on kinematic
criteria. We confirm that assessment: we obtain $UVW$ velocities of
$-36.5$, $-13.9$, and $-9.1$~\kms\ (with $U$ toward the Galactic
center), in good agreement with the mean values and dispersions for
the group of $-38 \pm 6$, $-17 \pm 6$, and $-11 \pm 12$~\kms\ given by
\cite{Zhao:09}. This circumstantial evidence would imply a composition
near solar, since the mean metallicity of the group appears to be
[Fe/H] $= -0.09$ with a scatter of 0.17 dex \citep{Zhao:09}.
Unfortunately our own spectroscopic material does not allow an
accurate determination of [m/H] because of the strong dependence of
metallicity on temperature over the narrow wavelength range available
(see \S\,\ref{sec:spectroscopy}). Other estimates of the photospheric
abundance scattered throughout the literature show very poor
agreement.  A rough determination by \cite{Miner:66} based on
photometry using narrow-band interference filters gave an overall
composition near solar for the combined light.  \cite{Boesgaard:71}
measured the Li abundance of Capella, and in the same study listed
also an iron abundance of [Fe/H] = +0.26. Few details of this
determination were given, aside from the fact that the equivalent
widths of the iron lines for each component were corrected for the
light contribution from the other star using the light ratio of
\cite{Wright:54}, which we now know to be reversed (see
\S\,\ref{sec:lightratio}). In their lithium study of Capella
\cite{Pilachowski:92} did not report an iron abundance, but pointed
out that the calcium abundance is essentially solar for both
components.  \cite{Randich:94} reported [Fe/H] $= -0.4$ for the
primary of Capella, and solar metallicity for the secondary.  They
speculated the discrepancy could be due to differences in
chromospheric activity, although they also noted that other evidence
goes against this.  Finally, a study of the coronal metallicities from
X-ray observations by \cite{Bauer:96} mentions a photospheric
metallicity corresponding to [Fe/H] = +0.27, and attributes this
determination to Mercki, Strobel \& Strobel (1986) without giving a
bibliographic reference. We are unable to trace this source in the
literature.}

The photospheric $^{12}$C/$^{13}$C isotope ratio has been measured in
the optical for the primary star by \cite{Tomkin:76}, who reported the
value $27 \pm 4$. As noted earlier, this study used the incorrect
light ratio of \cite{Wright:54} to subtract the contribution of the
secondary to the observed continuum, although we do not believe this
introduces a large error due to the differential nature of the
measurement.\footnote{We estimate the equivalent width measurements of
the CN lines reported by \cite{Tomkin:76} to be underestimated by
about 6--11\% due to this efffect.} This isotope ratio is a valuable
indicator of evolution.

A large difference in the lithium abundance between Capella~A and B
was first pointed out by \cite{Wallerstein:64}, and confirmed by
others. The hotter secondary has approximately two orders of magnitude
stronger lithium than the primary.  Measurements have been made by
\cite{Wallerstein:66}, \cite{Boesgaard:71}, \cite{Pilachowski:92},
\cite{Liu:93}, and \cite{Randich:94}, in which the first two are
affected by the use of Wright's light ratio, and the latter three
adopted effective temperatures somewhat different from ours. The
measurements by \cite{Pilachowski:92} appear to be the most reliable,
although the others are generally consistent when adjusted for the
modern light ratio.  Here we have used the \cite{Pilachowski:92}
equivalent width measurements for the \ion{Li}{1}~6708\,\AA\ feature
($25 \pm 2$~m\AA\ and $200 \pm 10$~m\AA\ for the primary and
secondary, respectively). We recalculated the abundances using the
models by \cite{Pavlenko:96}, accounting for the temperature and
gravity differences as well as non-LTE effects (not considered in the
original analysis). We obtain revised lithium abundances of $\log
N({\rm Li}) = 1.3 \pm 0.2$ for the primary and $\log N({\rm Li}) = 3.2
\pm 0.3$ for the secondary, in which the uncertainties include all
measurement errors as well as possible errors in the microturbulent
velocity following \cite{Pilachowski:92}.  To be conservative, the
uncertainties have been further increased by 0.1 dex to account for
slight extrapolations that were necessary in using the
\cite{Pavlenko:96} tables.

As an active binary system, Capella has been studied extensively in
the ultraviolet and X rays for decades using virtually every space
facility capable of observing at those wavelengths. It was in fact the
first X-ray detection of a stellar corona other than the Sun, made by
sounding rockets (\citealt{Catura:75}; see also \citealt{Fisher:64,
Ayres:95}). At ultraviolet wavelengths, \cite{Bohm-Vitense:92} have
presented evidence that reliable abundance ratios between carbon and
nitrogen can be determined for giant stars from measurements of the
emission fluxes of the \ion{C}{4}~$\lambda$1550.8 and
\ion{N}{5}~$\lambda$1238.8 lines in the lower transition layers between
the stellar chromosphere and the corona, and that these ratios show
good correspondence with the photospheric abundance ratios.  Emission
fluxes for these lines have been measured in Capella by a number of
authors. However, early observations did not clearly resolve the
contribution of the two components, of which the primary represents
only $\sim$10\%.  This was first achieved by \cite{Linsky:95} based on
high spectral resolution observations with HST. Using the fluxes they
reported, we have derived the C/N ratios for Capella and use them
below as diagnostics of evolution.

Abundance determinations for Capella have also been made by many
authors from X-ray observations of coronal lines. With current
instrumentation it is generally not possible to separate the spectral
contribution of the two components in X rays, as it is in the UV.
\cite{Ishibashi:06} and others have reported that the cool primary
often dominates the coronal emission in this spectral region, although
its flux is variable with time.  Others find a more nearly equal
contribution \citep[e.g.,][]{Linsky:98}. Therefore, any measurements
may refer mainly to the primary, but are likely to be contaminated by
the secondary. Most of these observations have revealed enhanced
nitrogen \citep[see, e.g.,][]{Mewe:01, Schmitt:02, Audard:03,
Argiroffi:03}.  Mean abundances averaged over all other elements from
these studies typically indicate a sub-solar composition, in
qualitative agreement with the photospheric determinations of
\cite{McWilliam:90}.  However, coronal metallicity measurements are
highly model-dependent \citep[see, e.g.,][]{Brickhouse:01}, and
individual values sometimes show a large scatter from author to
author.  Furthermore, in the Sun's coronal regions abundances are
known to depend on the first ionization potential (FIP) of the element
considered \citep[e.g.,][]{Feldman:07}.  In view of these
complications, we have preferred not to make use of these data here.
Nevertheless, the nitrogen enhancement constitutes an interesting
piece of chemical evidence for the evolved state of the primary, as
recognized by many authors, since it is a natural consequence of the
CNO cycle for stars that have already experienced first dredge-up (see
below).  Careful consideration of the FIP effect in Capella suggests
there may be even closer agreement between the overall coronal
abundance and the photospheric value, which we believe is worth noting
given our concerns expressed earlier about the latter.  We discuss
these coronal measurements and their patterns in
Appendix~\ref{sec:coronal}.  All other useful abundance determinations
for Capella described above are gathered in
Table~\ref{tab:abundances}.

\section{Absolute dimensions}
\label{sec:dimensions}

The astrometric-spectroscopic orbital solution in \S\,\ref{sec:orbit}
yields directly the absolute masses of the components. The relative
uncertainties (0.7\% and 0.5\%) represent a factor of 3 improvement
over those of \cite{Hummel:94}, which is critical for the comparison
with stellar evolution models. We also obtained the orbital parallax.
The formal uncertainty in the corresponding distance of $13.042 \pm
0.028$~pc is only 0.2\%. The absolute radii of the components follow
from the angular diameters and the distance, and are $R_{\rm A} =
11.87 \pm 0.56~R_{\sun}$ and $R_{\rm B} = 8.75 \pm
0.32~R_{\sun}$. Relative to the orbital separation, these values
correspond to 0.075 and 0.055, respectively, so the binary is well
detached.

Effective temperatures for the individual stars in Capella have been
estimated here in three different ways. A first determination relies
on our spectroscopy, and is described in \S\,\ref{sec:specnew}:
$T_{\rm eff}^{\rm A} = 4900 \pm 100$~K and $T_{\rm eff}^{\rm B} = 5710
\pm 100$~K.  A second method is that employed by \cite{Hummel:94}, who
made use of their angular diameter measurements along with the
apparent magnitudes and bolometric corrections to infer values of
$T_{\rm eff}^{\rm A} = 4940 \pm 50$~K and $T_{\rm eff}^{\rm B} = 5700
\pm 100$~K, nearly identical to ours.  We have updated that
calculation using the average angular diameters from
\S\,\ref{sec:angdiam}, together with apparent visual magnitudes for
the components as described below, and bolometric corrections $BC_V$
from \cite{Flower:96}.\footnote{To be consistent with the scale of the
bolometric corrections, the bolometric magnitude adopted here for the
Sun is $M_{\rm bol}^{\sun} = 4.732$. When combined with the tabulated
$BC_V$ corresponding to the solar temperature of $T_{\rm eff} =
5777$~K, this gives an apparent magnitude for the Sun that reproduces
the measured value of $V = -26.76 \pm 0.02$ as determined by
\cite{Stebbins:57} and \cite{Hayes:85}. See also the discussion by
\cite{Bessell:98}.\label{foot:mbolsun}} The results are $T_{\rm
eff}^{\rm A} = 4970 \pm 154$~K and $T_{\rm eff}^{\rm B} = 5687 \pm
130$~K, in which the uncertainties in $BC_V$ and in all other measured
quantities are included.  A third method to estimate individual
temperatures relies exclusively on photometry (color indices), and has
been applied by a number of authors over the years giving results
generally consistent with the above estimates. We return to this
technique below.  To our knowledge there are no other fundamentally
different $T_{\rm eff}$ estimates available, except for those one
might infer indirectly from the spectral types assigned to the
components. For example, \cite{Strassmeier:90} applied a spectrum
synthesis technique and found a good match to the primary and
secondary in the standard stars Pollux ($\beta$~Gem, \ion{K0}{3}) and
$\alpha$~Sge (\ion{G1}{3}). These classifications are roughly
consistent with our $T_{\rm eff}$ estimates.

The information on the absolute photometry for Capella and the light
ratios discussed earlier is collected in Table~\ref{tab:temps}. We use
it here to derive photometric estimates of $T_{\rm eff}$ for each
star.  The light ratios in the table are weighted averages of all
values near the $B$, $V$, $R$, $I$, $J$, $H$, and $K$ passbands,
respectively, and the combined-light magnitudes are taken from the
database of \cite{Mermilliod:97}. $R$ and $I$ magnitudes were
transformed to the Cousins system following \cite{Leggett:92}, and the
near-infrared magnitudes were placed on the 2MASS system using the
transformations of \cite{Carpenter:01}. Individual uncertainties are
taken as published. Color indices formed from these values are listed
in the second section of the table.  These are not strictly
independent, but they at least provide a sense of the consistency of
the measurements in different systems and the scatter one can expect
from the external calibrations applied in each case. Color/temperature
calibrations for giant stars by \cite{Ramirez:05} were used to derive
temperatures for each component as well as for the combined light
(third section of Table~\ref{tab:temps}). The metallicity adopted is
the value [m/H] $= -0.34 \pm 0.07$ based on the measurements by
\cite{McWilliam:90}, described in the previous section.  The
temperature uncertainties reported in the table account for all
photometric errors, the uncertainty in the assumed [m/H], and also the
scatter of each color/temperature relation.  Weighted average
temperatures computed from the seven indices are listed as well.  The
values for both Capella~A and B are in good agreement with the other
two determinations described previously.

The last line of Table~\ref{tab:temps} presents the weighted average
of the three independent $T_{\rm eff}$ determinations for each star,
based on the spectroscopy, the quantities \{$\Theta_{\rm LD}$, $V$,
$BC_V$\}, and photometry, respectively. To be conservative, the
uncertainty of the photometric values have been increased by adding
100~K in quadrature to the formal errors prior to taking the average,
in order to account for possible systematics in the color/temperature
calibrations. This follows the discussions of \cite{Ramirez:05} and
\cite{Casagrande:06} concerning our knowledge of the absolute
effective temperature scale. The final temperatures are $4920 \pm
70$~K and $5680 \pm 70$~K for the primary and secondary, respectively.

The very different rotational velocities of the components was already
evident to spectroscopic observers a century ago. The $v \sin i$
values have since been measured by many investigators, mostly by
traditional spectroscopic means but also with other methods such as
the differential speckle interferometry technique of \cite{Petrov:96}.
These estimates are collected in Table~\ref{tab:vsini} along with our
own. For the most part the more recent determinations agree fairly
well, considering the difficulty of the measurements.

The physical parameters for both components of Capella are summarized
in Table~\ref{tab:dimensions}. The luminosities were derived here from
the well determined absolute magnitudes and bolometric corrections
from \cite{Flower:96}. The uncertainties in $BC_V$ were propagated
from the error in $T_{\rm eff}$, and an additional conservative error
of 0.05~mag was added in quadrature. If we instead compute the
luminosities directly from the radii and temperatures, the values are
considerably more uncertain ($L_{\rm A} = 74.2 \pm 8.2~L_{\sun}$,
$L_{\rm B} = 71.5 \pm 6.1~L_{\sun}$), but are consistent with the
adopted estimates. The primary (cooler) star is the more luminous
bolometrically, but is the fainter one in the visible. Also included
in the table are the projected rotational velocities ($v_{\rm sync}
\sin i$) computed under the assumption that the stars have their spins
synchronized with the orbital motion and that the spin axes are
perpendicular to the orbital plane. We discuss these values in
\S\,\ref{sec:discussion}.

\section{Discussion}
\label{sec:discussion}

The key properties that determine the evolutionary state of the giants
in Capella are the masses. Prior to this study the values most often
adopted \citep[e.g.,][]{Nobuyuki:99} were those of \cite{Hummel:94},
$M_{\rm A} = 2.69 \pm 0.06~M_{\sun}$ and $M_{\rm B} = 2.56 \pm
0.04~M_{\sun}$, which rely on the velocity semi-amplitudes of
\cite{Barlow:93}. These masses are 9\% and 5\% larger, respectively,
than those in the present work.  As noted in \S\,\ref{sec:spechist},
our primary velocity semi-amplitude is not very different from other
determinations, but our secondary value is considerably smaller, and
this drives \emph{both} masses down. An independent check on the
accuracy of $K_{\rm B}$ can be made with the available astrometry
(specifically, the \hip\ observations), without using any secondary
velocities.  This is because the \hip\ measurements are on an absolute
frame of reference (ICRS) and therefore contain strong information on
the trigonometric parallax, and the parallax is related to $K_{\rm B}$
via eq.(\ref{eq:pxorb}). We carried out an orbital solution in which
our secondary velocities were given zero weight, and the result for
the secondary semi-amplitude is $K_{\rm B} = 26.01 \pm 0.62$~\kms.
This is considerably more uncertain than the spectroscopic value of
$26.260 \pm 0.087$~\kms, but is still perfectly consistent with it,
while at the same time being more than 2$\sigma$ away from the
determination by \cite{Barlow:93}. This suggests our masses for
Capella are more accurate than previously determined, in addition to
having smaller formal errors, and we proceed below to compare them
along with other observations against theory.
	
\subsection{Comparison with stellar evolution models}
\label{sec:evolution}

Detached binary systems such as Capella that are composed of two giant
stars and show double-lined spectra are rare, and they provide
important tests of models in a relatively short-lived phase of stellar
evolution. Their component masses are necessarily very close to each
other, and a precise measurement of the mass ratio $q$, as we provide
here, becomes critical to establishing their state of evolution
unambiguously.

The evolutionary status of Capella has been a subject of debate for
decades. While there is general consensus that the hotter secondary is
crossing the Hertzprung gap and approaching the base of the giant
branch (RGB), opinions have varied on the precise location of the
primary, in large part because of uncertainties in the masses as well
as the effective temperatures and luminosities used to place the star
on the H-R diagram. Capella is perhaps unique in that, in addition to
those properties, a wealth of other information is available to aid in
determining its evolutionary state, including the surface Li
abundances of both stars, the $^{12}$C/$^{13}$C isotope ratio for the
primary, C/N ratios, and activity indicators in the optical,
ultraviolet, and X rays.  The progenitors of Capella were late B- or
early A-type stars. When such stars leave the main sequence, they
develop convective envelopes that deepen significantly as they
approach the giant branch, mixing the outer layers with matter from
the interior partially processed through the CNO cycle. As a result,
fragile elements such as lithium are burned deeper in the star
decreasing the surface abundance of that element, and others such as
$^{13}$C and $^{14}$N that are created at the expense of $^{12}$C are
brought to the surface during the ``first dredge-up''. This causes a
dramatic reduction in the $^{12}$C/$^{13}$C ratio and in the C/N
ratio, both of which are measurable.  Thus, surface abundances contain
potentially important information on the evolutionary state of evolved
stars like Capella.

\cite{Iben:65} pioneered this approach by relying on early estimates
of the lithium abundance of both components \citep{Wallerstein:64,
Wallerstein:66} to conclude, based on his models, that the primary is
a core helium-burning star.  Compared to an alternate location on the
ascending giant branch, the ``clump'' phase also seems more likely
because it is longer-lived.\footnote{The predicted durations of the
different stages of evolution according to one of the models of
\cite{Claret:04} considered below (case~A) are as follows, for a star
with the mass of the primary. The main-sequence (MS) phase lasts
526~Myr. The crossing of the Hertzprung gap up to the point of minimum
luminosity at the base of the RGB is 7.3~Myr, or only 1.4\% of the MS
lifetime. The first ascent up to the helium flash lasts 5.7~Myr
(1.1\%), the subsequent descent to the luminosity minimum takes
16.2~Myr (3.1\%), and the clump phase is a more prolonged 89.8~Myr
(17.1\%).\label{foot:times}} This lifetime argument seems to have
weighed heavily in most of the other studies in which the measured
masses, temperatures and luminosities have been compared against
stellar evolution models, including the work of \cite{Barlow:93},
\cite{Hummel:94}, \cite{Schroder:97}, and \cite{Iwamoto:99}.  On the
other hand, \cite{Boesgaard:71} concluded based on her own Li
measurements, which differed from those of Wallerstein, that the
primary is not in such an advanced evolutionary state. Similarly,
\cite{Bagnuolo:89} found evidence in the small luminosity difference
between the stars that the primary is still at the beginning of the
RGB. They also argued for a much smaller difference in mass than
indicated by the measurements at the time, and indeed our present
determinations bring the mass ratio much closer to unity than implied
by the spectroscopy of \cite{Barlow:93}.  \cite{Ayres:83} also took
the view that the primary is not yet burning helium in its core based
on the high levels of chromospheric activity implied by their
ultraviolet observations, although the opposite conclusion was reached
by \cite{Ayres:88}.

The significantly improved parameters we have derived for Capella,
particularly the masses which are three times more precise than those
previously available, offer an opportunity to revisit the issue of its
evolutionary status. Among the many publicly available stellar
evolution models, we initially focused on those that allow some
flexibility in setting parameters such as the composition or the age
(for isochrones).  However, not all of these models extend past the
helium flash, which is necessary to explore the more advanced core
helium-burning phase for the primary, and thus we are somewhat limited
in our choices.

\begin{figure}
\epsscale{1.38}
\vskip -0.55in
{\hskip -0.33in \plotone{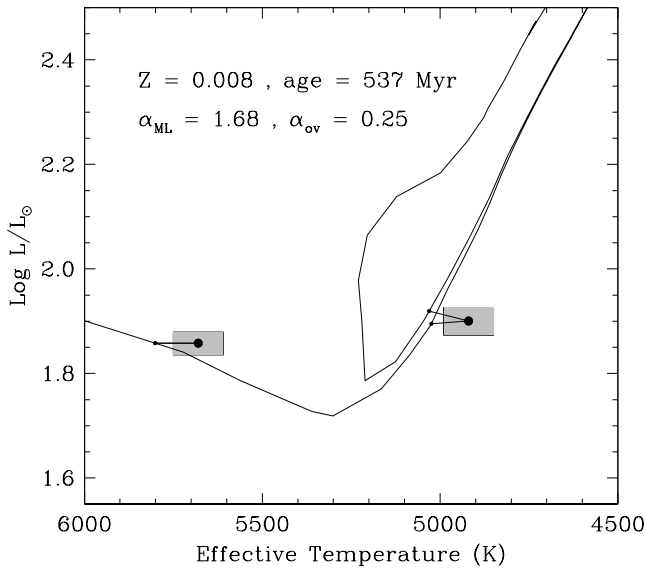}}
\vskip -0.3in

\figcaption[]{Isochrone from the series by \cite{Girardi:00} for an
age of $\tau = 537$~Myr and $Z = 0.008$ ([Fe/H] = $-0.38$) that
provides the best match to the observations of Capella ($M$, $T_{\rm
eff}$, $L$, and $R$ for both stars). Two different evolutionary states
for the primary provide similarly good fits: one on the ascending
branch, and the other on the descending branch prior to the beginning
of of the core helium-burning phase. The large filled circles and
shaded boxes correspond to the measured quantities for Capella and
their uncertainties, while small dots on the isochrone correspond to
the locations of the best fits. See Table~\ref{tab:modelsg}.
\label{fig:girardi}}

\end{figure}

As a starting point, we compare the observations against the widely
used set of stellar models by \cite{Girardi:00}, based on physics that
are now standard in most current models, including convective core
overshooting. The mixing length parameter is fixed in these models at
the value $\alpha_{\rm ML} = 1.68$, overshooting is set to
$\alpha_{\rm ov} = 0.25$, and the chemical composition adopted for the
Sun is $Z_{\sun} = 0.019$. Diffusion is not considered, although its
effect is completely negligible for Capella. Neither is mass loss due
to winds, which appears to be rather low according to most estimates
\citep{Drake:86, Katsova:98, Getman:99}.  For convenience we have
chosen to compare the observations against isochrones (thereby
imposing the constraint of coevality) computed using the web interface
provided by the authors.\footnote{
\url{http://stev.oapd.inaf.it/cgi-bin/cmd}~.}  In addition to $M$,
$L$, and $T_{\rm eff}$, we consider also the absolute radii since they
are determined independently from the temperatures and luminosities
and are of comparable precision. We explored a fine grid of ages and
chemical compositions, and used as a figure of merit the $\chi^2$
defined as
\begin{displaymath}
\chi^2 = \sum_{i=1}^{2}\left[\left({\Delta M\over \sigma_M}\right)_i^2 +
\left({\Delta T_{\rm eff}\over \sigma_{T_{\rm eff}}}\right)_i^2 +
\left({\Delta L\over \sigma_L}\right)_i^2 +
\left({\Delta R\over \sigma_R}\right)_i^2\right]~,
\end{displaymath}
in which the $\Delta$ quantities represent the differences between the
measurements and the models for each star ($i = 1, 2$). The best match
to the observations is achieved for a metallicity $Z = 0.008$
(corresponding to [Fe/H] = $-0.38$) and an age of $\tau =
537$~Myr. All four measured quantities are reproduced to within 1.4
times their nominal uncertainties (see Table~\ref{tab:modelsg},
case~A), and the primary star is located on the ascending branch.  A
similarly good fit to an isochrone of the same age and composition can
be found with the primary on the descending branch
(Table~\ref{tab:modelsg}, case~B), with only a slightly larger
difference with that star's measured effective temperature
(1.7$\sigma$).\footnote{We note that the mass \emph{ratio} for Capella
is better determined from our measurements (0.36\% relative error)
than the individual masses (0.73\% and 0.53\%, respectively). The best
fit with the primary on the ascending branch predicts a value of $q$
which is 2.2$\sigma$ larger than measured, while that on the
descending branch shows better agreement at the 0.3$\sigma$ level.}
These two fits are shown in Figure~\ref{fig:girardi}.  No satisfactory
match is found with the primary in the core helium-burning phase,
which has been favored by most previous investigators. Although the
detailed surface abundances are not included in the published tables
for these models, recent work by \cite{Bertelli:08} extending the same
series of calculations does provide typical values for some
elements. For a 2.5~$M_{\sun}$ star similar to the Capella primary and
a metallicity of $Z = 0.008$, the predicted $^{12}$C/$^{13}$C after
the first dredge-up is 19.0. This is lower than the estimate of $27
\pm 4$ by \cite{Tomkin:76}. On the other hand, the C/N ratios show
much closer agreement with theory.  The predicted value for the
primary at the end of first dredge-up is 0.67, which is only slightly
higher than the measured value of $0.57 \pm 0.06$. For the secondary,
which has not yet experienced first dredge-up and may therefore be
expected to still have its initial main-sequence value of 3.27, the
measured value is in fact very close: $3.30 \pm 0.16$.

\begin{figure}[t!]
\epsscale{2.28}
\vskip 0.51in
{\hskip -1.82in\plotone{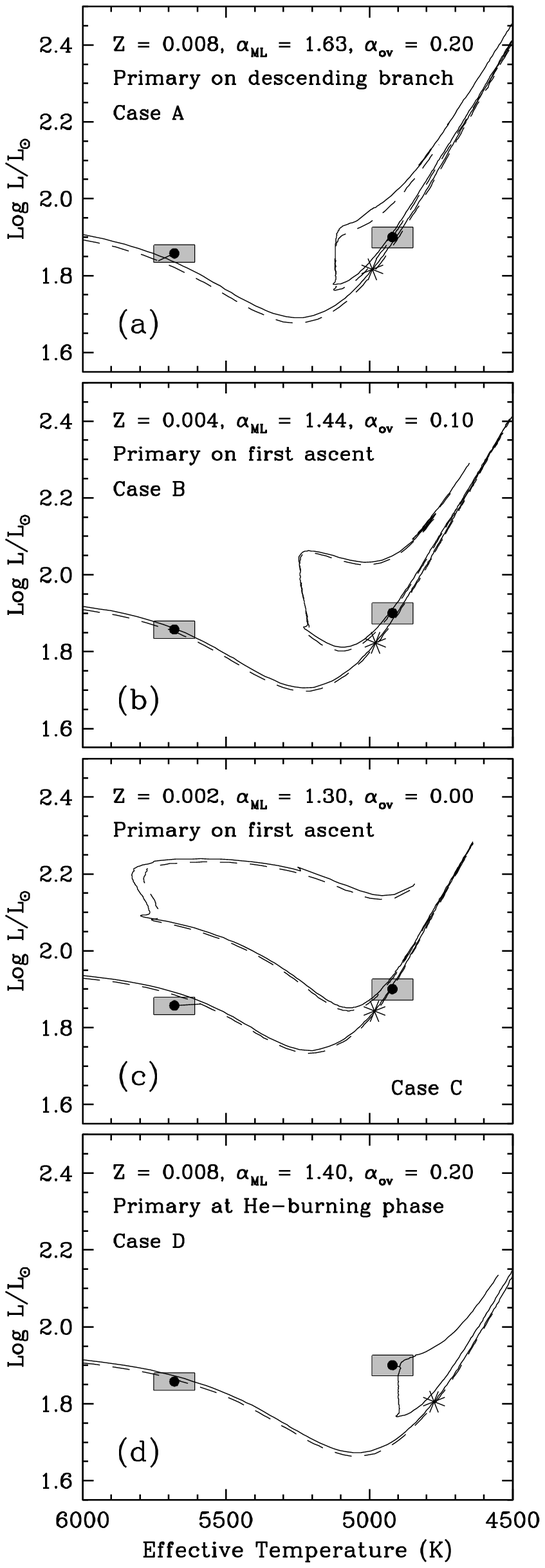}}
\vskip 1.04in

\figcaption[]{Evolutionary tracks for Capella A (solid) and B (dashed)
from the series of models by \cite{Claret:04}, for the exact masses
determined here. The measured temperature and luminosity of each
component are shown with the filled circles and error boxes.  A thin
line connects the measured positions with the best-fit location on
each track. Four different scenarios are considered, focusing on the
evolutionary status of the primary star. The parameters of each case
and the location of the primary are as labeled. The asterisk on the
primary tracks indicates the position where the models give a
$^{12}$C/$^{13}$C carbon isotope ratio exactly matching the measured
value. \label{fig:claret}}

\end{figure}

\begin{figure}
\epsscale{1.22}
\vskip -0.75in
{\hskip -0.1in\plotone{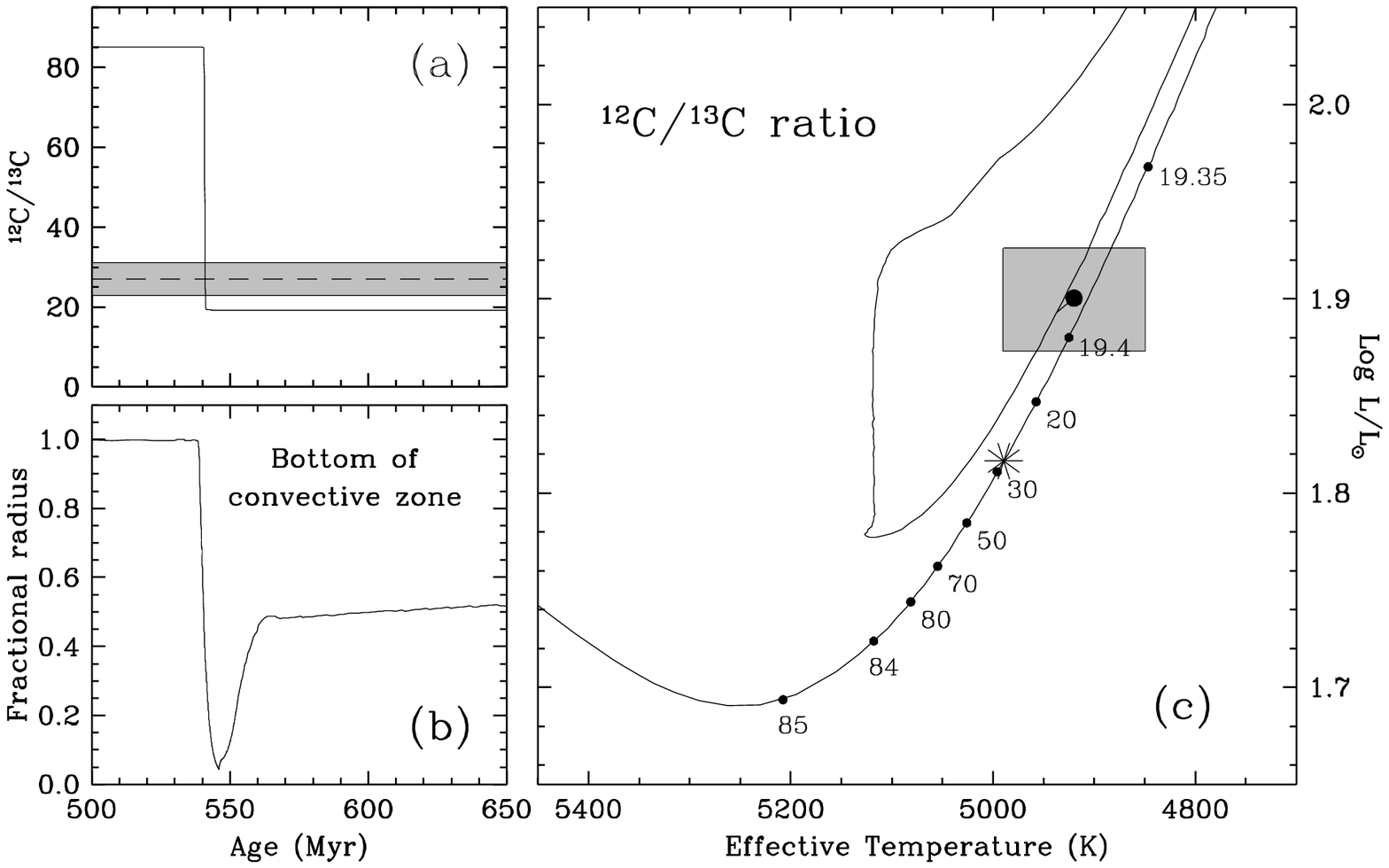}}
\vskip -0.5in

\figcaption[]{Carbon isotope ratio evolution for the primary of
Capella. (a) $^{12}$C/$^{13}$C ratio as a function of age according to
the Claret model in Figure~\ref{fig:claret}a ($Z = 0.008$,
$\alpha_{\rm ML} = 1.63$, $\alpha_{\rm ov} = 0.20$; case~A). The
measurement is indicated with the dashed line, and the shaded area
represents the 1$\sigma$ uncertainty. (b) Location of the bottom of
the convective zone in terms of the radius of the star, showing the
rapid deepening during the first dredge-up. (c) Close-up of the
evolutionary track for the measured mass of Capella A from
Fig.~\ref{fig:claret}a showing the change in the carbon isotope ratio
during the first ascent. The measured temperature and luminosity along
with the error box are indicated. An asterisk marks the location on
the track where the isotope ratio matches the measured value of $27
\pm 4$.\label{fig:carbon}}

\end{figure}

In order to provide more flexibility in exploring possible
evolutionary stages for the primary, and also to gain better access to
the detailed surface composition, we have considered a different set
of models by \cite{Claret:04} in which we can more easily vary not
only the overall metallicity but also the mixing length and
overshooting parameters.  Evolutionary tracks were computed for the
exact masses we measure. The best match is found for the same
metallicity as before ($Z = 0.008$, corresponding to [Fe/H] $= -0.33$
in these models, which adopt solar abundances from
\citealt{Grevesse:98}) and convective parameters $\alpha_{\rm ML} =
1.63$ and $\alpha_{\rm ov} = 0.20$. As indicated in
Figure~\ref{fig:claret}a, the primary is on the descending branch,
prior to the core helium-burning phase, and the age determined for the
two components (553~Myr) is virtually identical to within 0.2\%.
Details of the differences in $T_{\rm eff}$, $L$, and $R$ for each
star are given in Table~\ref{tab:modelsc} (case~A), along with the
goodness of fit ($\chi^2$).  Lithium abundance calculations from these
models are not up to date and have been found to give surface values
that are much lower than other calculations. They are not considered
further, although this has no impact on other calculations since Li is
a low abundance element with no energetic importance.  As in the
previous models, the predicted $^{12}$C/$^{13}$C isotope ratio for the
primary is found to be lower than measured (19.18 versus $27 \pm 4$,
nearly a 2$\sigma$ difference). At face value the measured ratio seems
inconsistent with any position for the primary of Capella other than
on the first ascent. This is illustrated in Figure~\ref{fig:carbon}.
The decrease in the isotope ratio shown in the top left panel is very
rapid and follows closely the deepening of the convective envelope
displayed on the lower left (first dredge-up). Values of
$^{12}$C/$^{13}$C are indicated on the evolutionary track in the right
panel. The ratio changes from 85 to 20 in less than 1~Myr, so there is
very little leeway to accommodate the measured value, which is
predicted to occur in this model at a considerably lower luminosity
and somewhat hotter temperature than observed, by 2.9$\sigma$ and
1.0$\sigma$, respectively (asterisk in Figure~\ref{fig:carbon}c).  The
alternative is an error in the measurement. We discuss this further
below.  On the other hand, the C/N ratios expected from these models
are quite consistent with the observed values, as seen in
Table~\ref{tab:modelsc}.

Given the disagreement of the previous models with $^{12}$C/$^{13}$C,
we explored other possible matches that place the primary on the
ascending branch, by varying both the mixing length and overshooting
parameters. One such solution has $Z = 0.004$ (corresponding to [Fe/H]
$= -0.63$), $\alpha_{\rm ML} = 1.44$, and a reduced overshooting of
$\alpha_{\rm ov} = 0.10$. The ages inferred for the two components
agree to within 1.0\%, and have a mean of 464~Myr (see
Figure~\ref{fig:claret}b). The carbon isotope ratio for the primary is
underpredicted by about the same amount as before
(Table~\ref{tab:modelsc}, case~B), and the location on the track that
matches the measured $^{12}$C/$^{13}$C value (asterisk) is only
marginally closer to the measured luminosity and temperature of
Capella A (2.7$\sigma$ and 0.8$\sigma$ discrepancies, respectively). A
similar fit with the primary also on the ascending branch can be
obtained with $\alpha_{\rm ML} = 1.30$ and no overshooting
($\alpha_{\rm ov} = 0.00$), but at an even lower metallicity of $Z =
0.002$ that is probably unrealistic.  The mean age in this case is
400~Myr, and the individual ages differ again by only 1.0\%
(Table~\ref{tab:modelsc}, case~C, and Figure~\ref{fig:claret}c). The
luminosity and temperature at which the primary model predicts a
carbon ratio matching the observation are 2.0$\sigma$ and 0.9$\sigma$
away from the measured values, respectively.  Setting aside the
$^{12}$C/$^{13}$C discrepancy for the moment, it is interesting to
note that the measured C/N ratio for the primary appears to rule out
all scenarios with that star on the ascending branch. While the
predictions for (C/N)$_{\rm B}$ are still consistent with the
observations, the expected (C/N)$_{\rm A}$ values are a factor of
$\sim$3 too high (Table~\ref{tab:modelsc}, cases~B and C).  Thus, the
chemical evidence for Capella seems contradictory regarding the
viability of the first ascent scenario.

We investigated also possible fits to the Claret models with the
primary in the core helium-burning phase. While previous researchers
have had no difficulty in finding satisfactory fits, in this case we
are unable to achieve a match to $M$, $R$, $T_{\rm eff}$, and $L$ as
good as in the scenarios discussed above. Evolutionary tracks with $Z
= 0.008$, $\alpha_{\rm ML} = 1.40$, and $\alpha_{\rm ov} = 0.20$ yield
ages for the stars that differ by 16.4\% (mean age of 589~Myr). The
carbon isotope ratio for the primary is not reproduced (see
Table~\ref{tab:modelsc}, case~D, and Figure~\ref{fig:claret}d), but
the C/N fractions are consistent with the measurements since the
primary is well past the first dredge-up.

\begin{figure}
\epsscale{1.28}
\vskip -0.35in
{\hskip -0.2in\plotone{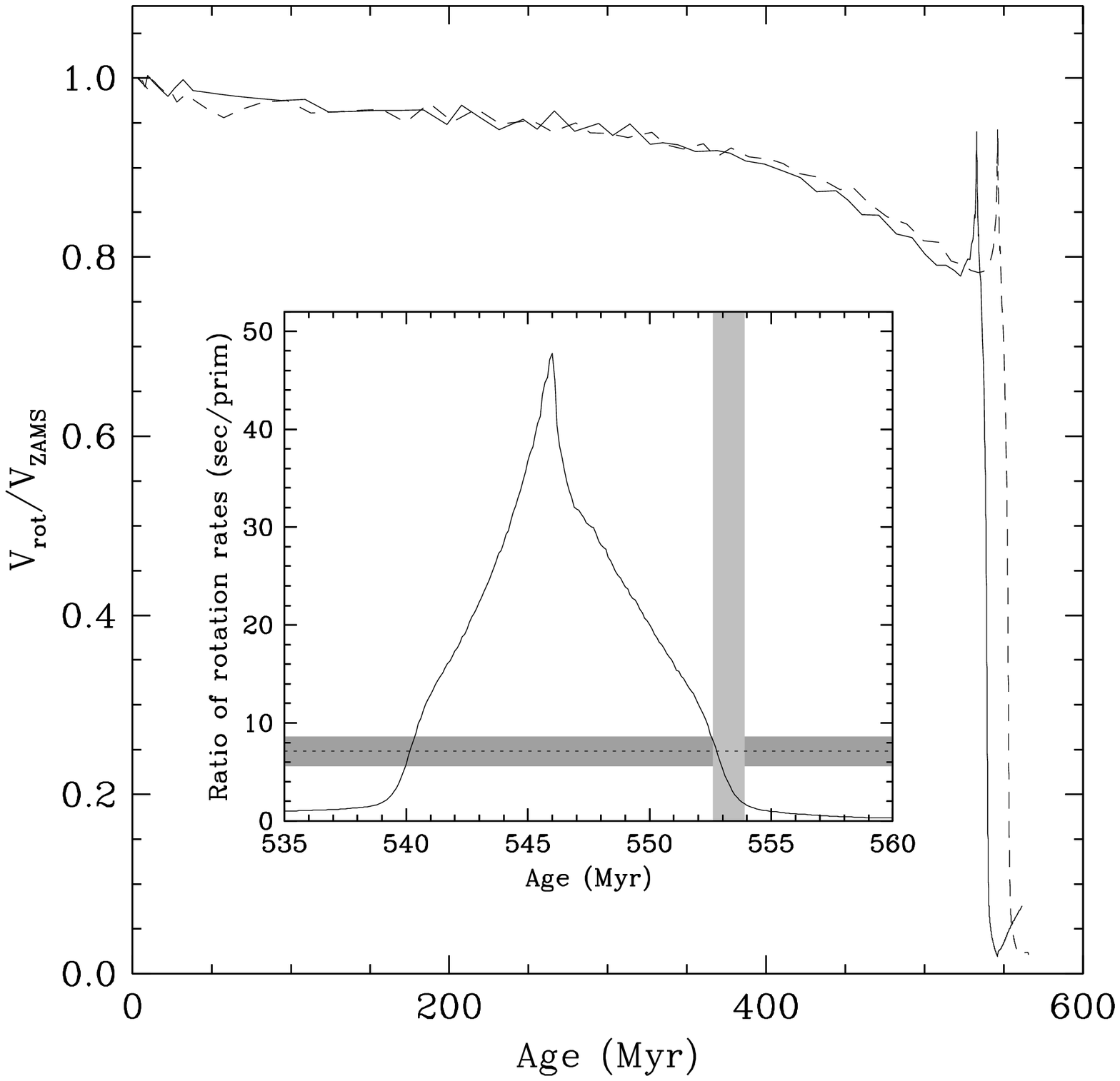}}
\vskip -0.02in

\figcaption[]{Evolution of the rotation rates of Capella A (solid
line) and B (dashed line) relative to their values on the ZAMS,
according to Claret models with $Z = 0.008$, $\alpha_{\rm ML} = 1.63$,
and $\alpha_{\rm ov} = 0.20$ (case~A). The inset shows the ratio of
these curves (secondary divided by primary) along with the measured
ratio of the projected rotational velocities ($v_{\rm B} \sin i /
v_{\rm A} \sin i = 7.1 \pm 1.5$; shaded horizontal band). The
theoretical ratio agrees with the measurement at the evolutionary age
of the system (shaded vertical band spanning the primary and secondary
ages from Table~\ref{tab:modelsc}). This suggests tidal forces within
the binary, which are not accounted for here, are
weak. \label{fig:rot}}

\end{figure}

None of the stellar evolution models considered so far include the
effects of rotation. Although the current rotation rates for Capella A
and B are slow or moderate, this was not the case for the
main-sequence progenitors, which were late B- or early A-type stars
spinning at typical equatorial velocities of 150--200~\kms.
Evolutionary tracks with rotation were computed for both components
following \cite{Claret:99} for the same set of parameters considered
in case~A above, under the assumption of rigid-body rotation. The
initial angular rotation rate $\omega_{\rm rot}$ was set by requiring
an approximate match between the predicted equatorial velocities at
the current age and the measured values (de-projected from $v \sin i$
by assuming the spin axes are aligned with the orbital axis; see
below). However, given that the rotation of the primary appears to be
synchronized with the orbital motion (\S\,\ref{sec:tidal}), one may
expect tidal forces to have had some effect on that star whereas the
secondary is farther from synchronization, and more likely to have
evolved as an isolated star.  Thus we only required a match to the
rotation of the secondary, and we adopted the same value of
$\omega_{\rm rot}$ for the primary on the grounds that the masses are
the same within 1\%, so all properties including rotation must have
been very similar to begin with.  The differences compared to
non-rotating models are most obvious on the main sequence, where
rotation renders the tracks some 500~K cooler, but are smaller toward
the later evolutionary stages. We find that the best match between the
rotating models and the observations is obtained for parameters nearly
identical to those found earlier, at an age only 5 Myr older. The
details of this comparison are included in Table~\ref{tab:modelsc},
case~E. The older age produced by these models compared to case~A is
reminiscent of the effect of overshooting, which extends the
main-sequence lifetime.  We also find that, although we did not
require a match, the predicted rotational velocity of the primary
reproduces the measured value exactly. This is an indication that the
effects of tidal forces in damping the rotation of the star are
relatively weak, given the long orbital period and relatively small
relative radii of the components ($R_{\rm A}/a \approx 0.075$ and
$R_{\rm B}/a \approx 0.055$, respectively). Another way of looking at
this is to examine the evolution of the rotation rate for each star
due solely to changes in the moment of inertia with time, under the
assumption of rigid-body rotation and conservation of angular momentum
without including tidal forces. These rotation rates relative to their
initial values on the zero-age main sequence (ZAMS) are depicted in
Figure~\ref{fig:rot} for both components as a function of time, for
the models with $\alpha_{\rm ML} = 1.63$.  The inset shows the ratio
of these two curves near the present age of the binary, which should
correspond to the actual ratio of the rotational velocities if the
stars started out with the same value on the ZAMS and if tidal forces
are unimportant.  There is in fact good agreement with the measured
ratio of $v_{\rm B} \sin i / v_{\rm A} \sin i = 7.1 \pm 1.5$
(horizontal shaded area) at the evolutionary age determined with these
models (vertical shaded area).

The comparisons made above against the stellar evolution models by
\cite{Girardi:00} and \cite{Claret:04} do not constitute very strong
tests of the predictive power of theory because a number of parameters
have been adjusted to produce a good fit to the observations of
Capella, namely, the overall composition, the mixing length, and
overshooting. We note that in both models the metallicity that seems
to be favored ($Z = 0.008$) is actually in very good agreement with
the spectroscopic value (\S\,\ref{sec:abundances}) derived from the
work of \cite{McWilliam:90}, despite our concerns about possible
biases in those measurements. Even if we accepted the latter as
accurate, and had \emph{imposed} that value on the models thereby
eliminating $Z$ as a free parameter, $\alpha_{\rm ML}$ and
$\alpha_{\rm ov}$ are still adjustable quantities. This is unavoidable
because of the phenomenological way in which convection and mixing are
treated in these ``standard'' models.

We have thus considered a third set of models based on the TYCHO
stellar evolution code \citep{Young:01a, Young:05} that implements
somewhat different prescriptions for some of the physical
processes. Predicted surface abundances are available for all the
elements in the network, including lithium, and rotation has not been
considered for the present work. The models calculated with the most
recent version of this code, TYCHO 8, have a partial implementation of
results from a new, non-local hydrodynamic description of convection
presented in \cite{Arnett:09}.  The key results implemented in this
version are the boundary conditions for convective zones. Convective
stability is evaluated based on the Richardson number ($Ri =
N^2/(\partial u/\partial r)^2$). This compares the potential energy in
stratification as measured by the Br\"unt-V\"ais\"al\"a frequency $N$
versus the kinetic energy in shear (including turbulent
velocities). Using the Richardson criterion results in larger
convective zones since thermodynamically stable regions in mixing
length theory can be hydrodynamically unstable. The complete velocity
integration described in \cite{Arnett:09} has not yet been implemented
for this work. Convective velocities are calculated according to
mixing length theory with a mixing length equal to the depth of the
convection zone, according to \cite{Arnett:09}. This eliminates
$\alpha_{\rm ML}$ as a free parameter. For very deep convective zones
the simulations suggest there may be a saturation in the mixing length
at a value of roughly 1.5--2.5 as dissipative processes become
important. We place this saturation near the lower limit at 1.64.  The
models also include wave-generated mixing from intermittent turbulence
and slow circulation according to \cite{Young:05}.  For most of the
star's history the convection is relatively shallow ($l_{\rm CZ} <
2h_p$, where $l_{\rm CZ}$ is the depth of the convection zone and
$h_p$ is the pressure scale height) except in the photosphere. In this
regime the convective description has no free parameters. On the RGB,
however, the convection zone has a very large density gradient over
its depth. This has two effects.  First, significant radiative losses
from a fluid parcel can occur over a single convective
traverse. Secondly, the sound speed drops with increasing radius and
decreasing density. Initially strongly subsonic flows can generate
shocks in the outer regions of the convective zone causing more
dissipation. Unfortunately, we do not yet have a satisfactory
quantitative theoretical description of this process, so we are forced
to parametrize it. We introduce a geometric parameter $g_{\rm ML}$
(compare to the geometric parameter introduced by
\cite{Bohm-Vitense:58} for mixing length) which describes the aspect
ratio of a convective fluid parcel. The value of the B\"{o}hm-Vitense
parameter gives a surface-to-volume ratio less than that of a
sphere. If we compare our parameter to the mixing length context,
values of 1 to 1000 are a reasonable range to account for true
geometric differences for large surface area turbulent flows,
radiative losses over large temperature gradients, and hydrodynamic
dissipative processes.

\begin{figure}
\epsscale{1.8}
\vskip -0.35in
{\hskip -0.98in\plotone{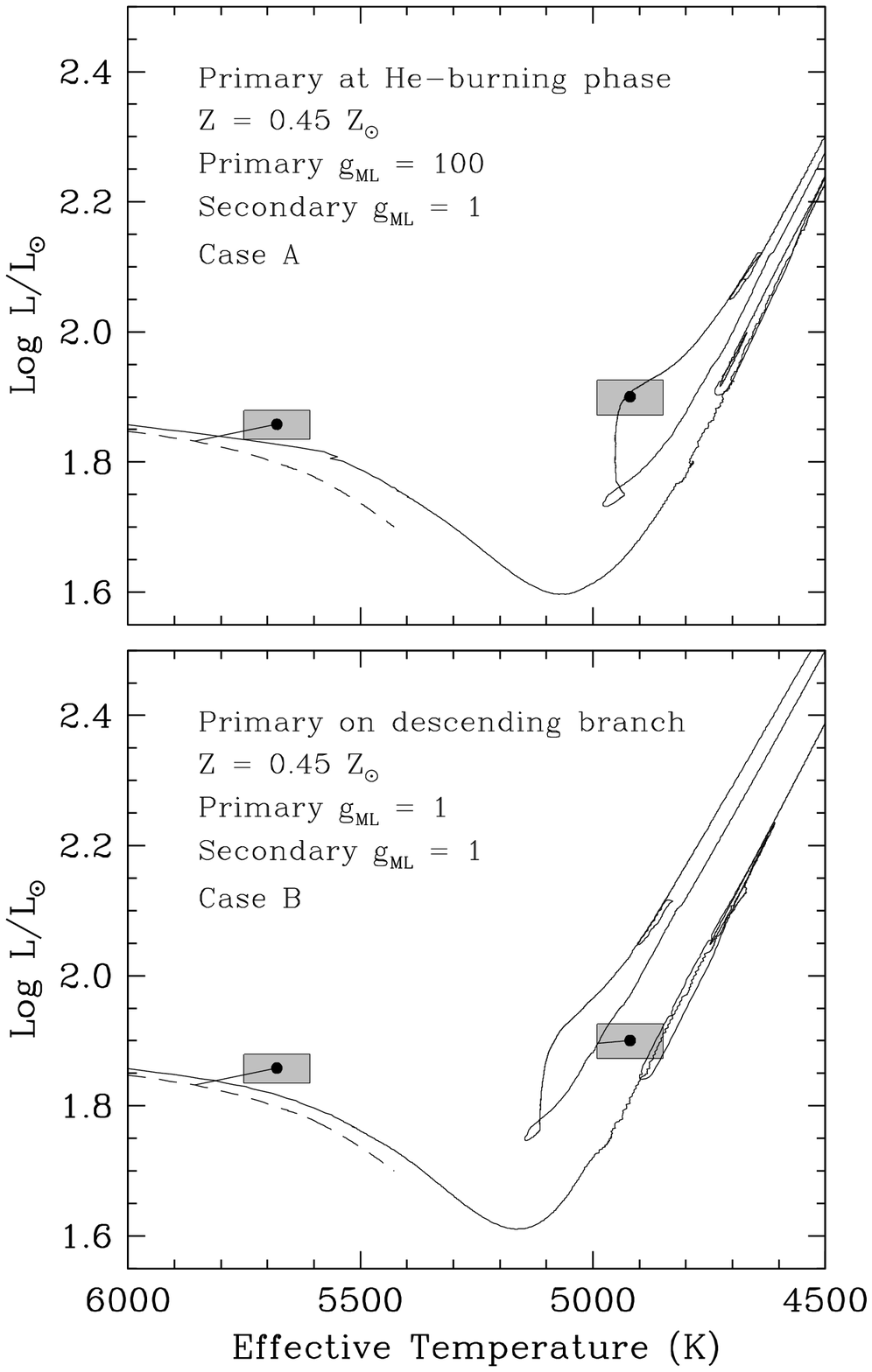}}
\vskip 0.6in
\figcaption[]{Evolutionary tracks for the measured masses of Capella
from the models of \cite{Young:01a} and \cite{Young:05}. Two different
scenarios are shown, focusing on the evolutionary state of the primary.
\label{fig:young}}

\end{figure}

Using these models in the same way as before, the best match to the
observations for Capella was found for evolutionary tracks with an
overall metallicity of $Z = 0.45 Z_{\sun}$ \citep[in which the solar
composition is adopted from][]{Asplund:05}, corresponding
approximately to [Fe/H] $= -0.35$. This is very similar to the
composition that gave the best fits in the previous sets of models.
For the secondary we used $g_{\rm ML} = 1$ since it is only beginning
to develop a deep convective envelope, and we are using a
parametrization instead of a quantitative physical description. A
modest increase may be supportable and provide a better fit.  For the
primary a value of $g_{\rm ML} = 100$ provides a reasonable match to
the global properties $M$, $R$, $T_{\rm eff}$, and $L$ with the star
on the red clump (Table~\ref{tab:modelsy}, case~A), giving a mean age
for the binary of $\sim$593 Myr and an age difference between the
components of about 6\%.  A value for $g_{\rm ML}$ of 1 gives a good
fit for the primary on the first descent of the RGB, but a poorer age
match with the secondary (see Table~\ref{tab:modelsy}, case~B).  An
intermediate value of $g_{\rm ML} = 10$ yields a less satisfactory
agreement, also on the first descent.  The above comparisons are shown
graphically in Figure~\ref{fig:young}. The C/N abundance ratios
predicted for the secondary in these two scenarios are quite
consistent with the measured value, while those for the primary are
more discrepant (see Table~\ref{tab:modelsy}).

The $^{12}$C/$^{13}$C ratios expected for the primary from the TYCHO
models (19.0--19.6, depending on the evolutionary stage) are still
significantly lower than the measured value of $27 \pm 4$, as found
before with the Claret models (18.9--19.2) as well as the
\cite{Bertelli:08} models ($\sim$19.0).  Other calculations indicate
very similar values \citep[e.g.,][]{Charbonnel:94}. It would appear,
therefore, that current theory systematically underestimates the
carbon isotope ratio in this case, regardless of where the primary is
located on the H-R diagram.  Unfortunately there is only a single
measurement of $^{12}$C/$^{13}$C for Capella~A, by \cite{Tomkin:76}. A
comparison of the results from that work for other stars that have
also been measured by others seems to indicate some systematic
differences, but it is difficult to tell which values are correct. For
example, measurements by \cite{Tomkin:76} of the four Hyades giants
$\gamma$~Tau, $\delta$~Tau, $\epsilon$~Tau, and $\theta^1$~Tau, which
have masses only slightly smaller than Capella, give $^{12}$C/$^{13}$C
ratios of 19, 23, 22, and 20, respectively, while \cite{Gilroy:89}
reported 24, 26, 26, and 27, all systematically higher. It is possible
that errors in the determination of the microturbulent velocity
$\xi_{\rm t}$ can explain these differences, since \cite{Keller:01}
and many others have shown that the $^{12}$C abundances are quite
sensitive to $\xi_{\rm t}$ because the $^{12}$CN lines from which they
are often measured are typically close to saturation.  A cursory
examination of measurements for other giant stars by different authors
indicates similar discrepancies, which leads us to believe the
uncertainties in some of these studies may be underestimated. A new
measurement of the $^{12}$C/$^{13}$C ratio for Capella~A would be
extremely helpful to settle this issue.

Table~\ref{tab:modelsy} shows that the surface lithium abundance
predicted for the secondary of Capella from the TYCHO models is in
agreement with the measurement, within the uncertainties, and is not
much reduced from its main-sequence value. The primary lithium
abundance, on the other hand, is larger than observed by about
3$\sigma$, regardless of the value of $g_{\rm ML}$. According to these
models $\log N({\rm Li})$ is not expected to dip much below 1.9 even
after the phase of core helium burning, as the star evolves up the
asymptotic giant branch. Thus, it appears lithium burning in Capella~A
has been more efficient than predicted by theory during the first
dredge-up, or there was additional burning afterwards. Unlike the case
of $^{12}$C/$^{13}$C, there are several independent measurements of
the \ion{Li}{1}~$\lambda$6708 line for the primary, and they display
rather good agreement. A possible explanation for the discrepancy is
that in the current TYCHO models rotational and convective/wave
sources of shear at the convective boundary are not correctly coupled.
Mixing at the base of the envelope in the models will be less than
that of the real star, and this could result in discrepancies in the
abundances of low-temperature burning products.

As seen from the comparisons above with three different sets of
evolutionary models, we are unable to find an evolutionary state for
the binary that gives complete consistency with all absolute
dimensions for both stars as well as all chemical indicators,
simultaneously. Partial matches are possible, indicating either
deficiencies in our understanding of stellar evolution, observational
errors in some quantities, or both.  Other evidence of the
evolutionary status of Capella has occasionally been considered by
some authors, including activity indicators in the UV and X rays. For
example, based on similarities with other stars, \cite{Ayres:83}
estimated that the primary may be responsible for as much as half of
the total 0.1--0.4 keV soft X-ray coronal emission from the system,
even though it only accounts for a small fraction of the far-UV
emission characteristic of the chromospheric layers.  A comparison
with the generally weaker coronal X-ray levels of clump giants led
them to conclude that Capella~A is crossing the Hertzprung gap for the
first time, and is yet to go through the first ascent. Although this
significant level of X-ray emission was subsequently confirmed by
others \citep[e.g.,][]{Linsky:98, Young:01b}, this picture seems
inconsistent with some of the chemical indicators discussed above.
Additionally, clump giants are known to display a wide range of
coronal emission levels \citep[see, e.g.,][]{Pizzolato:00}, perhaps
indicative of activity cycles analogous to that of the
Sun.\footnote{There may in fact be evidence that Capella itself is
undergoing such cycles.  \cite{Katsova:98} detected long-term
variations in the equivalent width of the chromospheric
\ion{He}{1}~10830~\AA\ line associated with the primary, with a
possible period of roughly 11 years. \cite{Raassen:07} found a similar
variability in X rays in the flux of the coronal Fe~XXI~128.73~\AA\
line, also associated mainly with the primary, coincidentally with an
apparent period also near 11 years.}  Virtually all UV and X-ray
studies after the mid 1980s have implicitly \emph{assumed} the primary
of Capella is a core-helium burning star, on the grounds that it would
be very unlikely to find both components of a binary in a very rapid
state of evolution such as the Hertzprung gap or the first ascent,
given that the secondary is already known to be a first crosser. This
could only happen if the mass ratio were very near unity, and that did
not appear to be the case until now.\footnote{The most recent
determination, by \cite{Barlow:93}, gave a mass difference of
5.2\%. In our analysis it is less than 1\%.}  \cite{Ayres:88} proposed
the primary may be in the early stages of core-helium burning, while
other giants that are even more active such as $\beta$~Cet and the
Hyades stars $\gamma$~Tau and $\theta^1$~Tau were considered to be in
a more advanced clump phase (``blue loop'').  Qualitative assessments
of the evolutionary state based on the overall activity level remain
problematic, and are not easily testable against models.

We conclude this section by pointing out that the detached nature of
Capella, and therefore its value for testing stellar evolution theory,
is beyond doubt. The size predicted for the primary at the tip of the
giant branch from any of the models considered above is less than
35~$R_{\sun}$, and would still fit comfortably inside its critical
Roche surface ($\sim$60~$R_{\sun}$). Thus, even if the primary were
already in the clump, mass transfer by Roche-lobe overflow would have
been avoided.

\subsection{Comparison with tidal theory}
\label{sec:tidal}

Capella is a particularly favorable case to study the rotational state
of the components. In addition to the measures of $v \sin i$ and the
radii, the rotational periods $P_{\rm rot}$ of both stars have been
measured as well. \cite{Shcherbakov:90} found that the radial velocity
of the chromospheric \ion{He}{1}~$\lambda$10830 absorption line
follows the velocity of the photospheric lines of the primary,
unambiguously identifying this activity feature with that star. The
equivalent width of the line was also seen to vary with a period
indistinguishable from the orbital period, which was interpreted as
the rotational signature of the primary. \cite{Katsova:98} confirmed
this with additional data, and measured a rotation period of
103.97~days. They were also able to detect a weaker component of the
\ion{He}{1} line associated with the secondary star. The radial
velocity of this feature again varied in phase with the photospheric
velocities for Capella~B, but its equivalent width was seen to vary
with a period of 8.25 days, representing the rotation of the
secondary. \cite{Strassmeier:01} detected the rotational signature of
both stars in their H$\alpha$ photometry, with periods of $106 \pm 3$
days for the primary and $8.64 \pm 0.09$ days for the secondary. These
observations make it very likely that the primary's rotation is
effectively synchronized with the orbital motion, while the secondary
is spinning more than 12 times more rapidly.  For the purpose of this
work we adopt a compromise value of $P_{\rm rot} = 8.5 \pm 0.2$ days
for Capella~B.

With these periods and the measured radii, the predicted rotational
velocities projected along the line of sight are $3.92 \pm 0.19$~\kms\
and $35.4 \pm 1.5$~\kms\ for the primary and secondary, respectively,
under the assumption that the spin axes are aligned with the orbital
axis. These predictions may be compared directly against the measured
$v \sin i$ values in Table~\ref{tab:vsini}. We note, however, that
most of those determinations --- including our own --- have ignored
the additional broadening that comes from macroturbulence. This can be
significant for giants and in our case is particularly important for
the primary, which is a slow rotator. The only two determinations of
$v_{\rm A} \sin i$ in the table that have accounted for
macroturbulence \citep{Fekel:86, Strassmeier:90} are both smaller than
ours, and average 4~\kms.  A giant star with the spectral type of
Capella~A is expected to have a macroturbulent velocity $\zeta_{\rm
RT}$ between 5 and 6~\kms\ \citep[see, e.g.,][]{Gray:82}, whereas the
synthetic templates we used in \S\,\ref{sec:specnew} to determine $v
\sin i$ have $\zeta_{\rm RT} = 1.5$~\kms. A correction for this
difference following \cite{Fekel:97} and \cite{Massarotti:08} brings
our $v \sin i$ value for the primary down from 6.5~\kms\ to about
5.0~\kms.  For the secondary this effect hardly matters, but a similar
correction assuming $\zeta_{\rm RT} \approx 7$~\kms\ leads to $v_{\rm
B} \sin i = 35.6$~\kms. These are the values we adopt in the following
(Table~\ref{tab:dimensions}). The agreement with the predicted
velocities, which is well within the errors, can be considered as an
indication that the absolute radii based on angular diameters are
accurate.

\begin{figure}
\epsscale{1.45}
\vskip -0.1in
{\hskip -0.45in\plotone{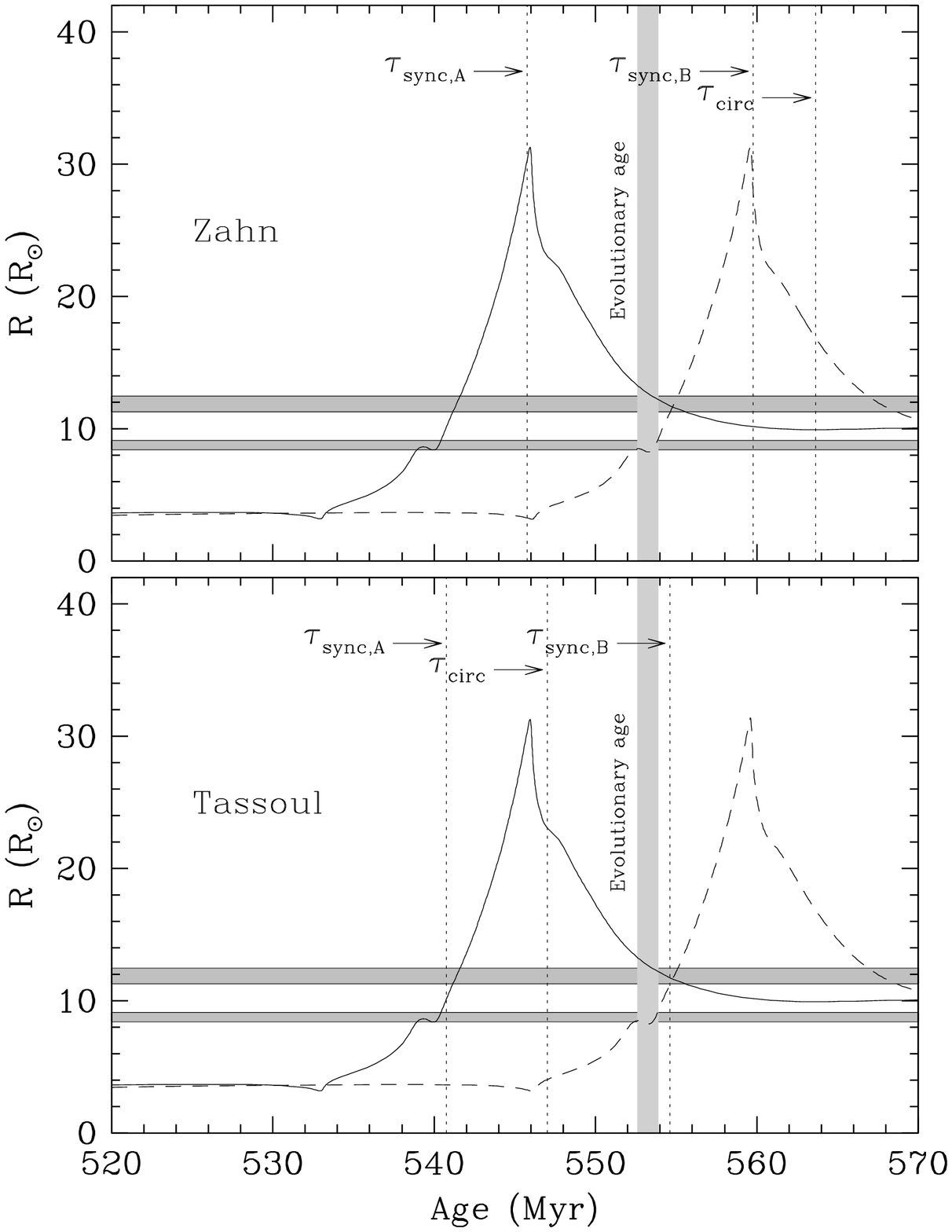}}
\vskip 0.5in

\figcaption[]{Critical times of synchronization ($\tau_{\rm sync,A}$
and $\tau_{\rm sync,B}$) and circularization ($\tau_{\rm circ}$) for
the components of Capella from tidal theory, shown in a diagram of
radius versus age. Solid and dashed lines represent the predicted
radii for the primary and secondary, respectively, according to Claret
models with $Z = 0.008$, $\alpha_{\rm ML} = 1.63$, and $\alpha_{\rm
ov} = 0.20$ (case~A). The measured radii are shown as horizontal
shaded regions, and the evolutionary age of the system is indicated
with a vertical shaded area. Calculations according to the tidal
theory of \citet[and references therein]{Zahn:92} are shown at the
top, and \citet[and references therein]{Tassoul:97} at the bottom.
\label{fig:tidal}}

\end{figure}

The binary system of Capella resembles in many respects that of the
76-day period eclipsing system TZ~For \citep{Andersen.etal:91}, which
is also composed of giants. Both have a circular orbit, and one
component in synchronous rotation while the other is
super-synchronous. This puzzling situation is of course explained by
tidal forces, which increase exponentially as the more massive
primaries evolve first and expand in size. The result is a sharp drop
in their rotation rates, and the damping of the orbital eccentricity.
We have computed the critical times of rotational synchronization and
orbital circularization for Capella according to the turbulent
dissipation and radiative damping mechanisms by \citet[and references
therein]{Zahn:92}, as well as the hydrodynamical mechanism of
\citet[and references therein]{Tassoul:97}. Note that we calculate
actual \emph{times} rather than time\emph{scales}, since the latter
would be completely meaningless for giant stars such as Capella that
have altered their structure drastically since the ZAMS, changing from
having radiative to convective envelopes. Thus we have integrated the
differential equations for the changes in the orbital eccentricity and
rotation rates along the evolutionary tracks following the
prescriptions of \cite{Claret:95, Claret:97}, to properly account for
the stellar properties during the radiative and convective phases of
evolution. The results obtained using the Claret non-rotating models
for Capella A and B for $Z = 0.008$, $\alpha_{\rm ML} = 1.63$, and
$\alpha_{\rm ov} = 0.20$ (case~A) are illustrated in
Figure~\ref{fig:tidal}. In this diagram the radii are plotted as a
function of age, and the times of synchronization and circularization
are indicated along with the evolutionary age of the system
($\bar{\tau} = 553.2$~Myr).  For the primary the theory of
\cite{Zahn:92} gives $\tau_{\rm sync,A} = 545.8$~Myr, and for the
secondary $\tau_{\rm sync,B} = 559.8$~Myr, both of which agree with
the observation that the primary already appears to have its rotation
synchronized with the orbital motion while the secondary is clearly
asynchronous. According to these models the spin of each star becomes
locked with the orbital motion when the components reach their maximum
size at the tip of the giant branch (Figure~\ref{fig:tidal},
top). Circularization occurs somewhat later, at $\tau_{\rm circ} =
563.6$~Myr. Strictly speaking, this is larger than the age of the
binary and the prediction is thus formally inconsistent with the
apparent circularity of the orbit, although perhaps not by much
considering the approximations in tidal theory\footnote{In particular,
the tidal equations used here are valid only for small departures from
synchronization, and do not include the effect that magnetic fields in
these active stars could have on angular momentum transport. The
secondary of Capella is rotating some 12 times faster than
synchronous, so the predictions should be taken with caution. An
additional approximation in our calculations is that the concomitant
changes in the orbital semimajor axis are ignored, although this is
likely to be a small effect for Capella.}. Similar calculations for
the mechanism of \cite{Tassoul:97} are shown in the bottom panel of
Figure~\ref{fig:tidal}, and indicate formal consistency with the
observations for all three critical times.

We have repeated the calculations according to both tidal theories for
the other scenarios considered earlier based on the Claret models, and
the agreement with the observations varies. The results are summarized
in Table~\ref{tab:tidal}. With one exception \citep[case~A and the
mechanism of][]{Tassoul:97}, in no other scenario are all three
critical times strictly consistent with reality, but the predictions
are so close to the best-fit evolutionary ages that the results are
very sensitive to the details of the models and the calculations. For
example, the mere addition of rotation to the first set of
evolutionary tracks examined above is enough to change the conclusions
entirely regarding tidal theory, even though it hardly changes the
stellar properties for the best fit. Thus, it is not possible to
establish which tidal theory is more appropriate for Capella. We note,
in this connection, that the theory of Tassoul has often been found to
be too efficient, at least for main-sequence stars, giving critical
times that are occasionally 1--2 orders of magnitude shorter than
those by Zahn. In Capella that is not the case. Given that the
critical times from these two theories tend toward each other in the
limit of long periods, this again is an indication that tidal forces
in this binary are relatively weak.

Another consequence of tidal forces in binaries is that the spin axes
of the stars tend towards alignment with the orbital axis. Because the
timescale for this is typically much shorter than for the other two
effects, at least for main-sequence stars, it is almost universally
assumed that the axes in binary systems are aligned. The simplifying
advantage is that one can then set the inclination of the rotation
axes equal to the orbital inclination, as we ourselves have done
above, for example to infer equatorial rotational velocities. We note
however, that this assumption may not always hold \citep[see, e.g.,
the case of $\lambda$~Vir;][]{Zhao:07}. The timescale for alignment
depends strongly on the ratio between the orbital and rotational
angular momenta. This ratio is large for Capella because of the long
orbital period, so the timescales for alignment are close to those for
synchronization.  As an example of such calculations, we use the
rotating Claret models with $Z = 0.008$, $\alpha_{\rm ML} = 1.63$, and
$\alpha_{\rm ov} = 0.20$ (case~E), and find that the alignment times
according to \cite{Tassoul:97} are 555.9~Myr for the primary and
563.6~Myr for the secondary. These are again so near the evolutionary
age of 558~Myr for this scenario that it is difficult to draw general
conclusions.  At face value the primary is formally predicted to have
reached alignment, while the secondary is still a few Myr from
reaching it.  The critical times for alignment according to
\cite{Zahn:92} are 562.3~Myr and 570.2~Myr, both of which are formally
greater than the age, implying that the spin axes are not necessarily
aligned.  In Capella there is an observational constraint on this
prediction, given by the rotation periods, $v \sin i$ values, and
absolute radii determined for the two components. As indicated earlier
in this section, there is very good agreement (particularly for the
more discriminating secondary) between the measured $v \sin i$ values
and the predicted rotational velocities, when the latter are projected
along the line of sight using the well-known orbital inclination
angle. This is evidence that the orbital axes are in fact closely
aligned with the orbit, and we conclude that tidal theory does not
clearly agree with the observations in this instance. 

\section{Concluding remarks}
\label{sec:finalremarks}

With our new spectroscopic observations of Capella and a global
orbital solution that uses all available astrometric and
radial-velocity measurements, the precision of the most important of
the physical properties of the stars --- their masses --- has been
improved threefold compared to previous values, and now stands at
0.7\% for the primary and 0.5\% for the secondary.  These are at the
level of the best available determinations from double-lined eclipsing
binaries \citep[see, e.g.,][]{Andersen:91}. As others have pointed
out, the mass \emph{ratio} is perhaps more important in this case
since it sets constraints on the differential evolution of the stars;
that quantity is now known to even higher precision (0.36\%).  Based
on the tests and crosschecks documented in this paper, it is our
belief that the \emph{accuracy} has also improved significantly. Our
dynamical masses differ from earlier results by 9\% and 5\%,
respectively, which has an enormous impact on the evolutionary state
of giants such as these.

Our goal here has been to bring to bear all relevant information on
the physical properties of Capella for a comprehensive discussion of
the system in terms of its evolution.  In addition to the masses, we
report significantly improved determinations of the absolute radii,
the effective temperatures, the luminosities, the rotational state of
the stars, and the chemical composition, including the overall [m/H]
index, the lithium abundance, and the $^{12}$C/$^{13}$C and C/N
ratios.  We know of no other evolved system with so much high-quality
information available. Sanity checks have been performed, whenever
possible, to assess the accuracy of these quantities. Our expectation
at the outset was then that we should be able to pinpoint the
evolutionary state of these giants, particularly the primary, and
validate current models of stellar evolution that seem to work so well
on the main sequence.  Instead, we find that there are now so many
constraints that no single model is able to satisfy them all
simultaneously, \emph{even when adjusting some of the free parameters
in theory}.

Three types of stellar evolution models have been considered, two
based on standard convection prescriptions \citep{Girardi:00,
Claret:04} and one with an alternative formulation \citep[TYCHO
evolution code;][]{Young:01a}.  We have also explored the role of
rotation. In most cases it is possible to obtain reasonably good
matches to $M$, $R$, $L$, and $T_{\rm eff}$ for both components at a
variety of evolutionary states for the primary, of which the clump
does not particularly provide one of the best matches. This is in
contrast with most previous studies, which overwhelmingly place the
primary in the core-helium burning phase, and is largely due to the
fact that the mass ratio is now much closer to unity than previously
believed (within 1\%, versus 5\% previously). This allows the
``evolutionary gap'' between the components to narrow considerably.
These four properties alone are therefore insufficient to discriminate
between different models or different states for Capella~A.

Chemical constraints are seen to be particularly powerful in this
case. We find that the overall metallicity [m/H] required for the
models to match the properties of the two stars is in excellent
agreement with the spectroscopic determination. However, the predicted
$^{12}$C/$^{13}$C values for the primary are systematically too low
(at the $\sim$2$\sigma$ confidence level), if the measurement is to be
taken at face value, and would require that component to be near the
base of the RGB. This is inconsistent with the well-determined
temperature and luminosity (or radius). Errors in the rather delicate
measurement of the carbon isotope ratio cannot be ruled out, and a new
determination would be highly beneficial. Similarly, models tend to
overestimate the apparently well-measured lithium abundance of the
primary (discrepancies at the $\sim$3$\sigma$ level), although they do
match that of the secondary, which is essentially the initial
value. The observed C/N ratios only permit the primary star to be on
the descending branch prior to the helium-burning phase, or at the
clump. Ascending branch scenarios are ruled out at the many-$\sigma$
level. A location on the descending branch would in fact be more
likely than on the first ascent, given that the descent takes about 3
times longer (see footnote~\ref{foot:times}).

It may well be that the primary is indeed a clump giant, but the fact
remains that current stellar evolution models fail to satisfy all of
the observational constraints when considered jointly, for both
components at the same time, and at a single age, whether the primary
is in the core-helium burning phase or at any other location.
Predictions are still very strongly dependent on the physics in the
models, underscoring our incomplete understanding of many of these
processes.  The $^{12}$C/$^{13}$C measurement suggests that convective
mixing in the primary is less efficient than predicted, while the
lithium determination appears to shows the opposite. It is also quite
possible that the masses, as precise as they already are, need to be
improved further by an order of magnitude given the sensitivity of the
predictions. If that is the case, then both better spectroscopy and
better astrometry will be required, since the latter contributes a
non-negligible $\sim$0.3\% to the present mass errors.

Tidal theory does not seem to fare much better. The circular orbit of
Capella along with the evidence for synchronous rotation of the
primary and asynchronous rotation of the secondary pose strong
constraints that are in general not quite met by the mechanisms of
\cite{Zahn:92} and \cite{Tassoul:97}. Similarly with the alignment of
the rotational axes with the orbital axis, which is observationally
constrained by the measurements of $R$, $v \sin i$, and $P_{\rm rot}$.
To be fair, however, the critical ages for circularization,
synchronization, and alignment inferred from these models for the
various scenarios considered are typically close to the evolutionary
age in each case, so the results may be sensitive to the details of
the calculations and are not conclusive.

Despite these challenges, much has been learned about Capella in its
more than 100 years of observational history. It has been a favorite
target of virtually every space facility and has provided, among many
other results, very important insight into the structure and other
properties of the coronae in active stars.  Because it is so bright,
nearby, and easy to observe, Capella remains a very important
benchmark for testing our understanding of stellar physics in the
advanced stages of evolution.

\acknowledgements

It is a pleasure to acknowledge J.\ Andersen for originally suggesting
we observe Capella spectroscopically, as well as J.\ Caruso, R.\ P.\
Stefanik, and J.\ M.\ Zajac for gathering most of the observations
used here, and R.\ J.\ Davis for maintaining the CfA echelle
database. L.\ Kaltenegger provided translations from the German
literature. We also thank the anonymous referee for a careful reading
of the manuscript and helpful suggestions. This work was partially
supported by NSF grant AST-0708229. The research has made use of the
SIMBAD database, operated at CDS, Strasbourg, France, of NASA's
Astrophysics Data System Abstract Service, of the Washington Double
Star Catalog maintained at the U.S.\ Naval Observatory, and of data
products from the Two Micron All Sky Survey (2MASS), which is a joint
project of the University of Massachusetts and the Infrared Processing
and Analysis Center/California Institute of Technology, funded by NASA
and the NSF.

\appendix

\section{The Greenwich visual observations of Capella}
\label{sec:greenwich}

The relatively long orbital period of Capella from the spectroscopic
studies of \cite{Newall:99} and \cite{Campbell:01} immediately
suggested to investigators from the beginning of the 20th century the
possibility that it might be detected as a visual binary, with a
separation somewhat under 0\farcs1, according to the best estimates at
the time.  Foreseeing the potential of interferometry,
\cite{MillerBarr:00} suggested it should be resolvable with
Michelson's ``interference apparatus'' on a large telescope, although
this would only come to happen two decades later. In the meantime,
visual binary observers lost no time in attempting the measurement,
even though predictions indicated the separation would be at or below
the resolution limit of available telescopes.\footnote{In such cases
the binary nature of the object is often obvious from the elongated
images, and experienced visual observers can still make rough
estimates of the angular separation, especially if the components are
of similar brightness, as in Capella \cite[see,
e.g.,][]{Aitken:64}. Even if estimates of $\rho$ are not possible, a
meaningful measure of the direction of this elongation (position
angle) can sometimes be made for systems that are considerably closer
than the nominal resolving power. In any case, under these
circumstances, the result can depend significantly on the observer's
mental disposition, not to mention the observing conditions.}
\cite{Aitken:00} described five separate attempts with the Lick
Observatory 36-inch Clark refractor (resolving power $\sim$0\farcs15),
and on only one of those dates (1900 February 24) did he indicate that
once or twice a slight elongation of the image was suspected, but that
repeated measurements of the position angle showed a spread of some
60\arcdeg, which he considered large enough to be a sign that no real
elongation was observed.  \cite{Hussey:00, Hussey:01} used the same
telescope to look at Capella on multiple occasions, but also reported
the images to be essentially round.

At about the same time, visual binary observers at the Greenwich
Observatory took an interest as well, and reported seeing the image
elongated with the 28-inch refractor (resolving power $\sim$0\farcs19)
on 1900 April 4. They even recorded a position angle measurement for
Capella.  Over the next two years they continued observing the star
regularly and making measurements of $\theta$, and on a few occasions
they even ventured to estimate $\rho$. More than 100 position-angle
measurements were made by 11 different observers, including the
Astronomer Royal \citep{Christie:00a, Christie:00b, Christie:02,
Christie:03}. These observations have been viewed with suspicion ever
since \citep[e.g.,][]{Burnham:06}. As pointed out by
\cite{Ashbrook:76}, it is difficult to see how ``a smaller instrument
at sea level could consistently do what the 36-inch Clark refractor on
Mount Hamilton found impossible''. No other visual observations
elsewhere appear to have been successful; the only other recorded
attempts, decades later by \cite{Wilson:39, Wilson:41}, saw no
elongation in Capella.

The most remarkable fact about the Greenwich observations, as
discussed by \cite{Ashbrook:76}, is their overall agreement with the
astrometric orbit of \cite{Merrill:22}, which was not published until
\emph{twenty years later}. Merrill himself had compared the Greenwich
measurements with his orbit, but other than noting some relatively
minor discrepancies, he made no judgement about them. \cite{Goos:08}
had also discussed these measurements in connection with his own
spectroscopic observations, but this was also long before the
astrometric orbit was known. Aside from these two studies, the
Greenwich observations appear to have been largely ignored.

The refined orbital solution derived in the present work enables us to
examine the accuracy of these measurements more closely. The
circumstances of each observation were reconstructed based on
published information.\footnote{One misprint was corrected for the
observation on 1900 November 9 \citep{Christie:00b}. The time of
observation was adjusted to correspond to the hour angle also reported
for the measurement. In all other cases we found reasonable agreement
between the times (GMT) and hour angles listed, within about 2 hours.}
Most of the position angle estimates were made during the daytime,
some at night, and a few during twilight, and all of them at hour
angles greater than about 2\case{1}{2} hours (east or west) because of
pointing limitations. A group of observations from 1900 April 23 to
May 11 were considered by \cite{Christie:00b} to be influenced by
color effects, and we have not considered them here.

The first peculiar coincidence is that the observers correctly
recorded the position angle of Capella to be decreasing with time
(retrograde motion on the sky), which, as noted by \cite{Ashbrook:76},
they could not have known in 1900 since the only information available
to them was that provided by the spectroscopic orbit. Secondly, the
observations correctly placed the ascending node in the first quadrant
as opposed to the third. Given the knowledge of the spectroscopic
orbit, this can only be determined unambiguously from visual
observations and if a difference in brightness is detected, which the
Greenwich observers apparently saw, as reported by
\cite{Christie:00b}. We now know that this difference is quite small
in the visible, and as pointed out earlier, observers many decades
later were still struggling with the difficulty of quadrant
determination (see \S\,\ref{sec:lightratio}). Furthermore, the actual
position angle of the node implied by these observations is very
roughly the correct value.  The modern determination gives $\Omega =
40\fdg4$, and the estimate reported by \cite{Christie:00b} after the
first year of observation is $\Omega = 80\arcdeg$. However, this value
may be influenced by the fact that the Greenwich observers allowed the
orbit to be eccentric, and in fact obtained a small but apparently
significant value of $e = 0.05$, whereas we now know the orbit is
essentially circular (see \S\,\ref{sec:orbit}).  Repeating the
calculation for a circular orbit, and holding the period and epoch of
nodal passage fixed at their known values, we obtain a revised
estimate of $\Omega = 57\arcdeg$ from the full sample of original
Greenwich measurements, which is even closer to the modern
value. Finally, the orbital inclination angle from this revised
solution is $i = 152\arcdeg$, also not far from the correct value of
$137\fdg2$.

\begin{figure}
\epsscale{1.25}
\vskip -1.1in
{\hskip -0.18in\plotone{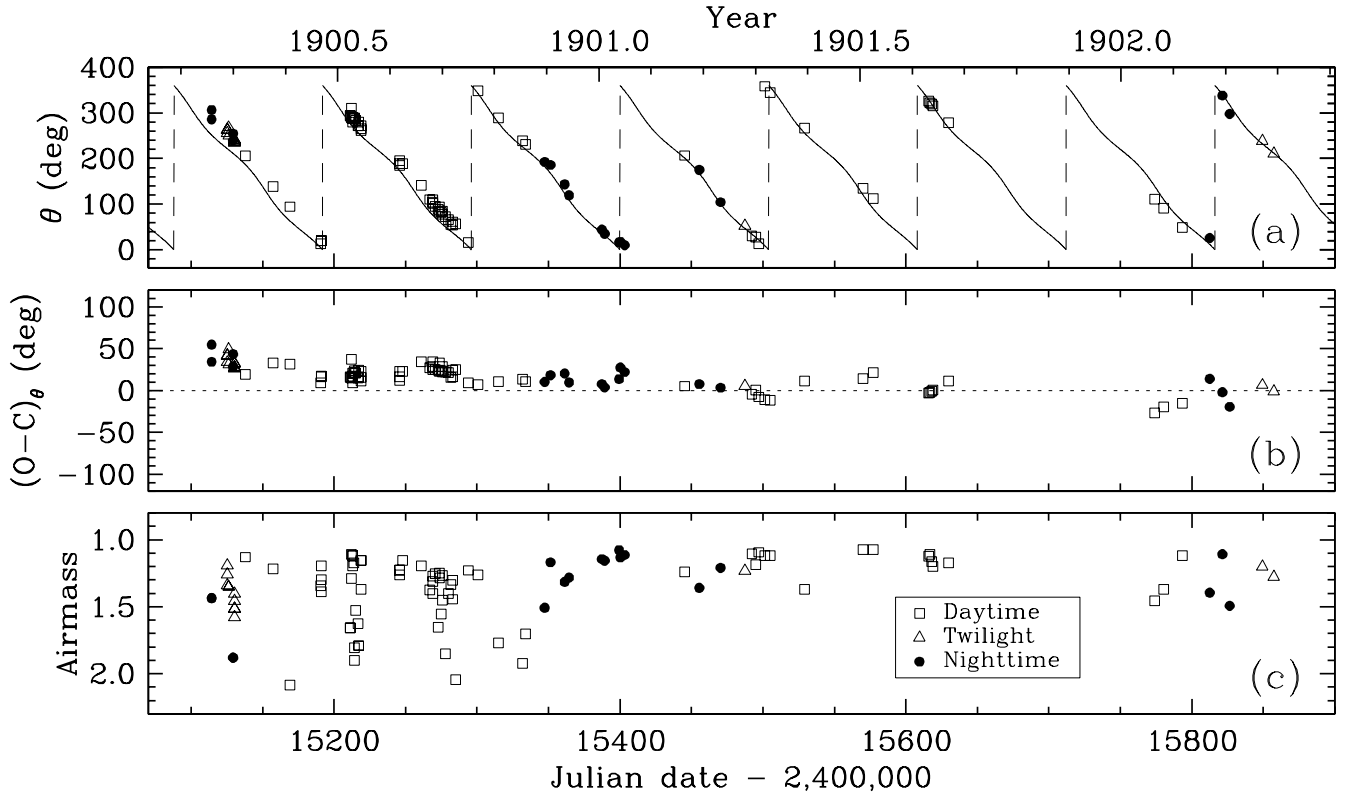}}
\vskip -2.1in

\figcaption[]{Greenwich estimates of the position angle of Capella
made from the elongated images. The different symbols indicate whether
the observation was made during the daytime, twilight, or nighttime.
(a) $\theta$ as a function of time, together with the computed orbit
from the solution in this paper (which excludes these measurements);
(b) $O\!-\!C$ residuals of the $\theta$ measurements from the computed
orbit; (c) Airmass of each observation.
\label{fig:greenwich}}

\end{figure}

In Figure~\ref{fig:greenwich}a we show the position angle measurements
as a function of time, compared with the prediction from our own
orbital solution. They cover 7 full cycles of the binary. The second
panel displays the $O\!-\!C$ residuals.  A decreasing trend with time
is clearly visible, although the magnitude of the residuals is quite
modest considering the difficulty of the measurements, which can
easily have errors of tens of degrees.  The bottom panel (airmass
versus time) shows that during the first few months the observers
often made the measurements at large airmasses, whereas later on they
observed closer to the zenith. It is also during those first few
months that the residuals display the largest positive
trend.\footnote{\cite{Merrill:22} had also noticed positive residuals
averaging +18\arcdeg\ from his orbit, and speculated that the position
angle of the node might be decreasing with time (see
\S\,\ref{sec:orbit}).}

The agreements described above with the modern orbit of Capella are
most intriguing.  While it is certainly true that these measurements
were made pushing the limits of what is possible for the visual
technique, and that some of the most experienced observers of the time
failed repeatedly to see any convincing elongation in the star's image
even using larger telescopes, the Greenwich observations are not
easily dismissed.  \cite{Ashbrook:76} suspected that some of the
measurements could have resulted --- consciously or subconsciously ---
from foreknowledge of what the motion of Capella should be, based on
the spectroscopic orbit that had just been published. The observers
would certainly have known the average daily rate of change of the
position angle, but nothing about variations in that rate throughout
the orbit due to the inclination angle, or the direction of motion,
and most certainly they could not have known in which quadrant the
ascending node would be. It is quite likely, however, that since the
period was known the measurements are not entirely independent, and
some may well be wild guesses by inexperienced observers. We note, for
example, that the quadrants of the individual observations as
published are all consistent with the modern orbit, with the exception
of two that need to be reversed.\footnote{Both are by the same
observer (T.\ Lewis).  Interestingly, on one date the quadrant was
stated in the observing log as being ambiguous, and on the other the
measurement was qualified as uncertain.} Rather than implying that the
observers were able to determine the correct orientation in every case
by detecting the slight difference in brightness, it seems more likely
that they adjusted their $\theta$ values by $\pm$180\arcdeg\ prior to
publication, as needed for consistency with the initial measurements.
This practice is still followed today, but indicates at the very least
that the Greenwich observers were paying close attention to the
agreement with the spectroscopic orbit. Three observers contributed
almost 85\% of all measures shown in Figure~\ref{fig:greenwich}. About
half of them were carried out by W.\ W.\ Bryant, and another third by
W.\ Bowyer and T.\ Lewis. No systematic differences are seen among
them. The very detailed observing notes they left indicate that they
also made similar measurements of other very close binaries on some of
the same nights on which they observed Capella, and to our knowledge
those results have not been called into question.

After more than a century the Greenwich observations remain a mystery.
H.\ F.\ Newall, who in his earlier spectroscopic study at that
Observatory had discovered the binary nature of Capella and encouraged
the visual observers to turn their attention to it \citep{Newall:00},
later expressed some doubts about those measurements in meetings of
the Royal Astronomical Society, but believed that at least some were
real \citep{Newall:08}.  Bryant himself, the most active observer,
took a similar view. In his own words \citep{Bryant:07}, ``though many
of the observations published are probably worthless, and may be
ascribed to optical or psychological causes and to want of experience
in judging of the suitability of observing conditions for what was
practically a unique observation, yet there is little doubt that a
certain proportion of the observations can be regarded as genuine.''

\section{Notes on the astrometric observations}
\label{sec:astrometrynotes}

We describe here the particulars of the astrometric measurements for
Capella, including corrections to some of the original sources, or
reasons that led us to reject some observations from the final
solution.

\noindent\emph{\cite{Anderson:20} and \cite{Merrill:22}}: The six
measurements reported by \cite{Anderson:20} were re-reduced and are
superseded in the paper by \cite{Merrill:22}. The P.A.\ for first of
these six (on JD $2,\!422,\!323.65$) is admittedly very rough, and has
been assigned an uncertainty of 10\arcdeg, following the
recommendation of the authors.  The last of those 6 measurements (on
JD $2,\!422,\!438.63$) was made during full daylight, and thus an
angular separation was not reported because it was affected by the
blue background that shifted the effective wavelength. The P.A.\
should be unaffected. The measurements reported in Table~II by
\cite{Merrill:22} are nightly averages, but smearing should be
insignificant given that the observing intervals were always less than
3 hours.  The relative uncertainties used here for all these
measurements account for the weights given by \cite{Merrill:22}, and
are scaled as described in \S\,\ref{sec:orbit}. We note in passing
that the dates of these observations have been translated into
Bessellian epochs incorrectly in the orbital studies of
\cite{McAlister:81} and \cite{Barlow:93}.  The values used by those
authors are exactly 0.5 days too small, possibly as a result of
overlooking the fact that prior to 1925 astronomical time (GMT) was
counted from noon, rather than midnight. The change in the P.A.\ and
separation of Capella over half a day can be up to $2\fdg4$ and
0.45~mas, respectively.

\noindent\emph{\cite{Kulagin:70}}: The original position angle for the
interferometric observation on 1968 February 18/19, listed as
$303\fdg5 \pm 1\fdg0$, is apparently incorrect \citep{McAlister:81},
and was changed by the latter author to $313\fdg5 \pm 1\fdg0$
following a private communication from W.\ S.\ Finsen. Even after this
change, the P.A.\ measurement gives a very large residual, and is
excluded from our solution. Nothing wrong is seen with the separation
measurement on this date.

\noindent\emph{\cite{Gezari:72}}: A speckle measurement of the angular
separation is given with no epoch. It is compared in Table~1 of the
original paper with predictions from the orbit by \cite{Merrill:22}
for JD $2,\!441,\!034.0$, which may then be taken as the date of
observation. This date is likely to be rounded off, however, since the
star was below the Palomar Observatory local horizon at the time. It
corresponds to 1971 March 23, which also disagrees with indications in
the text saying that ``observations were made during twilight on 11
dates in 1971 April, June, and October''. A later paper by the same
authors \citep[][see below]{Labeyrie:74} does report measurements from
the same site on 1971 March 23, so it seems likely they correspond to
the same observations, although the angular separation reported is
slightly different (possibly as a result of new reductions). We assume
here that the latter measurements supersede the original one by
\cite{Gezari:72}.

\noindent\emph{\cite{Labeyrie:74}}: The dates for the ten speckle
observations given here are only reported to 0.01 of a year, which is
insufficient given that orbital motion is $\sim$10\arcdeg\ and
$\sim$2.1 mas over this time. Previous investigators
\citep[e.g.,][]{McAlister:81, Barlow:93} have used the dates at face
value, not surprisingly with poor results. Furthermore, the first
seven measurements were reported by \cite{Labeyrie:74} with precisely
the same date (1971.23, presumably in Bessellian years), and were
therefore simply averaged by subsequent investigators. It is clear
from the text that the first four of these were taken on the night of
1971 March 23, probably during twilight at Palomar since the star was
setting.  We assign to these observations the date JD
$2,\!441,\!034.67$. This is unlikely to be in error by more than 2
hours, which is good enough for our purposes. Inspection of
measurements for other binaries from this source, as well as those of
Capella, indicates that the position angles are systematically too
small by several tens of degrees, a fact also noticed by others
(despite the date problem). This may be due to a calibration error in
these early speckle observations \citep{McAlister:81}, and we have
chosen not to use those angles here. The remaining observations of
\cite{Labeyrie:74} were likely taken a few nights later, but
sufficiently precise dates are unavailable so we have excluded them.

\noindent\emph{\cite{Blazit:77a}}: The results of these
interferometric measurements are taken from Figure~1. The date for the
last one (1977.207) is given to one more significant digit in the text
(1977.2066). Dates are assumed to be Bessellian years, more common in
binary star publications, and also because the assumption of Julian
years for the 1977.2066 measurement places the star below the
horizon. All dates are converted to Julian days for use in the present
work. The observation on 1977.125, which corresponds formally to JD
$2,\!443,\!189.6$, is considered to have been made 0.1 days earlier,
when the star was at a more plausible hour angle. This is well within
the uncertainty in the original date (0.001 years), which corresponds
to $\pm0.36$ days. We note that the change in $\theta$ over 0.36 days
is about 1\fdg6 on this date ($\rho$ only changes by 0.2 mas), so our
date correction is not entirely insignificant.  Uncertainties for the
$\theta$ and $\rho$ measurements are estimated graphically from
Figure~2 of this source.

\noindent\emph{\cite{McAlister:77}}: The position angles of these
speckle measurements were later slightly adjusted by the original
author.

\noindent\emph{\cite{Weigelt:78}}: Capella was unresolved in this
speckle measurement made at an unspecified date with the 1.8m
telescope at the Asiago Observatory (Italy).

\noindent\emph{\cite{McAlister:82a}}: The first of these speckle
measurements, which were made at the resolution limit of the 2.1m
telescope at Kitt Peak Observatory, gives large residuals both in
$\rho$ and $\theta$, and was excluded from the solution.

\noindent\emph{\cite{Hege:83}}: The original speckle observations
giving $\rho = 0\farcs042 \pm 0\farcs001$ and $\theta = 331\arcdeg\
\pm 2\arcdeg$ (quadrant reversed for the present work) were reduced
independently by \cite{Bagnuolo:83a}, who obtained $\rho = 0\farcs0420
\pm 0\farcs0015$ and $\theta = 333\fdg0 \pm 1\fdg2$. We have adopted
here the average of these determinations, with the more conservative
uncertainties. The date assigned to this observation (not precisely
given by either author) is JD $2,\!444,\!637.6$, which corresponds to
a reasonable hour angle at the Kitt Peak Observatory during twilight
on UT 1981 February~2.

\noindent\emph{\cite{Koechlin:83}}: The first three interferometric
measurements are apparently repeated from an earlier report by
\cite{Koechlin:79}, although they are slightly different in two
cases. We consider them to supersede the original measurements. The
date for the last of these three is also different from that given in
the original, and is the one used here.

\noindent\emph{\cite{McAlister:87}}: These speckle measurements were
later slightly corrected by the original authors.

\noindent\emph{\cite{McAlister:89}}: The speckle observation on JD
$2,\!446,\!895.62$ (Bessellian epoch 1987.2717) gives very large
residuals in both $\theta$ and $\rho$, and is excluded from our
solution.

\noindent\emph{\cite{Isobe:90}}: Capella was unresolved in this
speckle measurement made on 1988.8031 (Bessellian epoch) at the 2.12m
telescope at the San Pedro M\'artir Observatory (Mexico).

\noindent\emph{\cite{Hummel:94}}: Dates given in Table~2 of this
source have been converted from Bessellian years to Julian dates for
use in the present work. The interferometric visibilities on which the
results are based have not been published, but were condensed into and
published as $\theta$ and $\rho$ measurements for each night of
observation (at 8$^{\rm h}$ UT).

\noindent\emph{\cite{Baldwin:96}}: The P.A.\ and separation values
used in the present work were determined graphically from Figure~2 of
this source, since they were not given explicitly. The epoch of
observation for the first interferometric measurement is deduced from
the calendar date mentioned in the text, along with the time during
the night (2$^{\rm h}$ UT) from Figure~3. The second measurement is
assumed to be taken at a similar time during the night of 1995
September 29.

\noindent\emph{\cite{Young:02}}: Exact times of observation for these
direct-imaging HST measurements were not given in the original paper,
but have been recovered from the HST online archives. The position
angles are off by nearly 40\arcdeg\ for unknown reasons, and have not
been used here.

\noindent\emph{\cite{Kraus:05}}: The time of observation is taken to
be the middle of the interval given for each measurement in Table~2 of
this work.

\section{Coronal abundances in Capella}
\label{sec:coronal}

\begin{figure}
\epsscale{1.1}
\vskip -1.5in
{\hskip 0.2in\plotone{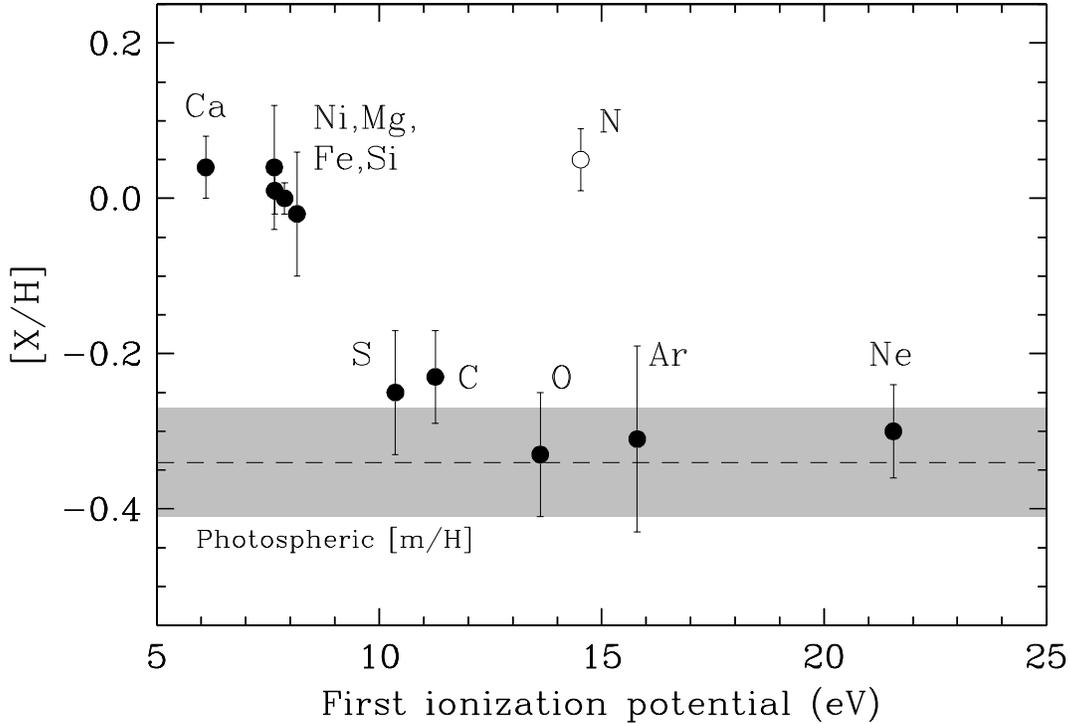}}
\vskip -1.1in

\figcaption[]{Measurements of the coronal abundances in Capella from
X-ray observations (combined flux of the two components) as a function
of the first ionization potential of each element. The values shown
are averages of several sources given in Table~\ref{tab:coronal}, and
are taken from the last column of that table.  Error bars represent
the error of the mean, and all measurements are referred to the solar
abundances as listed by \cite{Grevesse:98}. Nitrogen is marked with a
different symbol because it is expected to be enhanced due to
convective mixing in the primary star occurring at the first
dredge-up. The overall photospheric composition of Capella, [m/H] $=
-0.34 \pm 0.07$ \citep{McWilliam:90}, is indicated with the dashed
line and gray error box, and refers to the same solar abundances.
\label{fig:coronal}}
\vskip 0.1in
\end{figure}

There is a significant body of literature on X-ray observations of
Capella, which is not surprising given that it is the brightest steady
coronal source in the X-ray sky. The available abundance
determinations made from coronal emission-line fluxes have been
collected in Table~\ref{tab:coronal}, and have been adjusted to
conform to the standard solar abundances by \cite{Grevesse:98}, for
consistency with the photospheric determinations (see
\S\,\ref{sec:abundances}). Coronal abundances in the Sun are known to
depend on the first ionization potential (FIP), although the
large-scale element fractionation processes presumably involved are
not completely understood.  Elements having low FIP ($< 10$ eV) are
overabundant compared to those with high FIP ($> 10$ eV). We indicate
the FIP values in Table~\ref{tab:coronal}, along with the average
abundance for each element in the last column.\footnote{Some
systematic differences between studies are apparent, but their
investigation is beyond the scope of the present work.}  The FIP trend
has also been seen in some active stars, while others appear to show
the opposite effect. For Capella there has been some debate about this
issue, some authors claiming a mild FIP effect \citep{Brickhouse:01,
Audard:03, Argiroffi:03, Gu:06}, or even an inverse effect
\citep{Bauer:96}, and others claiming no effect at all
\citep{Favata:97, Audard:01, Mewe:01}. Some of this disagreement may
stem from the use of different solar abundances, which is why we have
homogenized all measurements here.  As indicated in
\S\,\ref{sec:abundances}, most authors agree that nitrogen is
overabundant. The average abundance of the remaining high-FIP elements
(C, O, Ar, Ne) is $-0.30 \pm 0.04$, while that of the low-FIP elements
(Ca, Ni, Mg, Fe, Si) is $+0.01 \pm 0.03$, although it is likely that
the true uncertainty is larger than indicated by these statistical
errors. Sulfur has been excluded since its first ionization potential
is close to the conventional dividing line of 10~eV.  The FIP effect
is thus clear in Capella: low-FIP elements are enhanced by a factor of
$\sim$2 relative to high-FIP elements (see Figure~\ref{fig:coronal}).
Furthermore, the evidence in the Sun indicates that high-FIP elements
show better average agreement with the photospheric composition
\citep[][and references therein]{Meyer:93, Feldman:07}. If the same
were true in Capella, then the coronal abundance as measured from the
high-FIP elements supports the photospheric determination of
\cite{McWilliam:90}, which gives an overall metallicity of [m/H] $=
-0.34 \pm 0.07$. Under the same assumption, the enhancement of N
relative to the high-FIP (or photospheric) composition would also be
about a factor of 2, or slightly higher.  Considering that this
measure is diluted to some degree by the flux contribution from the
(presumably normal composition) secondary, it appears to agree at
least qualitatively with model predictions by \cite{Bertelli:08},
which indicate a N enhancement of a factor of $\sim$3.1 for a post
first dredge-up star. The average C/N ratio from
Table~\ref{tab:coronal} (note that since both C and N have high FIP,
any residual FIP effect is more likely to cancel out) is approximately
0.52, though with a large uncertainty of at least 0.2--0.3 due to the
scatter in C. An independent and perhaps more accurate estimate was
reported by \cite{Schmitt:02} as C/N $= 0.91 \pm 0.04$.  The ratios
predicted by the \cite{Bertelli:08} models are 0.67 after first
dredge-up and 3.27 for the unevolved secondary, again roughly
consistent with the measured values for the combined flux. Other
models considered in \S\,\ref{sec:evolution} give similar values.

Although a somewhat clearer and consistent picture than one might
glean from a casual look at the literature seems to emerge from the
above regarding coronal abundances in Capella, uncertainties are still
rather large and our understanding of heating and element separation
in the outer layers of stellar atmospheres is still limited.

\clearpage

\begin{deluxetable}{lcrrccccc}
\tabletypesize{\scriptsize}
\tablewidth{0pc}
\tablecaption{CfA radial velocity measurements of Capella.\label{tab:cfarvs}}
\tablehead{\colhead{HJD} & 
\colhead{} & 
\colhead{$RV_{\rm A}$} & 
\colhead{$RV_{\rm B}$} & 
\colhead{$\sigma_{\rm A}$} & 
\colhead{$\sigma_{\rm B}$} & 
\colhead{$(O\!-\!C)_{\rm A}$} & 
\colhead{$(O\!-\!C)_{\rm B}$} & 
\colhead{} \\
\colhead{\hbox{~~(2,400,000$+$)~~}} & 
\colhead{Julian Year} & 
\colhead{(\kms)} & 
\colhead{(\kms)} & 
\colhead{(\kms)} & 
\colhead{(\kms)} & 
\colhead{(\kms)} & 
\colhead{(\kms)} & 
\colhead{Phase}}
\startdata
 46475.6179\dotfill & 1986.121 & 47.98 &  8.93 & 0.61 & 1.21 & $-$0.42 & $-$1.53 & 0.8781 \\
 46494.5383\dotfill & 1986.173 & 53.18 &  5.27 & 0.68 & 1.37 & $-$0.65 & $+$0.30 & 0.0600 \\
 50388.7228\dotfill & 1996.834 &  3.65 & 55.02 & 0.54 & 1.08 & $-$0.01 & $-$0.62 & 0.4963 \\
 50391.7233\dotfill & 1996.843 &  4.42 & 56.57 & 0.43 & 0.86 & $+$0.45 & $+$1.25 & 0.5251 \\
 50438.8498\dotfill & 1996.972 & 55.98 &  4.68 & 0.56 & 1.12 & $+$0.57 & $+$1.31 & 0.9781 \\ [-1.5ex]
\enddata
\tablecomments{Table~\ref{tab:cfarvs} is presented in its entirety in the
electronic edition of the \emph{Astrophysical Journal}. A portion is
shown here for guidance regarding its form and content.}
\end{deluxetable}

\clearpage

\begin{landscape}
\begin{deluxetable*}{lccccccccccc}
\tabletypesize{\scriptsize}
\tablewidth{0pc}
\tablecaption{Spectroscopic orbital solutions from published velocities for Capella.\label{tab:historicalorbits}}

\tablehead{\colhead{} & \colhead{$\gamma$} & \colhead{Prim/Sec offset} & \colhead{$K_{\rm A}$} &
 \colhead{$K_{\rm B}$} & \colhead{$T$} & \colhead{$\sigma_{\rm A}$} & \colhead{$\sigma_{\rm B}$} &
 \colhead{} & \colhead{} & \colhead{Time span} \\
\colhead{~~~~~~~~~~~~~~~~~~Source~~~~~~~~~~~~~~~~~~} & \colhead{(\kms)} & \colhead{(\kms)} &
 \colhead{(\kms)} & \colhead{(\kms)} & \colhead{(HJD$-2,\!400,\!000$)} &
 \colhead{(\kms)} & \colhead{(\kms)} & \colhead{$N_{\rm A}$} & \colhead{$N_{\rm B}$} & \colhead{(days)}}
\startdata
\cite{Campbell:01}\dotfill & +30.15~$\pm$~0.14 & \nodata & 25.89~$\pm$~0.18 & \nodata & 14865.54~$\pm$~0.14 & 0.77 & \nodata & 31 & \nodata & 1486.0 \\
\cite{Newall:00}\dotfill   & +27.19~$\pm$~0.55 & \nodata & 26.89~$\pm$~0.83 & \nodata & 14969.34~$\pm$~0.48 & 2.20 & \nodata & 23 & \nodata & \phn121.0 \\
\cite{Goos:08}\dotfill     & +30.67~$\pm$~0.27 &    0    & 25.83~$\pm$~0.37 & 36.23~$\pm$~1.45 & 17466.50~$\pm$~0.25 & 1.56 & 5.41 & 35 & 19 & 1004.3 \\
\cite{Sanford:22}\dotfill  & +29.66~$\pm$~0.08 &    0    & 27.06~$\pm$~0.20 & 34.34~$\pm$~1.49 & 23085.33~$\pm$~0.14 & 0.27 & 3.36 & 7 & 4 & \phn\phn55.9 \\
\cite{Struve:39}\dotfill   & +31.44~$\pm$~0.17 &    0    & 26.02~$\pm$~0.24 & 31.09~$\pm$~0.82 & 28180.59~$\pm$~0.15 & 1.01 & 3.10 & 36 & 16 & \phn753.7 \\
\cite{Struve:53}\tablenotemark{a}\dotfill   & +29.30~$\pm$~0.12 & \nodata & 26.46~$\pm$~0.17 & \nodata & 31821.46~$\pm$~0.12 & 0.83 & \nodata & 45 & \nodata & 8142.9 \\
\cite{Wright:54}\tablenotemark{b}\dotfill   & \nodata & \nodata & \multicolumn{2}{c}{53.23~$\pm$~0.63} & 34214.36~$\pm$~0.50 & \multicolumn{2}{c}{2.42} & \multicolumn{2}{c}{19} & \phn802.9 \\
\cite{Batten:75}\tablenotemark{c}\dotfill   & +29.50~$\pm$~0.15 & \nodata & 26.10~$\pm$~0.20 & \nodata & 42119.27~$\pm$~0.15 & 0.64 & \nodata & 18 & \nodata & \phn 274.0 \\
\cite{Shen:85}\tablenotemark{d}\dotfill     & +29.57~$\pm$~0.14 & +0.84~$\pm$~0.45 & 26.35~$\pm$~0.19 & 29.75~$\pm$~0.59 & 44927.73~$\pm$~0.11 & 0.86 & 2.58 & 42 & 39 & 1534.8 \\
\cite{Beavers:86}\dotfill  & +29.87~$\pm$~0.13 &    0    & 26.14~$\pm$~0.18 & 30.43~$\pm$~0.50 & 45031.92~$\pm$~0.12 & 1.33 & 3.85 & 105 & 91 & 1902.9 \\
\cite{Shcherbakov:90}\tablenotemark{e}\dotfill & +31.66~$\pm$~0.59 & \nodata & 25.48~$\pm$~0.83 & \nodata & 46593.65~$\pm$~0.51 & 2.92 & \nodata & 29 & \nodata & \phn957.6 \\
\cite{Shcherbakov:90}\tablenotemark{f}\dotfill & +28.79~$\pm$~0.87 & \nodata & 23.0~$\pm$~1.1 & \nodata & 46594.46~$\pm$~0.86 & 3.62 & \nodata & 24 & \nodata & \phn957.6 \\
\cite{Batten:91}\dotfill   & +30.50~$\pm$~0.12 &    0    & 26.29~$\pm$~0.17 & 27.06~$\pm$~0.46 & 42119.37~$\pm$~0.11 & 0.46 & 1.31 & 15 & 15 & \phn211.7 \\
\cite{Barlow:93}\dotfill   & +29.14~$\pm$~0.10 &    0    & 25.96~$\pm$~0.11 & 27.70~$\pm$~0.41 & 44615.96~$\pm$~0.12 & 0.47 & 1.38 & 23 & 14 & 4532.6 \\
\cite{Katsova:98}\tablenotemark{f}\dotfill  & +30.36~$\pm$~0.47 & $-$6.44~$\pm$~0.95 & 22.43~$\pm$~0.61 & 35.83~$\pm$~1.21 & 48568.57~$\pm$~0.34 & 2.77 & 2.64 & 40 & 16 & 1562.5 \\
CfA observations (this work)\dotfill   & +29.66~$\pm$~0.04 & +0.27~$\pm$~0.08 & 25.96~$\pm$~0.04 & 26.27~$\pm$~0.09 & 50857.21~$\pm$~0.03 & 0.46 & 0.91 & 162 & 162 & 5037.2 \\ [-1.5ex]
\enddata

\tablecomments{The following radial-velocity measurements were
rejected in computing these solutions: JD $2,\!445,\!002.6506$
\citep[][primary]{Shen:85}; $2,\!444,\!930.825$, $2,\!445,\!366.634$
\citep[][primary]{Beavers:86}; $2,\!445,\!315.781$,
$2,\!445,\!315.793$, $2,\!445,\!320.718$
\citep[][secondary]{Beavers:86}; $2,\!446,\!345.71$ \citep[][primary
from He~I~$\lambda$10830 measurements]{Shcherbakov:90};
$2,\!449,\!242.545$, $2,\!449,\!315.281$, $2,\!449,\!316.458$,
$2,\!449,\!316.480$ \citep[][primary]{Katsova:98}; and $2,\!449,\!443.384$
\citep[][secondary]{Katsova:98}.}

\tablenotetext{a}{The date of first of the observations on 1951
January 2 has been corrected, following \cite{Heintz:75}.}

\tablenotetext{b}{The measurements were published only in the form of
velocity \emph{differences}, and dates were given modulo the orbital
period (original dates not available).}

\tablenotetext{c}{The velocities measured by these authors were
superseded by those of \cite{Batten:91}, who re-measured most of the
original photographic plates taken by \cite{Batten:75}.}

\tablenotetext{d}{We consider the velocities measured by these authors
to be superseded by those reported by \cite{Beavers:86}, who added one
more year of observations.}

\tablenotetext{e}{The elements reported here are based on velocities
measured from photospheric features in Capella.}

\tablenotetext{f}{The elements reported here are based on velocities
measured from the He~I~$\lambda$10830 line, which is not
photospheric.}

\end{deluxetable*}
\clearpage
\end{landscape}

\begin{deluxetable*}{lcrrccc}
\tabletypesize{\scriptsize}
\tablewidth{0pc}
\tablecaption{Historical radial velocity measurements of the primary component of Capella.\label{tab:historicalrvs}}
\tablehead{
\colhead{JD} & 
\colhead{} & 
\colhead{$RV_{\rm A}$} & 
\colhead{$\sigma_{\rm A}$} & 
\colhead{$(O\!-\!C)_{\rm A}$} & 
\colhead{} &
\colhead{} \\
\colhead{\hbox{~~(2,400,000$+$)~~}} & 
\colhead{Julian Year} & 
\colhead{(\kms)} & 
\colhead{(\kms)} & 
\colhead{(\kms)} & 
\colhead{Phase} &
\colhead{Source}
}
\startdata
 13804.0360\dotfill & 1896.671 &  36.50 &  0.75 &  $-$0.72 &  0.7939 &  1 \\
 13820.0050\dotfill & 1896.715 &  53.90 &  0.75 &  $-$0.84 &  0.9474 &  1 \\
 13837.0030\dotfill & 1896.761 &  50.40 &  0.75 &  $+$0.31 &  0.1108 &  1 \\
 13839.0290\dotfill & 1896.767 &  47.00 &  0.75 &  $-$0.90 &  0.1303 &  1 \\
 13876.8650\dotfill & 1896.870 &   4.30 &  0.75 &  $+$0.15 &  0.4940 &  1 \\[-1.5ex]
\enddata
\tablecomments{Table~\ref{tab:historicalrvs} is presented in its entirety in the
electronic edition of the \emph{Astrophysical Journal}. A portion is
shown here for guidance regarding its form and content.}
\tablerefs{(1) \cite{Campbell:01}; (2) \cite{Newall:00}; (3) \cite{Goos:08};
(4) \cite{Struve:39}; (5) \cite{Struve:53}; (6) \cite{Beavers:86};
(7) \cite{Shcherbakov:90}; (8) \cite{Batten:91}; (9) \cite{Barlow:93}.}
\end{deluxetable*}


\begin{deluxetable*}{lccccccccccc}
\tabletypesize{\scriptsize}
\tablewidth{0pc}
\tablecaption{Interferometric measurements of Capella published in
polar coordinates.\label{tab:interferometry_polar}}
\tablehead{
\colhead{JD} & 
\colhead{} & 
\colhead{$\theta$} & 
\colhead{$\rho$} & 
\colhead{$\sigma_{\theta}$} & 
\colhead{$\sigma_{\rho}$} & 
\colhead{$(O\!-\!C)_{\theta}$} & 
\colhead{$(O\!-\!C)_{\rho}$} & 
\colhead{} &
\colhead{} \\
\colhead{\hbox{~~(2,400,000$+$)~~}} & 
\colhead{Julian Year} & 
\colhead{(\arcdeg)} & 
\colhead{(mas)} & 
\colhead{(\arcdeg)} & 
\colhead{(mas)} & 
\colhead{(\arcdeg)} & 
\colhead{(mas)} & 
\colhead{Phase} &
\colhead{Source}
}
\startdata
 22323.65\dotfill & 1919.996 &  148.0  &  42.8 &  10.00 &  0.54 &  $-$8.66 &  $+$1.16  &  0.696  & 1 \\
 22368.63\dotfill & 1920.119 &  \phn\phn5.0  &  46.9 &  \phn1.73 &  0.80 &  $+$1.46 &  $-$0.35  &  0.129  & 1 \\
 22369.65\dotfill & 1920.122 &  \phn\phn1.0  &  46.2 &  \phn1.44 &  0.67 &  $+$0.94 &  $-$0.16  &  0.138  & 1 \\
 22370.66\dotfill & 1920.125 &  356.4  &  45.4 &  \phn1.44 &  0.67 &  $-$0.08 &  $-$0.09  &  0.148  & 1 \\
 22399.63\dotfill & 1920.204 &  242.0  &  51.7 &  \phn1.73 &  0.80 &  $+$0.91 &  $+$0.00  &  0.427  & 1 \\[-1.5ex]
\enddata
\tablecomments{Table~\ref{tab:interferometry_polar} is presented in
its entirety in the electronic edition of the \emph{Astrophysical
Journal}. A portion is shown here for guidance regarding its form and
content.}
\tablerefs{(1) \cite{Merrill:22}; (2) \cite{Kulagin:70}; (3) \cite{Blazit:77a}; (4) \cite{Hummel:94}}
\end{deluxetable*}


\begin{deluxetable*}{lccccccccccc}
\tabletypesize{\scriptsize}
\tablewidth{0pc}
\tablecaption{Interferometric measurements of Capella published in
Cartesian coordinates.\label{tab:interferometry_xy}}
\tablehead{
\colhead{JD} & 
\colhead{} & 
\colhead{$\Delta x$} & 
\colhead{$\Delta y$} & 
\colhead{$\sigma_{\Delta x}$} & 
\colhead{$\sigma_{\Delta y}$} & 
\colhead{$(O\!-\!C)_{\Delta x}$} & 
\colhead{$(O\!-\!C)_{\Delta y}$} & 
\colhead{} &
\colhead{} \\
\colhead{\hbox{~~(2,400,000$+$)~~}} & 
\colhead{Julian Year} & 
\colhead{(mas)} & 
\colhead{(mas)} & 
\colhead{(mas)} & 
\colhead{(mas)} & 
\colhead{(mas)} & 
\colhead{(mas)} & 
\colhead{Phase} &
\colhead{Source}
}
\startdata
 43773.6980\dotfill & 1978.723 & $+$40.0 &  +19.4  &   6.70 &  0.76 &   $-$8.07 & $-$0.49 &  0.904 & 1 \\
 43792.6460\dotfill & 1978.775 & $+$14.3 &  +50.8  &   6.70 &  1.26 &   $-$1.09 & $+$0.21 &  0.086 & 1 \\
 43798.6290\dotfill & 1978.792 & \phn$-$2.8 &  +47.9  &   4.02 &  1.26 &   $-$1.00 & $+$0.17 &  0.143 & 1 \\
 43805.6100\dotfill & 1978.811 & $-$23.8 &  +36.4  &   2.68 &  0.51 &   $-$2.39 & $-$0.28 &  0.210 & 1 \\
 43813.5880\dotfill & 1978.832 & $-$34.2 &  +16.8  &   5.36 &  1.26 &   $+$4.85 & $+$0.48 &  0.287 & 1 \\[-1.5ex]
\enddata
\tablecomments{Table~\ref{tab:interferometry_xy} is presented in
its entirety in the electronic edition of the \emph{Astrophysical
Journal}. A portion is shown here for guidance regarding its form and
content.}
\tablerefs{(1) \cite{Koechlin:79}; (2) \cite{Koechlin:83}; (3)
\cite{Baldwin:96}; (4) \cite{Kraus:05}.}
\end{deluxetable*}

\clearpage

\begin{deluxetable*}{lccccccccccc}
\tabletypesize{\scriptsize}
\tablewidth{0pc}
\tablecaption{Speckle and imaging measurements of Capella.\label{tab:speckle}}
\tablehead{
\colhead{JD} & 
\colhead{} & 
\colhead{$\theta$} & 
\colhead{$\rho$} & 
\colhead{$\sigma_{\theta}$} & 
\colhead{$\sigma_{\rho}$} & 
\colhead{$(O\!-\!C)_{\theta}$} & 
\colhead{$(O\!-\!C)_{\rho}$} & 
\colhead{} &
\colhead{} \\
\colhead{\hbox{~~(2,400,000$+$)~~}} & 
\colhead{Julian Year} & 
\colhead{(\arcdeg)} & 
\colhead{(mas)} & 
\colhead{(\arcdeg)} & 
\colhead{(mas)} & 
\colhead{(\arcdeg)} & 
\colhead{(mas)} & 
\colhead{Phase} &
\colhead{Source}
}
\startdata
 41034.6700\dotfill & 1971.224 & \nodata &  53.00 &  \nodata  &  1.36 &  \nodata  &   $-$0.85 &   0.572 & 1 \\
 41034.6700\dotfill & 1971.224 & \nodata &  54.00 &  \nodata  &  1.36 &  \nodata  &   $+$0.15 &   0.572 & 1 \\
 41034.6700\dotfill & 1971.224 & \nodata &  55.00 &  \nodata  &  1.36 &  \nodata  &   $+$1.15 &   0.572 & 1 \\
 41034.6700\dotfill & 1971.224 & \nodata &  56.00 &  \nodata  &  1.36 &  \nodata  &   $+$2.15 &   0.572 & 1 \\
 42643.1500\dotfill & 1975.628 &   35.00 &  53.00 &  5.76 &  3.40 &     $+$3.75 &   $-$2.78 &   0.035 & 1 \\[-1.5ex]
\enddata
\tablecomments{Table~\ref{tab:speckle} is presented in
its entirety in the electronic edition of the \emph{Astrophysical
Journal}. A portion is shown here for guidance regarding its form and
content.}
\tablerefs{
(1) \cite{Labeyrie:74}; 
(2) \cite{Blazit:77b}; 
(3) \cite{McAlister:77};
(4) \cite{McAlister:78};
(5) \cite{McAlister:79};
(6) \cite{Balega:77};
(7) \cite{McAlister:80};
(8) \cite{Morgan:80};
(9) \cite{McAlister:82a};
(10) \cite{McAlister:82b};
(11) \cite{McAlister:83};
(12) \cite{Dudinov:82};
(13) \cite{Hege:83} and \cite{Bagnuolo:83a};
(14) \cite{Balega:84};
(15) \cite{McAlister:87};
(16) \cite{Balega:85};
(17) \cite{Hartkopf:00};
(18) \cite{Balega:87};
(19) \cite{Bonneau:86};
(20) \cite{McAlister:89};
(21) \cite{McAlister:93};
(22) \cite{McAlister:90};
(23) \cite{Hartkopf:92};
(24) \cite{Horch:99};
(25) \cite{Young:02} (direct imaging rather than speckle);
(26) \cite{Scardia:00};
(27) \cite{Balega:04}.
}
\end{deluxetable*}


\begin{deluxetable*}{lccccc}
\tabletypesize{\scriptsize}
\tablewidth{0pc}
\tablecaption{{\it Hipparcos\/} measurements of Capella and corresponding $O\!-\!C$ residuals.\label{tab:hipparcos}}
\tablehead{\colhead{HJD} & \colhead{} & \colhead{$\Delta v$} & \colhead{$\sigma_{\Delta v}$} & \colhead{$O\!-\!C$} & \colhead{} \\
\colhead{\hbox{~~(2,400,000$+$)~~}} & \colhead{Julian Year} & \colhead{(mas)} & \colhead{(mas)} & \colhead{(mas)} & \colhead{Phase}}
\startdata
 47921.1897\dotfill &   1990.079  &  $-$6.53 &    3.54 &   $-$7.11 &   0.7749 \\
 47921.2922\dotfill &   1990.079  &  $-$0.61 &    2.57 &   $-$1.19 &   0.7759 \\
 47933.1656\dotfill &   1990.111  &  $-$2.90 &    1.85 &   $-$0.99 &   0.8901 \\
 47933.2030\dotfill &   1990.111  &  $-$1.76 &    3.50 &   $+$0.16 &   0.8904 \\
 47965.5793\dotfill &   1990.200  &  $-$0.71 &    1.68 &   $+$0.87 &   0.2017 \\[-1.5ex]
\enddata
\tablecomments{Table~\ref{tab:hipparcos} is presented in its entirety
in the electronic edition of the \emph{Astrophysical Journal}. A
portion is shown here for guidance regarding its form and content.}
\end{deluxetable*}


\begin{deluxetable*}{ll@{~~~~~~}c}
\tablewidth{0pc}
\tablecaption{Summary of data sets used in the orbital solution for Capella, along with typical uncertainties.\label{tab:datasets}}
\tablehead{\colhead{~~~~~~~~~~~~~~~~~~~~~Source~~~~~~~~~~~~~~~~~~~~~} & 
\colhead{Measurements} & \colhead{$\sigma$\tablenotemark{a}}}
\startdata
\noalign{\vskip -4pt}
\sidehead{Spectroscopic observations} \\
\noalign{\vskip -9pt}
~~\cite{Campbell:01}\dotfill & 31 measures of $RV_{\rm A}$ & 0.74 \\
~~\cite{Newall:00}\dotfill & 23 measures of $RV_{\rm A}$ & 2.12 \\
~~\cite{Goos:08}\dotfill & 35 measures of $RV_{\rm A}$ & 1.52 \\
~~\cite{Struve:39}\dotfill & 36 measures of $RV_{\rm A}$ & 0.97 \\
~~\cite{Struve:53}\dotfill & 45 measures of $RV_{\rm A}$ & 0.91 \\
~~\cite{Beavers:86}\dotfill & 105 measures of $RV_{\rm A}$ & 1.27 \\
~~\cite{Shcherbakov:90}\dotfill & 29 measures of $RV_{\rm A}$ & 3.14 \\
~~\cite{Batten:91}\dotfill & 15 measures of $RV_{\rm A}$ & 0.48 \\
~~\cite{Barlow:93}\dotfill & 23 measures of $RV_{\rm A}$ & 0.45 \\
~~CfA (this paper)\dotfill & 162 measures of $RV_{\rm A}$ and $RV_{\rm B}$ & 0.44~,~0.89 \\
\noalign{\vskip -4pt}
\sidehead{Astrometric observations} \\
\noalign{\vskip -9pt}
~~\cite{Merrill:22}\dotfill & 16 measures of $\theta$ and 15 of $\rho$ & 1.50~,~0.62 \\
~~\cite{Kulagin:70}\dotfill & 11 measures of $\theta$ and 12 of $\rho$ & 1.10~,~1.52 \\
~~\cite{Blazit:77a}\dotfill & 3 measures of $\theta$ and $\rho$ & 5.98~,~3.39 \\
~~\cite{Koechlin:79,Koechlin:83}\dotfill & 25 measures of $\Delta x$ and $\Delta y$ & 2.67~,~0.67 \\
~~\cite{Hummel:94}\dotfill & 40 measures of $\theta$ and $\rho$ & 0.13~,~0.11 \\
~~\cite{Baldwin:96}\dotfill & 2 measures of $\Delta x$ and $\Delta y$ & 0.93~,~1.57 \\
~~\hip\ \citep{Perryman:97}\dotfill & 41 measures of $\Delta v$ & 2.82 \\
~~\cite{Young:02}\dotfill & 4 measures of $\rho$ & 2.19 \\
~~\cite{Kraus:05}\dotfill & 5 measures of $\Delta x$ and $\Delta y$ & 0.26~,~0.19 \\
~~Speckle observers\dotfill & 47 measures of $\theta$ and 53 of $\rho$ & 1.18~,~1.19 \\[-1.5ex]
\enddata
\tablenotetext{a}{RMS residual of unit weight from our global fit, in
the natural units of each type of observation (\kms\ for $RV_{\rm A}$
and $RV_{\rm B}$, degrees for $\theta$, and mas for $\rho$, $\Delta
x$, $\Delta y$, and $\Delta v$).}
\end{deluxetable*}

\clearpage

\begin{deluxetable*}{lc}
\tabletypesize{\scriptsize}
\tablecolumns{2}
\tablewidth{0pc}
\tablecaption{Orbital solution for Capella.\label{tab:elements}}
\tablehead{
\colhead{~~~~~~~~~~~~~~~~~~~~~~~~~~~~~Parameter~~~~~~~~~~~~~~~~~~~~~~~~~~~~~~} & \colhead{Value}}
\startdata
\noalign{\vskip -4pt}
\sidehead{Adjusted quantities} \\
\noalign{\vskip -9pt}
~~~~$P$ (days)\dotfill                         &  104.02173~$\pm$~0.00022\phn\phn \\
~~~~$a''$ (mas)\dotfill                        &  56.408~$\pm$~0.025\phn        \\
~~~~$i$ (deg)\dotfill                          &  137.212~$\pm$~0.051\phn\phn  \\
~~~~$\Omega_{\rm J2000}$ (deg)\dotfill         &  40.421~$\pm$~0.064\phn       \\
~~~~$T$ (HJD$-$2,400,000)\dotfill              &  47528.514~$\pm$~0.016\phm{2222} \\
~~~~$\gamma$ (\kms)\dotfill                    &  $+29.653$~$\pm$~0.035\phn\phs   \\
~~~~$K_{\rm A}$ (\kms)\dotfill                 &  26.005~$\pm$~0.036\phn       \\
~~~~$K_{\rm B}$ (\kms)\dotfill                 &  26.260~$\pm$~0.087\phn       \\
~~~~$a''_{\rm phot}$ (mas)\dotfill             &  2.08~$\pm$~0.69              \\
~~~~$\Delta\alpha^*$ (mas)\dotfill             &  $-$0.26~$\pm$~0.80\phs       \\
~~~~$\Delta\delta$ (mas)\dotfill               &  $-$0.30~$\pm$~0.57\phs       \\
~~~~$\Delta\mu_{\alpha}^*$ (mas~yr$^{-1}$)\dotfill & $+$0.32~$\pm$~0.99\phs    \\
~~~~$\Delta\mu_{\delta}$ (mas~yr$^{-1}$)\dotfill   & $-$0.08~$\pm$~0.59\phs    \\
~~~~$f_{\rho}$\tablenotemark{a}\dotfill        &  1.0400~$\pm$~0.0035          \\
~~~~$\Delta_{\rm AB}$ (\kms)\tablenotemark{b}\dotfill           &  $+$0.267~$\pm$~0.079\phs        \\
~~~~$\Delta_1$ $\langle$CfA $-$ \cite{Campbell:01}$\rangle$ (\kms)\dotfill    &  $-$0.48~$\pm$~0.14\phs \\
~~~~$\Delta_2$ $\langle$CfA $-$ \cite{Newall:00}$\rangle$ (\kms)\dotfill      &  $+$2.12~$\pm$~0.45\phs \\
~~~~$\Delta_3$ $\langle$CfA $-$ \cite{Goos:08}$\rangle$ (\kms)\dotfill        &  $-$0.93~$\pm$~0.26\phs \\
~~~~$\Delta_4$ $\langle$CfA $-$ \cite{Struve:39}$\rangle$ (\kms)\dotfill      &  $-$1.82~$\pm$~0.17\phs \\
~~~~$\Delta_5$ $\langle$CfA $-$ \cite{Struve:53}$\rangle$ (\kms)\dotfill      &  $+$0.34~$\pm$~0.14\phs \\
~~~~$\Delta_6$ $\langle$CfA $-$ \cite{Beavers:86}$\rangle$ (\kms)\dotfill     &  $-$0.37~$\pm$~0.13\phs \\
~~~~$\Delta_7$ $\langle$CfA $-$ \cite{Shcherbakov:90}$\rangle$ (\kms)\dotfill &  $-$2.37~$\pm$~0.59\phs \\
~~~~$\Delta_8$ $\langle$CfA $-$ \cite{Batten:91}$\rangle$ (\kms)\dotfill      &  $-$0.90~$\pm$~0.13\phs \\
~~~~$\Delta_9$ $\langle$CfA $-$ \cite{Barlow:93}$\rangle$ (\kms)\dotfill      &  $+$0.50~$\pm$~0.10\phs \\
\sidehead{Derived quantities} \\
\noalign{\vskip -9pt}
~~~~$M_{\rm A}$ ($M_{\sun}$)\dotfill           &  2.466~$\pm$~0.018           \\
~~~~$M_{\rm B}$ ($M_{\sun}$)\dotfill           &  2.443~$\pm$~0.013           \\
~~~~$q\equiv M_{\rm B}/M_{\rm A}$\dotfill      &  0.9903~$\pm$~0.0036          \\
~~~~$a$ ($10^6$ km)\dotfill                    & 110.06~$\pm$~0.23\phn\phn     \\
~~~~$a$ (AU)\dotfill                           & 0.7357~$\pm$~0.0015           \\
~~~~$\pi_{\rm orb}$ (mas)\dotfill              &  76.67~$\pm$~0.17\phn         \\
~~~~Distance (pc)\dotfill                      &  13.042~$\pm$~0.028\phn       \\
~~~~$\mu_{\alpha}^*$ (mas~yr$^{-1}$)\dotfill   &  $+$75.84~$\pm$~0.99\phn\phs  \\
~~~~$\mu_{\delta}$ (mas~yr$^{-1}$)\dotfill     & $-$427.21~$\pm$~0.59\phn\phn\phs \\
~~~~$(\ell_{\rm B}/\ell_{\rm A})_{H_p}$\dotfill & 1.171~$\pm$~0.057            \\[-1.5ex]
\enddata
\tablenotetext{a}{Scale factor for the angular separation measurements by
\cite{Merrill:22} and \cite{Kulagin:70}.}
\tablenotetext{b}{Primary/secondary offset for the CfA velocities, in
the sense $\langle$primary \emph{minus} secondary$\rangle$.}
\end{deluxetable*}

\clearpage

\begin{deluxetable*}{lcccc}
\tablewidth{0pc}
\tabletypesize{\small}
\tablecaption{Relative brightness measurements for Capella.\label{tab:lightratio}}
\tablehead{
\colhead{} &
\colhead{Wavelength} &
\colhead{Bandwidth} &
\colhead{} \\
\colhead{~~~~~~~~~~~~~~~~~~~~~~~~~Source~~~~~~~~~~~~~~~~~~~~~~~~~} & 
\colhead{($\mu$m)} &
\colhead{($\mu$m)} &
\colhead{$\ell_{\rm cool}/\ell_{\rm hot}$}}
\startdata
  \cite{Young:02}\dotfill              & 0.1400 &  0.0200  &   0.059~$\pm$~0.012  \\
  \cite{Young:02}\dotfill              & 0.2750 &  0.0500  &   0.25~$\pm$~0.04   \\
  \cite{Bagnuolo:88}\tablenotemark{a}\dotfill           & 0.4108 &  0.0223  &   0.61~$\pm$~0.02   \\
  \cite{Griffin:86}\dotfill            & 0.4425 &  0.0988  &   \phm{\tablenotemark{b}}0.69~$\pm$~0.05\tablenotemark{b}  \\
  \cite{Hummel:94}\dotfill             & 0.4500 &  0.0200  &   0.77~$\pm$~0.07   \\
  \cite{Bagnuolo:88}\tablenotemark{a}\dotfill           & 0.4669 &  0.0202  &   0.82~$\pm$~0.02   \\
  CfA spectroscopy (this paper)\dotfill& 0.5188 &  0.0045  &   0.677~$\pm$~0.023  \\
  \cite{Hege:83}\tablenotemark{c}\dotfill  & 0.5200 &  0.0100  &   0.64~$\pm$~0.03   \\
  \hip\ (this paper)\dotfill           & 0.5275 &  0.2378  &   0.854~$\pm$~0.041  \\
  \cite{Bagnuolo:88}\tablenotemark{a}\dotfill           & 0.5478 &  0.0253  &   0.93~$\pm$~0.02   \\
  \cite{Bagnuolo:88}\tablenotemark{a}\dotfill           & 0.5478 &  0.0253  &   0.87~$\pm$~0.02   \\
  \cite{Bagnuolo:88}\tablenotemark{a}\dotfill           & 0.5478 &  0.0253  &   0.92~$\pm$~0.02   \\
  \cite{Bagnuolo:88}\tablenotemark{a}\dotfill           & 0.5478 &  0.0253  &   0.91~$\pm$~0.02   \\
  \cite{Hummel:94}\dotfill             & 0.5483 &  0.0200  &   0.87~$\pm$~0.04   \\
  Frost \& Rushforth (1979, priv.\ comm.)\tablenotemark{d}\dotfill  & 0.55?   &  \nodata &   0.65      \\
  \cite{Bagnuolo:83a}\dotfill          & 0.5500 &  0.0100  &   0.82~$\pm$~0.02   \\
  \cite{Bagnuolo:83b}\dotfill          & 0.5520 &  0.0200  &   0.89~$\pm$~0.01   \\
  \cite{Griffin:86}\dotfill            & 0.5544 &  0.0898  &   \phm{\tablenotemark{b}}0.87~$\pm$~0.05\tablenotemark{b} \\
  \cite{Strassmeier:90}\dotfill        & 0.6430 &  0.0084  &   \phm{\tablenotemark{b}}0.91~$\pm$~0.05\tablenotemark{b} \\
  \cite{Hummel:94}\dotfill             & 0.8103 &  0.0200  &   1.05~$\pm$~0.05   \\
  \cite{Baldwin:96}\dotfill            & 0.8300 &  0.0400  &   \phm{\tablenotemark{b}}1.11~$\pm$~0.05\tablenotemark{b} \\
  \cite{Young:99}\dotfill              & 1.3    &  0.16    &   1.38~$\pm$~0.06   \\
  \cite{Kraus:05}\dotfill              & 1.65   &  0.3     &   1.49~$\pm$~0.10   \\
  \cite{DiBenedetto:91}\dotfill        & 2.15   &  0.11    &   1.54~$\pm$~0.35  \\[-1.5ex]
\enddata
\tablenotetext{a}{Measurements have not been transformed to the standard Str\"omgren system.}
\tablenotetext{b}{Uncertainty not originally reported, and assigned arbitrarily for this work.}
\tablenotetext{c}{Supersedes the result of \cite{Cocke:83}.}
\tablenotetext{d}{Cited by \cite{Hege:83}. The passband is uncertain.}
\end{deluxetable*}


\begin{deluxetable*}{lcc}
\tablewidth{0pc}
\tablecaption{Limb-darkened angular diameter measurements for
Capella.\label{tab:angdiam}}
\tablehead{
\colhead{} &
\colhead{$\Theta_{\rm LD}^{\rm A}$} &
\colhead{$\Theta_{\rm LD}^{\rm B}$} \\
\colhead{~~~~~~~~~~~~~~~~~~~~Source\tablenotemark{a}~~~~~~~~~~~~~~~~~~~~} & 
\colhead{(mas)} &
\colhead{(mas)}
}
\startdata
\cite{DiBenedetto:91}\dotfill &  9.5~$\pm$~2.3   & 6.18~$\pm$~0.42 \\
\cite{Hummel:94}\dotfill      &  8.46~$\pm$~0.10 & 6.32~$\pm$~0.30 \\
\cite{Kraus:05}\dotfill       &  9.14~$\pm$~0.62 & 5.92~$\pm$~0.82 \\
\noalign{\vskip 4pt}
\hline
\noalign{\vskip 4pt}
Weighted average\dotfill      &  8.47~$\pm$~0.40 & 6.24~$\pm$~0.23 \\[-1.5ex]
\enddata
\tablenotetext{a}{The original measurements have been homogenized by
uniformly applying limb-darkening coefficients by \cite{vanHamme:93}.}
\end{deluxetable*}

\clearpage

\begin{deluxetable*}{lccc}
\tablewidth{0pc}
\tabletypesize{\scriptsize}

\tablecaption{Summary of photospheric abundance determinations for
Capella.\label{tab:abundances}}

\tablehead{
\colhead{~~~~~Element or ratio~~~~~} &
\colhead{Primary\tablenotemark{a}} &
\colhead{Secondary} &
\colhead{Source}
}
\startdata
${\rm [Si/H]}$\dotfill & $-0.32$ (3) & \nodata & 1 \\
${\rm [Ca/H]}$\dotfill & $-0.26$ (2) & \nodata & 1 \\
${\rm [Sc/H]}$\dotfill & $-0.73$ (1) & \nodata & 1 \\
${\rm [Ti/H]}$\dotfill & $-0.28$ (8) & \nodata & 1 \\
${\rm [V/H]}$\dotfill  & $-0.09$ (2) & \nodata & 1 \\
${\rm [Fe/H]}$\dotfill & $-0.20 \pm 0.22$\phs & \nodata & 1 \\
${\rm [Co/H]}$\dotfill & $-0.02$ (2) & \nodata & 1 \\
${\rm [Ni/H]}$\dotfill & $-0.63$ (7) & \nodata & 1 \\
${\rm [Sr/H]}$\dotfill & $+1.18$: (1) & \nodata & 1 \\
${\rm [Y/H]}$\dotfill  & $-0.20$ (1) & \nodata & 1 \\
${\rm [Zr/H]}$\dotfill & $-0.77$ (1) & \nodata & 1 \\
${\rm [La/H]}$\dotfill & $-0.32$ (1) & \nodata & 1 \\
${\rm [Nd/H]}$\dotfill & $-0.27$ (1) & \nodata & 1 \\
${\rm [m/H]}$\tablenotemark{b}\dotfill  & $-0.34 \pm 0.07$\phs & \nodata & 1 \\
$\log N({\rm Li})$\dotfill &  $1.3 \pm 0.2$ & $3.2 \pm 0.3$ & 2 \\
$^{12}$C/$^{13}$C\dotfill & $27 \pm 4$\phn & \nodata & 3 \\
C/N\tablenotemark{c}\dotfill    & $0.57 \pm 0.06$ & $3.30 \pm 0.16$ & 4 \\[-1.5ex]
\enddata

\tablenotetext{a}{Values in parentheses represent the number of
spectral lines measured by \cite{McWilliam:90}. All abundances
relative to hydrogen have been converted to the scale of
\cite{Grevesse:98}.}  

\tablenotetext{b}{Average metallicity excluding [Sr/H].}

\tablenotetext{c}{Ratios derived from transition layer lines, believed
to be representative of the photospheric values (see text).
Uncertainties for the line fluxes of C\,IV\,$\lambda$1550.8 and
N\,V\,$\lambda$1238.8 used to compute the C/N ratios were not reported
by \cite{Linsky:95} separately for the two components; we have
somewhat arbitrarily assumed here the same \emph{relative} uncertainty
as the ratio calculated from the total flux of each line
\citep[][Table~3]{Linsky:95}, and then doubled that of the primary on
account of its fainter lines (see \S\,\ref{sec:abundances}).}

\tablerefs{
(1) \cite{McWilliam:90}, converted to the abundance scale of \cite{Grevesse:98};
(2) \cite{Pilachowski:92}, recomputed here to account for temperature
    and gravity differences, and NLTE effects.
(3) \cite{Tomkin:76};
(4) \cite{Linsky:95};
}
\end{deluxetable*}

\clearpage

\begin{deluxetable*}{lcccc}
\tabletypesize{\scriptsize}
\tablewidth{0pc}
\tablecaption{Apparent magnitudes, color indices, and inferred
effective temperatures for Capella.\label{tab:temps}}

\tablehead{
\colhead{~~~~~~~~~~~~~Passband~~~~~~~~~~~~~} & 
\colhead{$\ell_{\rm cool}/\ell_{\rm hot}$\tablenotemark{a}} & 
\colhead{Combined light\tablenotemark{b}} & 
\colhead{Primary star} & 
\colhead{Secondary star}
}
\startdata
\noalign{\vskip -4pt}
\sidehead{Apparent magnitudes} \\
\noalign{\vskip -7pt}
~~~$B$ (mag)\dotfill                   & \phm{~(4)}0.715~$\pm$~0.058~(4)  & \phm{~(24)}0.870~$\pm$~0.018~(24)   & 1.820~$\pm$~0.055     & 1.456~$\pm$~0.041 \\
~~~$V$ (mag)\dotfill                   & \phm{~(9)}0.888~$\pm$~0.011~(9)  & \phm{~(24)}0.073~$\pm$~0.014~(24)   & 0.892~$\pm$~0.016     & 0.763~$\pm$~0.015 \\
~~~$R_{\rm C}$ (mag)\dotfill           & \phm{~(1)}0.91~$\pm$~0.05~(1)    & \phm{~(3)}$-$0.346~$\pm$~0.031~(3)\phs & 0.459~$\pm$~0.044     & 0.357~$\pm$~0.042 \\
~~~$I_{\rm C}$ (mag)\dotfill           & \phm{~(2)}1.080~$\pm$~0.035~(2)  & \phm{~(3)}$-$0.734~$\pm$~0.018~(3)\phs & $-$0.022~$\pm$~0.025\phs  & 0.061~$\pm$~0.026 \\
~~~$J_{\rm 2}$ (mag)\dotfill           & \phm{~(1)}1.38~$\pm$~0.06~(1)    & \phm{~(2)}$-$1.337~$\pm$~0.090~(2)\phs & $-$0.745~$\pm$~0.092\phs  & $-$0.396~$\pm$~0.094\phs \\
~~~$H_{\rm 2}$ (mag)\dotfill           & \phm{~(1)}1.49~$\pm$~0.10~(1)    & \phm{~(1)}$-$1.57~$\pm$~0.18~(1)\phs   & $-$1.02~$\pm$~0.19\phs    & $-$0.58~$\pm$~0.19\phs \\
~~~$K_{\rm 2}$ (mag)\dotfill           & \phm{~(1)}1.54~$\pm$~0.35~(1)    & \phm{~(2)}$-$1.795~$\pm$~0.082~(2)\phs & $-$1.25~$\pm$~0.14\phs    & $-$0.78~$\pm$~0.17\phs \\
\sidehead{Color indices} \\
\noalign{\vskip -7pt}
~~~$B\!-\!V$ (mag)\dotfill             & \nodata                & 0.797~$\pm$~0.023        & 0.928~$\pm$~0.057     & 0.693~$\pm$~0.044 \\
~~~$V\!-\!R_{\rm C}$ (mag)\dotfill     & \nodata                & 0.419~$\pm$~0.034        & 0.433~$\pm$~0.047     & 0.406~$\pm$~0.045 \\
~~~$V\!-\!I_{\rm C}$ (mag)\dotfill     & \nodata                & 0.807~$\pm$~0.023        & 0.914~$\pm$~0.030     & 0.702~$\pm$~0.030 \\
~~~$(R\!-\!I)_{\rm C}$ (mag)\dotfill   & \nodata                & 0.388~$\pm$~0.036        & 0.481~$\pm$~0.051     & 0.296~$\pm$~0.049 \\
~~~$V\!-\!J_{\rm 2}$ (mag)\dotfill     & \nodata                & 1.410~$\pm$~0.091        & 1.637~$\pm$~0.093     & 1.159~$\pm$~0.095 \\
~~~$V\!-\!H_{\rm 2}$ (mag)\dotfill     & \nodata                & 1.65~$\pm$~0.18          & 1.91~$\pm$~0.19       & 1.35~$\pm$~0.19 \\
~~~$V\!-\!K_{\rm 2}$ (mag)\dotfill     & \nodata                & 1.868~$\pm$~0.083        & 2.14~$\pm$~0.14       & 1.55~$\pm$~0.18 \\
\sidehead{Temperatures (K) inferred from the color indices, assuming [m/H] $= -0.34 \pm 0.07$} \\
\noalign{\vskip -7pt}
~~~$B\!-\!V$\dotfill                   & \nodata                & 5138~$\pm$~76\phn\phn    & 4862~$\pm$~124\phn    & 5391~$\pm$~128\phn \\
~~~$V\!-\!R_{\rm C}$\dotfill           & \nodata                & 5219~$\pm$~199\phn       & 5142~$\pm$~258\phn    & 5294~$\pm$~263\phn \\
~~~$V\!-\!I_{\rm C}$\dotfill           & \nodata                & 5224~$\pm$~81\phn\phn    & 4927~$\pm$~85\phn\phn & 5582~$\pm$~120\phn \\
~~~$(R\!-\!I)_{\rm C}$\dotfill         & \nodata                & \phm{\tablenotemark{c}}5238~$\pm$~94\tablenotemark{c}\phn\phn       & 4748~$\pm$~195\phn    & out of range \\
~~~$V\!-\!J_{\rm 2}$\dotfill           & \nodata                & 5244~$\pm$~159\phn       & 4861~$\pm$~144\phn    & \phm{\tablenotemark{c}}5806~$\pm$~119\tablenotemark{c}\phn \\
~~~$V\!-\!H_{\rm 2}$\dotfill           & \nodata                & 5423~$\pm$~254\phn       & 5092~$\pm$~223\phn    & 5890~$\pm$~232\phn \\
~~~$V\!-\!K_{\rm 2}$\dotfill           & \nodata                & 5212~$\pm$~105\phn       & 4905~$\pm$~146\phn    & 5657~$\pm$~244\phn \\
~~~Weighted average\dotfill            & \nodata                & 5207~$\pm$~40\phn\phn    & 4910~$\pm$~53\phn\phn & 5610~$\pm$~83\phn\phn \\
\sidehead{Final temperatures (K)} \\
\noalign{\vskip -7pt}
~~~Spectroscopy (\S\,\ref{sec:specnew})\dotfill & \nodata       & \nodata                  & 4900~$\pm$~100\phn    & 5710~$\pm$~100\phn \\
~~~$\Theta_{\rm LD}$, $V$, and $BC_V$ (\S\,\ref{sec:dimensions})\dotfill & \nodata & \nodata & 4970~$\pm$~154\phn  & 5687~$\pm$~130\phn \\
~~~Photometry\tablenotemark{d}\dotfill & \nodata                & 5207~$\pm$~108\phn       & 4910~$\pm$~113\phn    & 5610~$\pm$~130\phn \\
~~~Weighted average\dotfill            & \nodata                & \nodata                  & 4917~$\pm$~67\phn\phn & 5677~$\pm$~68\phn\phn \\[-1.5ex]
\enddata

\tablenotetext{a}{The number of individual values used to form these
weighted averages (from Table~\ref{tab:lightratio}) is indicated in
parentheses.}

\tablenotetext{b}{The number of individual measurements used in each
of these weighted averages is indicated in parentheses. All red and
near-infrared photometry was transformed to the standard system in
column~1 as described in the text.}

\tablenotetext{c}{Color index is slightly out of the range of validity
of the color/temperature calibration used, but the extrapolation is
not considered serious.}

\tablenotetext{d}{Uncertainties increased by adding 100~K in quadrature
to account for possible systematics (see text).}

\end{deluxetable*}

\clearpage

\begin{deluxetable*}{lc@{~~~~~}c}
\tablewidth{0pc}
\tablecaption{Estimates of the rotational broadening for the components of
Capella.\label{tab:vsini}}
\tablehead{
\colhead{} &
\colhead{$v_{\rm A} \sin i$} &
\colhead{$v_{\rm B} \sin i$} \\
\colhead{~~~~~~~~~~~~~~~~~~~~Source~~~~~~~~~~~~~~~~~~~~} &
\colhead{(\kms)} &
\colhead{(\kms)}
}
\startdata
\cite{Herbig:55}\dotfill      &  $< 15$ & $\approx$85  \\
\cite{Franklin:59}\dotfill    &  \nodata & $\sim$10--12\phm{$\sim$} \\
\cite{Boesgaard:71}\dotfill   &  \nodata & $\sim$10--12\phm{$\sim$} \\
\cite{Ayres:80}\dotfill       &  \nodata & $\leq$30 \\
\cite{Shen:85}\dotfill        &  9~$\pm$~5 & 35~$\pm$~4\phn \\
\cite{Griffin:86}\dotfill     &  7~$\pm$~(2--3) & 36~$\pm$~(1--2)\phn \\
\cite{Fekel:86}\dotfill       &  5~$\pm$~2 & 36~$\pm$~3\phn \\
\cite{Huisong:87}\dotfill     &  14~$\pm$~3\phn & \nodata \\
\cite{Strassmeier:90}\dotfill &  3 & 36 \\
\cite{Batten:91}\dotfill      &  \nodata & 36 \\
\cite{Randich:94}\dotfill     &  $\leq$7 & 35 \\
\cite{deMedeiros:95}\tablenotemark{a}\dotfill  &  \nodata & 32.7~$\pm$~3.3\phn \\
\cite{Petrov:96}\dotfill      &  8.5~$\pm$~1.0 & \nodata \\
This paper\tablenotemark{b}\dotfill            &  6.5~$\pm$~1.0 & 36.0~$\pm$~1.5\phn \\[-1.5ex]
\enddata

\tablenotetext{a}{The measurement presumably refers to the hot star. A
10\% uncertainty has been adopted following \cite{deMedeiros:99}.}

\tablenotetext{b}{This refers to the total line broadening (see text).}
\end{deluxetable*}


\begin{deluxetable*}{lc@{~~~~}c}
\tablecolumns{3}
\tablewidth{0pc}
\tablecaption{Physical parameters of Capella.\label{tab:dimensions}}
\tablehead{
\colhead{~~~~~~~~~~~~~~~~Parameter~~~~~~~~~~~~~~~~~} & 
\colhead{Primary} &
\colhead{Secondary}
}
\startdata
~~~~Mass ($M_{\sun}$)\dotfill              &  2.466~$\pm$~0.018 & 2.443~$\pm$~0.013  \\
~~~~$q\equiv M_{\rm B}/M_{\rm A}$\dotfill  & \multicolumn{2}{c}{0.9903~$\pm$~0.0036} \\
~~~~$a$ ($10^6$ km)\dotfill                & \multicolumn{2}{c}{110.06~$\pm$~0.23\phn\phn}  \\
~~~~$a$ (AU)\dotfill                       & \multicolumn{2}{c}{0.7357~$\pm$~0.0015}   \\
~~~~$\pi_{\rm orb}$ (mas)\dotfill          & \multicolumn{2}{c}{76.67~$\pm$~0.17\phn}  \\
~~~~Distance (pc)\dotfill                  & \multicolumn{2}{c}{13.042~$\pm$~0.028\phn}   \\
~~~~Radius ($R_{\sun}$)\dotfill            & 11.87~$\pm$~0.56\phn & 8.75~$\pm$~0.32 \\
~~~~$\log g$ (cgs)\dotfill                 & 2.681~$\pm$~0.041 & 2.942~$\pm$~0.032 \\
~~~~$T_{\rm eff}$ (K)\dotfill              & 4920~$\pm$~70\phn\phn & 5680~$\pm$~70\phn\phn \\
~~~~Luminosity ($L_{\sun}$)\tablenotemark{a}\dotfill        & 79.5~$\pm$~4.8\phn & 72.1~$\pm$~3.6\phn \\
~~~~$BC_V$ (mag)\dotfill                     & $-$0.334~$\pm$~0.063\phs & $-$0.098~$\pm$~0.052\phs \\
~~~~$M_V$ (mag)\dotfill                    & 0.315~$\pm$~0.017 & 0.186~$\pm$~0.016 \\
~~~~$v \sin i$ (\kms)\tablenotemark{b}\dotfill              & 5.0~$\pm$~1.0 & 35.6~$\pm$~1.5\phn \\
~~~~$v_{\rm sync} \sin i$ (\kms)\dotfill   & 3.92~$\pm$~0.19 & 2.89~$\pm$~0.11 \\
~~~~Age (Myr)\tablenotemark{c}\dotfill                      & \multicolumn{2}{c}{540--590} \\[-1.5ex]
\enddata
\tablenotetext{a}{Computed from $V$, $\pi_{\rm orb}$, and $BC_V$ from
\cite{Flower:96}, adopting $M_{\rm bol}^{\sun} = 4.732$ (see
footnote~\ref{foot:mbolsun}).}
\tablenotetext{b}{Accounts for the broadening produced by
macroturbulence (see \S\,\ref{sec:dimensions}).}
\tablenotetext{c}{The values given represent the range from all models
giving good age agreement between the primary and secondary from
Table~\ref{tab:modelsg}, Table~\ref{tab:modelsc}, and
Table~\ref{tab:modelsy}, for $Z \approx 0.008$.}
\end{deluxetable*}

\clearpage

\begin{deluxetable*}{lcccccc}
\tabletypesize{\scriptsize}
\tablewidth{0pc}
\tablecaption{Best fit stellar evolution models from \cite{Girardi:00}
compared with the observations for Capella\label{tab:modelsg}}
\tablehead{
\colhead{} & 
\colhead{Age} & 
\colhead{Mass} & 
\colhead{Temperature} & 
\colhead{Radius} & 
\colhead{Luminosity} &
\colhead{}
\\
\colhead{~~~~~~Star~~~~~~} & 
\colhead{(Myr)} & 
\colhead{($M_{\sun}$)} & 
\colhead{(K)} & 
\colhead{($R_{\sun}$)} & 
\colhead{($L_{\sun}$)} &
\colhead{$\chi^2$}
}
\startdata
\noalign{\vskip -1pt}
\multicolumn{7}{c}{{\bf Case A:} $Z = 0.008$, $\alpha_{\rm ML} = 1.68$, $\alpha_{\rm ov} = 0.25$, no rotation} \\
\multicolumn{7}{c}{Components constrained to have the same age; primary on the first ascent} \\
\noalign{\vskip 4pt}
Primary\dotfill   &  537.0 & 2.457 & 5016 & 11.99 & 80.5 & 2.22 \\
  & & $-0.5\sigma$ & $+1.4\sigma$ & $+0.2\sigma$ & $+0.2\sigma$ & \\
Secondary\dotfill &  537.0 & 2.453 & 5723 &  8.55 & 69.3 & 1.97 \\
  & & $+0.8\sigma$ & $+0.6\sigma$ & $-0.6\sigma$ & $-0.8\sigma$ & \\
\noalign{\vskip 4pt}
\hline
\noalign{\vskip 3pt}
\multicolumn{7}{c}{{\bf Case B:} $Z = 0.008$, $\alpha_{\rm ML} = 1.68$, $\alpha_{\rm ov} = 0.25$, no rotation} \\
\multicolumn{7}{c}{Components constrained to have the same age; primary on the descending branch} \\
\noalign{\vskip 4pt}
Primary\dotfill   &  537.0 & 2.474 & 5039 & 11.94 & 81.2 & 3.24 \\
  & & $+0.4\sigma$ & $+1.7\sigma$ & $-0.1\sigma$ & $+0.4\sigma$ & \\
Secondary\dotfill &  537.0 & 2.453 & 5723 &  8.55 & 69.3 & 1.97 \\
  & & $+0.8\sigma$ & $+0.6\sigma$ & $-0.6\sigma$ & $-0.8\sigma$ & \\
\noalign{\vskip 4pt}
\hline
\noalign{\vskip 3pt}
\multicolumn{7}{c}{Measured values} \\
\noalign{\vskip 4pt}
Primary\dotfill   &         & $2.466 \pm 0.018$ & $4920 \pm 70$ & $11.87 \pm 0.56$ & $79.5 \pm 4.8$ &  \\
Secondary\dotfill &         & $2.443 \pm 0.013$ & $5680 \pm 70$ & \phn$8.75 \pm 0.32$ & $72.1 \pm 3.6$ & \\[-1.5ex]
\enddata
\tablecomments{Differences compared to the model, in the sense
$\langle$model minus observation$\rangle$, are indicated in units of
the observational uncertainty $\sigma$. The measured values are given at the
bottom of the table for reference.}
\end{deluxetable*}


\begin{deluxetable*}{lcccccccc}
\tabletypesize{\scriptsize}
\tablewidth{0pc}
\tablecaption{Best fit Claret stellar evolution models compared with
the observations for Capella\label{tab:modelsc}}
\tablehead{
\colhead{} & 
\colhead{Age} & 
\colhead{Mass} & 
\colhead{Temperature} & 
\colhead{Radius} & 
\colhead{Luminosity} &
\colhead{} & 
\colhead{} & 
\colhead{} 
\\
\colhead{~~~~~~Star~~~~~~} & 
\colhead{(Myr)} & 
\colhead{($M_{\sun}$)} & 
\colhead{(K)} & 
\colhead{($R_{\sun}$)} & 
\colhead{($L_{\sun}$)} &
\colhead{$^{12}$C/$^{13}$C} & 
\colhead{C/N} & 
\colhead{$\chi^2$}
}
\startdata
\noalign{\vskip -1pt}
\multicolumn{9}{c}{{\bf Case A:} $Z = 0.008$, $\alpha_{\rm ML} = 1.63$, $\alpha_{\rm ov} = 0.20$, no rotation} \\
\multicolumn{9}{c}{Mean age $\bar{\tau} = 553.2$ Myr, $\Delta\tau = 1.3$ Myr = 0.2\%; primary on the descending branch} \\
\noalign{\vskip 4pt}
Primary\dotfill   &  553.9 & 2.466 & 4939 & 12.16 & 77.8 & 19.18   & 0.64 & 0.48 \\
  & &               & $+0.3\sigma$ & $+0.5\sigma$ & $-0.4\sigma$ & $-2.0\sigma$ & $+1.2\sigma$ & \\
Secondary\dotfill &  552.6 & 2.443 & 5736 &  8.49 & 69.0 & 85.01   & 3.27 & 2.02 \\
  & &               & $+0.8\sigma$ & $-0.8\sigma$ & $-0.9\sigma$ & \nodata & $-0.2\sigma$ & \\
\noalign{\vskip 4pt}
\hline
\noalign{\vskip 3pt}
\multicolumn{9}{c}{{\bf Case B:} $Z = 0.004$, $\alpha_{\rm ML} = 1.44$, $\alpha_{\rm ov} = 0.10$, no rotation} \\
\multicolumn{9}{c}{Mean age $\bar{\tau} = 464.0$ Myr, $\Delta\tau = 4.5$ Myr =  1.0\%; primary on the first ascent} \\
\noalign{\vskip 4pt}
Primary\dotfill   &  461.7 & 2.466 & 4917 & 12.20 & 76.9 & 19.13   & 1.73 & 0.64 \\
  & &               & $-0.1\sigma$ & $+0.6\sigma$ & $-0.5\sigma$ & $-2.0\sigma$ & $+19.3\sigma$ & \\
Secondary\dotfill &  466.2 & 2.443 & 5694 &  8.78 & 71.6 & 85.01   & 3.27 & 0.06 \\
  & &               & $+0.2\sigma$ & $+0.1\sigma$ & $-0.1\sigma$ & \nodata  & $-0.2\sigma$ & \\
\noalign{\vskip 4pt}
\hline
\noalign{\vskip 3pt}
\multicolumn{9}{c}{{\bf Case C:} $Z = 0.002$, $\alpha_{\rm ML} = 1.30$, $\alpha_{\rm ov} = 0.00$, no rotation} \\
\multicolumn{9}{c}{Mean age $\bar{\tau} = 400.0$ Myr, $\Delta\tau = 4.0$ Myr =  1.0\%; primary on the first ascent} \\
\noalign{\vskip 4pt}
Primary\dotfill   &  398.0 & 2.466 & 4936 & 12.17 & 77.7 & 18.88   & 1.93 & 0.48 \\
  & &               & $+0.2\sigma$ & $+0.5\sigma$ & $-0.4\sigma$ & $-2.0\sigma$ & $+22.7\sigma$ &\\
Secondary\dotfill &  402.0 & 2.443 & 5646 &  9.16 & 75.4 & 85.01   & 3.27 & 2.72 \\
  & &               & $-0.5\sigma$ & $+1.3\sigma$ & $+0.9\sigma$ & \nodata & $-0.2\sigma$ & \\
\noalign{\vskip 4pt}
\hline
\noalign{\vskip 3pt}
\multicolumn{9}{c}{{\bf Case D:} $Z = 0.008$, $\alpha_{\rm ML} = 1.40$, $\alpha_{\rm ov} = 0.20$, no rotation} \\
\multicolumn{9}{c}{Mean age $\bar{\tau} = 588.6$ Myr, $\Delta\tau = 96.5$ Myr = 16.4\%; primary in the clump} \\
\noalign{\vskip 4pt}
Primary\dotfill   &  636.8 & 2.466 & 4892 & 12.43 & 78.2 & 19.18   & 0.64 & 1.24 \\
  & &               & $-0.4\sigma$ & $+1.0\sigma$ & $-0.3\sigma$ & $-2.0\sigma$ & $+1.2\sigma$ &\\
Secondary\dotfill &  540.3 & 2.443 & 5674 &  8.92 & 72.9 & 85.01   & 3.27 & 0.33 \\
  & &               & $-0.1\sigma$ & $+0.5\sigma$ & $+0.2\sigma$ & \nodata & $-0.2\sigma$ & \\
\noalign{\vskip 4pt}
\hline
\noalign{\vskip 3pt}
\multicolumn{9}{c}{{\bf Case E:} $Z = 0.008$, $\alpha_{\rm ML} = 1.63$, $\alpha_{\rm ov} = 0.20$, with rotation} \\
\multicolumn{9}{c}{Mean age $\bar{\tau} = 558.0$ Myr, $\Delta\tau = 1.1$ Myr = 0.2\%; primary on the descending branch} \\
\noalign{\vskip 4pt}
Primary\dotfill   &  558.6 & 2.466 & 4938 & 12.16 & 77.7 & 19.24   & 0.64 & 0.48 \\
  & &               & $+0.3\sigma$ & $+0.5\sigma$ & $-0.4\sigma$ & $-1.9\sigma$ & $+1.2\sigma$ & \\
Secondary\dotfill &  557.5 & 2.443 & 5731 &  8.53 & 69.4 & 85.01   & 3.27 & 1.56 \\
  & &               & $+0.7\sigma$ & $-0.7\sigma$ & $-0.8\sigma$ & \nodata & $-0.2\sigma$ & \\
\noalign{\vskip 4pt}
\hline
\noalign{\vskip 3pt}
\multicolumn{9}{c}{Measured values} \\
\noalign{\vskip 4pt}
Primary\dotfill   &         & $2.466 \pm 0.018$ & $4920 \pm 70$ & $11.87 \pm 0.56$ & $79.5 \pm 4.8$ &  $27 \pm 4$ & $0.57 \pm 0.06$ \\
Secondary\dotfill &         & $2.443 \pm 0.013$ & $5680 \pm 70$ & \phn$8.75 \pm 0.32$ & $72.1 \pm 3.6$ & \nodata & $3.30 \pm 0.16$ \\[-1.5ex]
\enddata
\tablecomments{Differences compared to the model, in the sense
$\langle$model minus observation$\rangle$, are indicated in units of
the observational uncertainty $\sigma$. The measured values are given at the
bottom of the table for reference.}
\end{deluxetable*}

\clearpage

\begin{landscape}
\begin{deluxetable*}{lcccccccccc}
\tabletypesize{\scriptsize}
\tablewidth{0pc}
\tablecaption{Best fit TYCHO stellar evolution models compared with
the observations for Capella\label{tab:modelsy}}
\tablehead{
\colhead{} & 
\colhead{Age} & 
\colhead{Mass} & 
\colhead{Temperature} & 
\colhead{Radius} & 
\colhead{Luminosity} &
\colhead{$\log N({\rm Li})$} & 
\colhead{} &
\colhead{} &
\colhead{} 
\\
\colhead{~~~~~~Star~~~~~~} & 
\colhead{(Myr)} & 
\colhead{($M_{\sun}$)} & 
\colhead{(K)} & 
\colhead{($R_{\sun}$)} & 
\colhead{($L_{\sun}$)} &
\colhead{($\log N({\rm H}) = 12$)} & 
\colhead{$^{12}$C/$^{13}$C} & 
\colhead{C/N} &
\colhead{$\chi^2$}
}
\startdata

\noalign{\vskip -1pt}
\multicolumn{10}{c}{{\bf Case A:} $Z = 0.45 Z_{\sun}$; primary $g_{\rm ML} = 100$; secondary $g_{\rm ML} = 1$, no rotation} \\
\multicolumn{10}{c}{Mean age $\bar{\tau} = 593.4$ Myr, $\Delta\tau = 34.2$ Myr = 5.8\%; primary in the clump} \\
\noalign{\vskip 4pt}
Primary\dotfill   &  610.5 & 2.466 & 4935 & 12.15 & 78.3 &   1.87  & 19.60   & 0.66 & 0.35 \\
  & &               & $+0.2\sigma$ & $+0.5\sigma$ & $-0.2\sigma$ & $+2.8\sigma$ & $-1.8\sigma$ & $+1.5\sigma$ & \\
Secondary\dotfill &  576.3 & 2.443 & 5736 &  8.18 & 64.7 &   3.01  & 82.39   & 3.11 & 8.03 \\
  & &               & $+0.8\sigma$ & $-1.8\sigma$ & $-2.1\sigma$ & $-0.6\sigma$ & \nodata & $-1.2\sigma$ & \\
\noalign{\vskip 4pt}
\hline
\noalign{\vskip 4pt}
\multicolumn{10}{c}{{\bf Case B:} $Z = 0.45 Z_{\sun}$; primary $g_{\rm ML} = 1$; secondary $g_{\rm ML} = 1$, no rotation} \\
\multicolumn{10}{c}{Mean age $\bar{\tau} = 537.4$ Myr, $\Delta\tau = 77.8$ Myr = 14.5\%; primary on the descending branch} \\
\noalign{\vskip 4pt}
Primary\dotfill   &  498.5 & 2.466 & 4989 & 11.91 & 78.7 &   1.94  & 18.96   & 0.69 & 1.01 \\
  & &               & $+0.8\sigma$ & $+0.1\sigma$ & $-0.2\sigma$ & $+3.2\sigma$ & $-2.0\sigma$ & $+2.0\sigma$ & \\
Secondary\dotfill &  576.3 & 2.443 & 5736 &  8.18 & 64.7 &   3.01  & 82.39   & 3.11 & 8.03 \\
  & &               & $+0.8\sigma$ & $-1.8\sigma$ & $-2.1\sigma$ & $-0.6\sigma$ & \nodata & $-1.2\sigma$ & \\
\noalign{\vskip 3pt}
\hline
\noalign{\vskip 3pt}
\multicolumn{10}{c}{Measured values} \\
\noalign{\vskip 4pt}
Primary\dotfill   &         & $2.466 \pm 0.018$ & $4920 \pm 70$ & $11.87 \pm 0.56$ & $79.5 \pm 4.8$ &   $1.3 \pm 0.2$   &  $27 \pm 4$  & $0.57 \pm 0.06$  & \\
Secondary\dotfill &         & $2.443 \pm 0.013$ & $5680 \pm 70$ & \phn$8.75 \pm 0.32$ & $72.1 \pm 3.6$ &   $3.2 \pm 0.3$  & \nodata & $3.30 \pm 0.16$ & \\[-1.5ex]
\enddata
\tablecomments{Differences compared to the model, in the sense
$\langle$model minus observation$\rangle$, are indicated in units of
the observational uncertainty $\sigma$. The measured values are given at the
bottom of the table for reference.}
\end{deluxetable*}
\clearpage
\end{landscape}

\begin{deluxetable*}{lcccc}
\tabletypesize{\small}
\tablewidth{0pc}
\tablecaption{Critical times from tidal theory, according
to models by Claret\label{tab:tidal}}
\tablehead{
\colhead{} & \colhead{$\tau_{\rm sync,A}$} & \colhead{$\tau_{\rm sync,B}$} & 
\colhead{$\tau_{\rm circ}$} & \colhead{Consistency with} \\
\colhead{~~~~~Tidal theory~~~~~} & \colhead{(Myr)} & \colhead{(Myr)} & 
\colhead{(Myr)} & \colhead{observations\tablenotemark{a}} }
\startdata
\noalign{\vskip -1pt}

\multicolumn{5}{c}{{\bf Case A:} $Z = 0.008$, $\alpha_{\rm ML} = 1.63$, $\alpha_{\rm ov} = 0.20$, no rotation} \\
\multicolumn{5}{c}{$\bar{\tau} = 553.2$ Myr; primary on the descending branch} \\
\noalign{\vskip 4pt}
Zahn\dotfill     & 545.8 & 559.8 & 563.6 & yes/yes/no \\
Tassoul\dotfill  & 540.8 & 554.6 & 547.0 & yes/yes/yes \\
\noalign{\vskip 4pt}
\hline
\noalign{\vskip 3pt}
\multicolumn{5}{c}{{\bf Case B:} $Z = 0.004$, $\alpha_{\rm ML} = 1.44$, $\alpha_{\rm ov} = 0.10$, no rotation} \\
\multicolumn{5}{c}{$\bar{\tau} = 464.0$ Myr; primary on the first ascent} \\
\noalign{\vskip 4pt}
Zahn\dotfill     & 464.5 & 471.0 & 469.9 & no/yes/no \\
Tassoul\dotfill  & 460.3 & 466.7 & 466.7 & yes/yes/no \\
\noalign{\vskip 4pt}
\hline
\noalign{\vskip 3pt}
\multicolumn{5}{c}{{\bf Case C:} $Z = 0.002$, $\alpha_{\rm ML} = 1.30$, $\alpha_{\rm ov} = 0.00$, no rotation} \\
\multicolumn{5}{c}{$\bar{\tau} = 400.0$ Myr; primary on the first ascent} \\
\noalign{\vskip 4pt}
Zahn\dotfill     & 401.8 & 407.4 & 406.4 & no/yes/no \\
Tassoul\dotfill  & 396.3 & 401.8 & 400.9 & yes/yes/no \\
\noalign{\vskip 4pt}
\hline
\noalign{\vskip 3pt}
\multicolumn{5}{c}{{\bf Case D:} $Z = 0.008$, $\alpha_{\rm ML} = 1.40$, $\alpha_{\rm ov} = 0.20$, no rotation} \\
\multicolumn{5}{c}{$\bar{\tau} = 588.6$ Myr; primary in the clump} \\
\noalign{\vskip 4pt}
Zahn\dotfill     & 537.0 & 540.8 & 542.0 & yes/no/yes \\
Tassoul\dotfill  & 532.1 & 539.5 & 538.3 & yes/no/yes \\
\noalign{\vskip 4pt}
\hline
\noalign{\vskip 3pt}
\multicolumn{5}{c}{{\bf Case E:} $Z = 0.008$, $\alpha_{\rm ML} = 1.63$, $\alpha_{\rm ov} = 0.20$, with rotation} \\
\multicolumn{5}{c}{$\bar{\tau} = 558.0$ Myr; primary on the descending branch} \\
\noalign{\vskip 4pt}
Zahn\dotfill     & 561.0 & 568.8 & 574.1 & no/yes/no \\
Tassoul\dotfill  & 555.9 & 563.6 & 562.3 & yes/yes/no \\[-1.8ex]
\noalign{\vskip 4pt}
\enddata
\tablecomments{$\tau_{\rm sync,A}$ and $\tau_{\rm sync,B}$ are the
synchronization times, $\tau_{\rm circ}$ is the circularization
time, and $\bar{\tau}$ represents the mean age of the components from
Table~\ref{tab:modelsc}.}
\tablenotetext{a}{Consistency with the observations for each of
$\tau_{\rm sync,A}$, $\tau_{\rm sync,B}$, and $\tau_{\rm circ}$ is
`yes' if $\tau_{\rm sync,A} \le \tau_{\rm A}$ (synchronized primary),
$\tau_{\rm sync,B} \ge \tau_{\rm B}$ (asynchronous secondary), and
$\tau_{\rm circ} \le \bar{\tau}$ (circular orbit).}
\end{deluxetable*}

\clearpage

\begin{landscape}
\begin{deluxetable*}{lccccccccccccc}
\tablewidth{0pc}
\tabletypesize{\scriptsize}
\tablecaption{Coronal abundances for Capella.\label{tab:coronal}}
\tablehead{
\colhead{} &
\colhead{FIP\tablenotemark{a}} &
\colhead{} &
\colhead{} &
\colhead{} &
\colhead{} &
\colhead{} &
\colhead{} &
\colhead{Aud03} &
\colhead{Aud03} &
\colhead{Arg03} &
\colhead{Arg03} &
\colhead{} &
\colhead{} \\
\colhead{~~~Element~~~} &
\colhead{(eV)} &
\colhead{Dra94} &
\colhead{Bri96} &
\colhead{Whi96} &
\colhead{Bau96} &
\colhead{Fav97} &
\colhead{Bri01} &
\colhead{APEC} &
\colhead{MEKAL} &
\colhead{2000} &
\colhead{2001} &
\colhead{Gu06} &
\colhead{Mean}
}
\startdata
${\rm [C/H]}$\dotfill   & 11.3     & \nodata & \nodata & \nodata                  &\nodata  & \nodata & \nodata  & $-$0.39     & $-$0.36     & $-$0.09 & $-$0.12 & $-$0.20 & $-$0.23 \\
${\rm [N/H]}$\dotfill   & 14.5     & $<$ $+$0.06 & \nodata & \nodata              &\nodata  & \nodata & \nodata  & $-$0.06     & $-$0.05     & $+$0.08 & $+$0.09 & $+$0.17 & $+$0.05 \\
${\rm [O/H]}$\dotfill   & 13.6     & $-$0.79 & $-$0.13\tablenotemark{b} & $-$0.07  &\nodata  & \nodata & \nodata  & $-$0.44     & $-$0.39     & $-$0.30 & $-$0.26 & $-$0.25 & $-$0.33 \\
${\rm [Ne/H]}$\tablenotemark{c}\dotfill  & 21.6     & $<$ $-$0.41 & \nodata & $-$0.16              &\nodata  & \nodata & $-$0.58  & $-$0.22     & $-$0.42     & $-$0.30 & $-$0.30 & $-$0.15 & $-$0.30 \\
${\rm [Mg/H]}$\dotfill  & \phn7.6     & $+$0.01 & \nodata & $-$0.17                  &\nodata  & \nodata & $+$0.13  & $+$0.05     & $+$0.02     & $-$0.05 & $-$0.02 & $+$0.08 & $+$0.01 \\
${\rm [Si/H]}$\dotfill  & \phn8.2     & $-$0.07 & $+$0.30 & $-$0.57                  &\nodata  & \nodata & $+$0.11  & $+$0.08     & $+$0.02     & $-$0.06 & $-$0.02 & $+$0.06 & $-$0.02 \\
${\rm [S/H]}$\dotfill   & 10.4     & $-$0.26 & $-$0.06 & $-$0.47                  &\nodata  & \nodata & $-$0.08  & \phm{:}$-$0.69:    & \phm{:}$-$0.94:    & \nodata & \nodata & $-$0.36 & $-$0.25 \\
${\rm [Ar/H]}$\dotfill  & 15.8     & \nodata & $+$0.54: & \nodata                 &\nodata  & \nodata & \nodata  & $<$ $-$0.06 & $<$ $-$0.56 & \nodata & \nodata & $-$0.31 & $-$0.31 \\
${\rm [Ca/H]}$\dotfill  & \phn6.1     & \nodata & \nodata & \nodata                  &\nodata  & \nodata & \nodata  & \phm{:}$-$0.54:    & $<$ $-$1.00 & \nodata & \nodata & $+$0.04 & $+$0.04 \\
${\rm [Fe/H]}$\dotfill  & \phn7.9     & $-$0.17 & $+$0.11 & $+$0.00                  &\phm{:}$-$0.53: & $+$0.00 & $+$0.03  & $-$0.04     & $-$0.05     & $+$0.01 & $+$0.01 & $+$0.07 & $+$0.00 \\
${\rm [Ni/H]}$\dotfill  & \phn7.6     & \nodata & $+$0.26 & $-$0.17                  &\nodata  & \nodata & \phm{:}$-$0.42: & $-$0.30     & $+$0.04     & $+$0.16 & $+$0.20 & $+$0.12 & $+$0.04 \\[-1.5ex]
\enddata

\tablenotetext{a}{First ionization potential. The separation between
low FIP and high FIP is made at 10 eV. Sulfur is near the dividing
line, and is considered neither low nor high.}

\tablenotetext{b}{Weighted average of two determinations.}

\tablenotetext{c}{Strictly speaking, the Ne abundance refers to the
coronal value in the Sun, rather than the photospheric value, since
there are no Ne lines in the solar spectrum.}

\tablecomments{Individual determinations have all been referred to the
solar abundances of \cite{Grevesse:98}. Upper limits and uncertain
values marked with `:' are not considered in any averages. Sources of
abundance determinations are as follows:
Dra94 = \cite{Drake:94};
Bri96 = \cite{Brickhouse:96};
Whi96 = \cite{White:96};
Bau96 = \cite{Bauer:96};
Fav97 = \cite{Favata:97};
Bri01 = \cite{Brickhouse:01};
Aud03 = \cite{Audard:03}, which supersedes the determinations by
\cite{Audard:01}. Results listed are from two independent analyses
(APEC, MEKAL);
Arg03 = \cite{Argiroffi:03}. Results listed are from two independent
sets of observations (2000, 2001);
Gu06 = \cite{Gu:06}. Results listed are from a combined RGS+HETGS
analysis.}

\end{deluxetable*}
\clearpage
\end{landscape}

\end{document}